\documentclass[aps,rmp,twocolumn]{revtex4}

\usepackage{times}
\usepackage{graphicx}
\usepackage{amsmath}
\usepackage{color}
\newcommand{\EQ}[1]{Eq.~(\ref{eq:#1})}

\newcommand{\FIG}[1]{Fig.~\ref{fig:#1}}
\newcommand{\TAB}[1]{Tab.~\ref{tab:#1}}

\newcommand{\SNP}{SNP}

\definecolor{drab}{rgb}{0.59, 0.44, 0.09}

\definecolor{celestialblue}{rgb}{0.29, 0.59, 0.82}

\definecolor{purple}{rgb}{0.459,0.109,0.538}

\definecolor{deepsaffron}{rgb}{1.0, 0.6, 0.2}

\newcommand{\bq}{\begin{equation}}
\newcommand{\eq}{\end{equation}}
\newcommand{\bn}{\begin{eqnarray}}
\newcommand{\en}{\end{eqnarray}}
\newcommand{\xsat}{\bar{x}}
\newcommand{\xavg}{\hat{x}}

\begin{document}
\title{In-vivo mutation rates and fitness landscape of HIV-1}

\author{Fabio Zanini$^{1,2}$, Vadim Puller$^{1}$, Johanna Brodin$^{3}$, Jan Albert$^{3,4}$, Richard A.~Neher,$^{1\ast}$}
\affiliation{$^{1}$Max Planck Institute for Developmental Biology, 72076 T\"ubingen, Germany
$^{2}$Department of Bioengineering, Stanford University, Stanford, USA
$^{3}$Department of Microbiology, Tumor and Cell Biology, Karolinska Institute, Stockholm, Sweden\\
$^4$Department of Clinical Microbiology, Karolinska University Hospital, Stockholm, Sweden\\}
\date{\today}

\begin{abstract}
Mutation rates and fitness costs of deleterious mutations are difficult to
measure \textit{in vivo} but essential for a quantitative understanding 
of evolution. Using whole genome deep sequencing data from longitudinal
samples during untreated HIV-1 infection, we estimated mutation rates and fitness costs in HIV-1 from the temporal dynamics of genetic variation.
At approximately neutral sites, mutations accumulate with a rate of
$1.2\times 10^{-5}$ per site per day, in agreement with the rate measured in cell cultures.
The rate from \texttt{G} to \texttt{A} is largest, followed by the other transitions
\texttt{C} to \texttt{T}, \texttt{T} to \texttt{C}, and \texttt{A} to \texttt{G}, while transversions are more rare.
At non-neutral sites, most mutations reduce virus replication; using a model of mutation selection balance, we estimated the fitness cost of mutations at every site in the HIV-1 genome.
About half of all nonsynonymous mutations have large fitness costs (greater than 10\%),
while most synonymous mutations have costs below 1\%. The cost of synonymous mutations is especially low in most of gag and pol, while much higher costs are observed in important RNA structures and regulatory regions.
The intrapatient fitness cost estimates are consistent across multiple patients, suggesting that the deleterious part of the fitness landscape is universal and explains a large fraction of global HIV-1 group M diversity.
\end{abstract}
\maketitle

\section*{Introduction}
HIV-1 evolves rapidly within individual hosts: mutations allow it to evade immune predation but can also impair viral replication.
Genetic changes arise during reverse transcription, during forward transcription by the human RNA polymerase II, or are caused by the innate immune system \cite{mansky_lower_1995,abram_nature_2010, cuevas_extremely_2015,malim_apobec_2009}. These changes are the source of genetic diversity, from which selection amplifies beneficial variants and filters deleterious mutations. Characterization of the mutation rate matrix and the genome wide landscape of fitness effects is a prerequisite a quantitative understanding of the evolutionary dynamics of HIV and for rational design of both vaccines and resistance proof drugs.

The majority of mutations are deleterious, some mutations are neutral and have little or no effect, and a minority of mutations are beneficial. While beneficial mutations rapidly spread through the virus population within a patient, deleterious mutations stay at low frequency in a balance between mutation and selection. 
Beneficial mutations are often patient-specific and mediate escape from cytotoxic T-lymphocytes (CTL) and neutralizing antibodies \citep{walker_t-cell_2012,goonetilleke_first_2009,bar_early_2012}.
At the same time, substitutions in response to immune selection are expected to lower intrinsic viral fitness; host-specific adaptation is a trade-off between immune evasion and fitness costs of escape mutations.

Since HIV-1 proteins serve the same function in different hosts, the landscape of fitness costs might be expected to be similar in different hosts. 
However, the effect of a particular mutation can depend on other sites in the genome -- a phenomenon known as epistasis \citep{de_visser_empirical_2014}.
Epistasis and interaction between mutation has been observed as compensatory evolution after CTL escape \citep{schneidewind_transmission_2009} or as covariation of amino acids \citep{carlson_phylogenetic_2008,dahirel_coordinate_2011}.
While epistasis is clearly an important aspects of protein fitness landscapes, it is expected to be only a weak effect at short evolutionary distances: \citet{doud_site-specific_2015} have shown that the majority of mutation effects tend to be conserved in mildly diverged influenza virus proteins.
Since sequences from the same HIV-1 subtype differ at only about 10\% of amino acids \citep{li_integrated_2015}, the majority of residues with which a given amino acid interacts will be conserved and the fitness effects of mutations are expected to be similar across HIV strains.
Consistent with such a universal fitness landscape, reversion of CTL escape mutations upon transmission to a new host is common \citep{li_rapid_2007,friedrich_reversion_2004,leslie_hiv_2004} and has been quantified during transmission \citep{carlson_selection_2014} and during chronic infection \citep{zanini_population_2016}.

Two main approaches to estimate fitness costs have been pursued. First, the cost of individual mutations can be quantified by competing mutant and wild-type viruses in cell culture \citep{martinez-picado_hiv-1_2008,parera_hiv-1_2007}. Similar measurements of replication capacity are done routinely for drug resistance testing \citep{petropoulos_novel_2000} and have been used to infer the fitness landscape of the HIV-1 protease and reverse transcriptase \citep{hinkley_systems_2011}. Recently, high-throughout methods have been developed to identify the amino acid preferences or fitness costs at every position in a protein \citep{thyagarajan_inherent_2014,acevedo_mutational_2014,rihn_uneven_2015}.
An alternative approach is to estimate the fitness landscape indirectly from large global collections of sequences \citep{dahirel_coordinate_2011,ferguson_translating_2013}, under the key assumption that high fitness variants are at high frequency in the global HIV-1 population. Either approach has limitations: whereas cell culture experiments are not sensitive to small costs (below 5\%), models based on cross-sectional data are confounded by immune escape because they cannot differentiate between selective sweeps and absence of functional constraints.

Here, we estimate the fitness landscape and the rates and spectrum of mutations of HIV-1 using whole genome deep sequencing data from longitudinal samples \citep{zanini_population_2016}.
In contrast to previous efforts, we determine fitness costs from the \textit{in vivo} intrapatient balance of mutation and selection against deleterious variants. Our estimates are most sensitive for small and moderate costs (between 0.1\% and 10\%), not affected by patterns of immune escape, and not restricted to one single protein: we estimated fitness costs at almost every position of the HIV-1 genome. This direct analysis from intrapatient diversity data can be used to quantify the relationship between sequence conservation across the HIV-1 pandemic and direct fitness costs of mutations.

\section*{Results}
We previously reported whole genome deep sequencing of HIV-1 RNA from 6-12 samples from 9 untreated patients \citep{zanini_population_2016}. RNA was reverse transcribed and amplified in six overlapping fragments and sequenced to high coverage on an Illumina MiSeq. 
Depending on template input, minor variation at frequencies down to 0.3\% could be detected and frequencies could be reliably measured down to about 1\% (see \citet{zanini_population_2016} and Methods below). 
For some of the analyses below, we include one additional patient (p7) described in \citep{brodin_establishment_2016}, for a total of 82 plasma samples. Eight of the ten patients were infected with subtype B, one with subtype C, and one with subtype CRF01\_AE.

We first discuss how we estimated the \textit{in-vivo} mutation rate matrix of HIV-1 from the accumulation of mutations at approximately neutral sites. We then show how these rates can be used to establish a quantitative correspondence between fitness costs and global diversity at non-neutral sites, and present site specific fitness cost estimates of mutations at almost every site in the HIV-1 genome.

\subsection*{Neutral mutation rate matrix}
Mutations at neutral sites accumulate freely over the time of infection and the average genetic distance from the founder sequence of later samples increases linearly with the time since infection. This rate of divergence at neutral sites is precisely the \textit{in vivo} mutation rate \citep{kimura_evolutionary_1968}. 
(Deleterious mutations, in contrast, accumulate more slowly and we will use this saturation to estimate their fitness costs.)

\begin{figure}[htb]
\centering
\includegraphics[width=0.98\columnwidth]{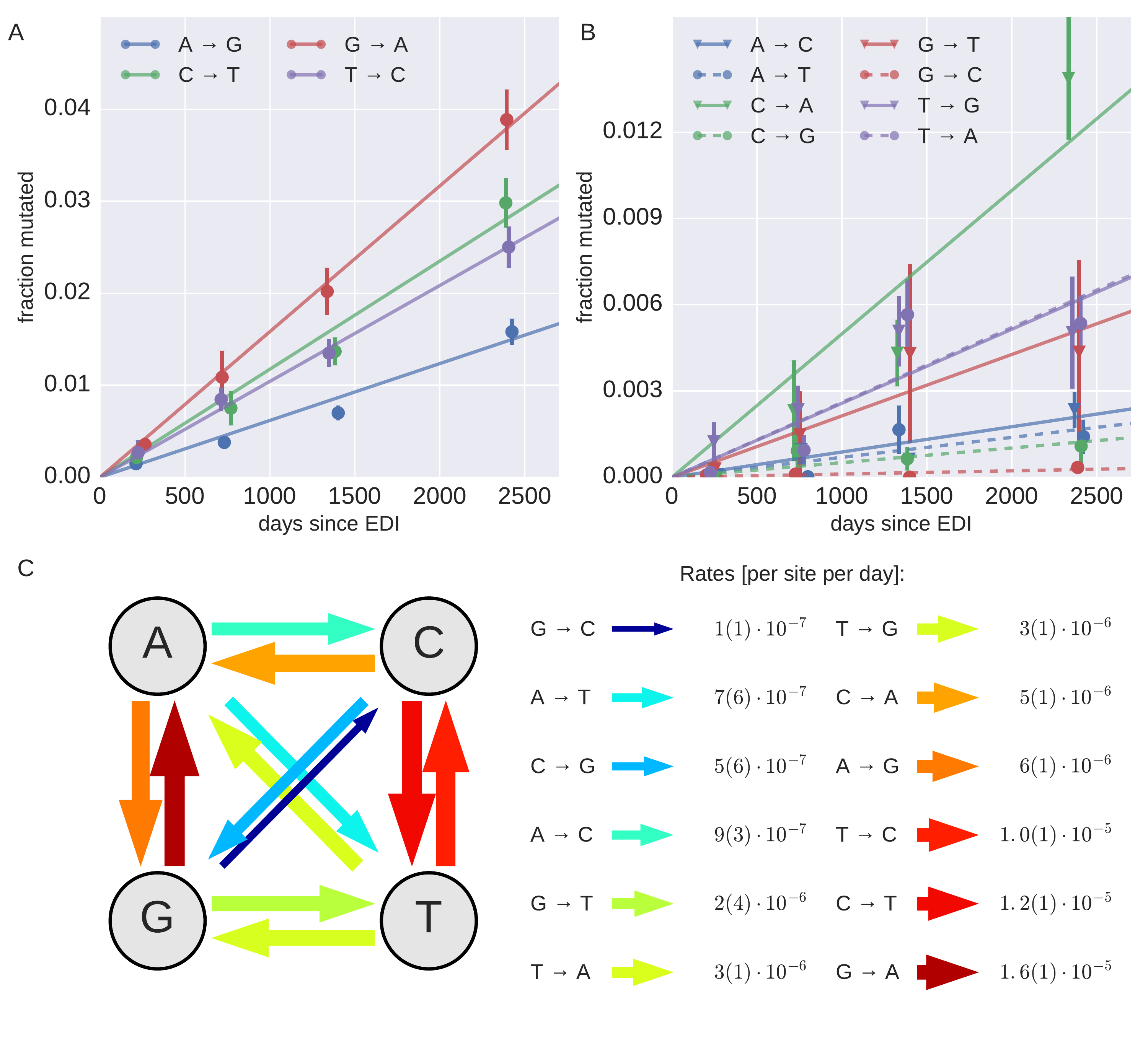}
\caption{{\bf Accumulation of approximately neutral mutations over time.} Panels A\&B show the accumulation of mutations at approximately neutral sites over time averaged over patients p1, p2, p5, p6, p8, p9, p11, for transitions (A) and transversions (B). EDI: estimated date of infection. (C) The slope of the individual regression lines in panel A\&B provide estimates of the in vivo mutation rates. 
Error bars for the estimates, indicated in parenthesis as uncertainties over the last significant digit, are standard deviations over 100 patient bootstraps.}
\label{fig:mutation_rate}
\end{figure}

To estimate the neutral mutation rate, it is crucial to identify a set of positions at which mutations are approximately neutral -- otherwise the mutation rate will be underestimated.
We selected a set of synonymous mutations that (i) are not part of known RNA secondary structures or overlapping reading frames, (ii) are globally unconserved (diversity $>0.3$ bits), (iii) are outside gp120 which has been shown to be sensitive to synonymous mutations and recoding \citep{zanini_quantifying_2013,vabret_large-scale_2014}, and (iv) align to HXB2. 
\FIG{mutation_rate}A and B show the average divergence from the approximate virus founder sequence in this neutral set, for all 12 nucleotide substitutions. We pooled data from patients p1, p2, p5, p6, p8, p9, p11 (those with early samples and without suspected dual infection); the error bars indicate standard deviations over patient bootstraps. 
The data confirm that divergence increases linearly, suggesting that our criteria for approximate neutrality succeed to identify a set of sites that are not strongly affected by selection. We can estimate the mutation rate matrix by linear regression -- indicated by straight lines. 
Transition rates are about 5-fold higher than transversions, while the total mutation rate per site is about $1.2 \cdot 10^{-5}$ per site and day. The highest rate is \texttt{G$\to$A}, while the lowest rates are estimated to be those between Watson-Crick binding partners. The smallest rates cannot by measured accurately because the corresponding mutations are hardly observed. 
The estimated rates are insensitive to the exact criteria used to select the set of neutral positions (see \FIG{mutation_rate_sensitivity}).

The estimated matrix (\FIG{mutation_rate}C) agrees well with previous estimates of HIV-1 mutation rates obtained using lacZ assays in cell culture \cite{abram_nature_2010,mansky_lower_1995}, see \FIG{mutation_rate_comparison}. This quantitative agreement suggests that the average properties of mutations to HIV-1 depend little on the host cell. To obtain sufficient statistics, we measure the rate averaging across many sites; the mutation rates at single sites are known to depend on local sequence context \citep{abbotts_mechanism_1993, lewis_uniquely_1999}. 

The high \texttt{G$\to$A} rate might be partially due to the effect of human deaminases such as APOBEC3G. 
But the \texttt{G$\to$A} rate is consistent with the rate estimated by \citet{abram_nature_2010} who produced virus  with an APOBEG3G negative cell line such that APOBEG3G probably only makes a minor contribution.
We do not observe rates as high as estimated from integrated proviral DNA \citep{cuevas_extremely_2015}. 
The high rate is likely due to the contribution of heavily hypermutated genomes and is discussed below.

Similarly, the functional latent reservoir of HIV-1 is unlikely to bias our estimates of mutations rates. In a recent study of proviral DNA in the same patients, we found that the latent reservoir is an accurate snapshot of the HIV-1 diversity circulating in the year prior to the sample \citep{brodin_establishment_2016}. Hence we don't expect that the accumulation of diversity is delayed in substantial ways by contributions from reactivated latent virus.

\subsection*{Landscape of fitness costs in the HIV-1 genome}
In contrast to neutral mutations, deleterious mutations reduce the replication rate of viruses carrying them. As a result, they accumulate less rapidly. The temporal dynamics of their frequency $x(t)$ is roughly described by
\begin{equation}\label{eq:Langevin}
\frac{d}{dt}x(t) = \mu - sx(t) + \xi(x, t)
\end{equation}
where $\mu$ and $s$ are the mutation rate and fitness cost specific to the \SNP{} in question, respectively \citep{haldane_effect_1937,haigh_accumulation_1978}. The last term $\xi(x,t)$ describes stochastic effects including genetic drift and selection on linked \SNP{}s at other loci in the genome. Frequent recombination within HIV-1 populations \citep{neher_recombination_2010,zanini_population_2016} reduces the effects of linked selection such that \EQ{Langevin} can be a useful approximation.
Depending on whether linked selection or genetic drift dominates the stochastic component, the absolute value of $\xi(x,t)$ is in average proportional to $x$ or $\sqrt{x}$, respectively \citep{Kimura_stochastic_1955,neher_genetic_2013}. By definition, the average of $\xi$ is zero.

Starting with a genetically monomorphic population, the \textit{average} trajectory of a \SNP{} frequency is given by
\begin{equation} \label{eq:sat}
\langle x\rangle = \frac{\mu}{s}(1-e^{-st})
\end{equation}
and saturates at $\xsat = \mu/s$ after a time of order $s^{-1}$ \citep{haldane_effect_1937}. If an appropriate average of the data is available, the fitness cost $s$ can be estimated both from the approach to saturation and the level of saturation $\mu/s$.
Linear accumulation of neutral mutations is recovered in the limit $s\to 0$.
This approach has been generalized to complex fitness landscapes \citep{seifert_framework_2015}.

\EQ{sat} describes the \textit{average} trajectory, but trajectories of individual SNPs are noisy.
To make progress, trajectories at many sites or in many samples need to be averaged in ways that preserves important features of the fitness landscape.

\subsubsection*{Relationship of global conservation and fitness costs}
\begin{figure}
\centering
\includegraphics[width=0.5\textwidth]{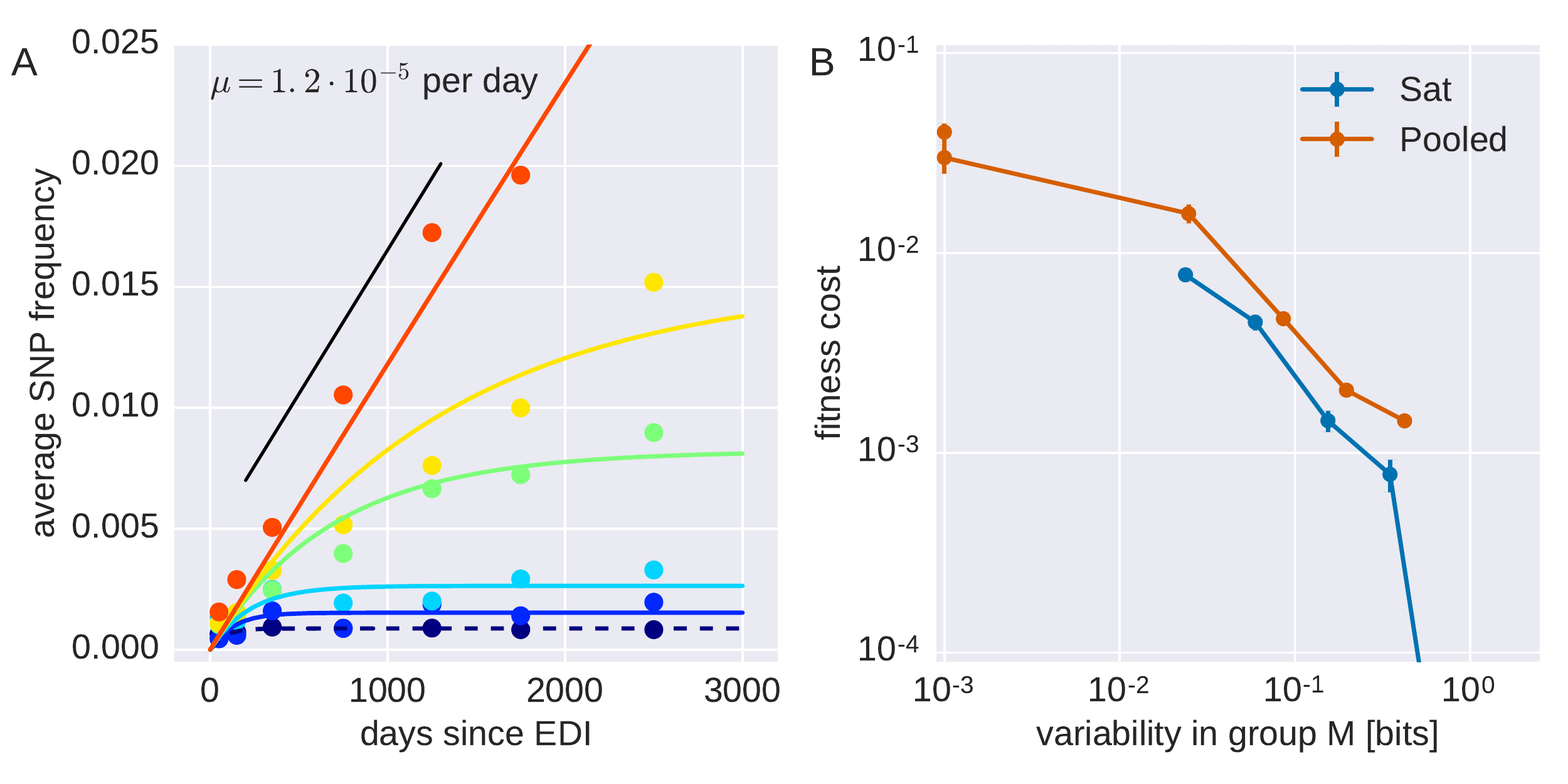}
\caption{{\bf Average intrapatient fitness cost within quantiles of global HIV-1 group M diversity.} (A) Average derived SNP frequencies (1 - frequency of the ancestral state) saturate fast at positions in the conserved quantiles (blue), while intrapatient diversity keeps increasing in variable quantiles (yellow to red). The initial slope is the mutation rate $1.2 \cdot 10^{-5}$ per site per day. The solid lines show fits of \EQ{sat} to the binned data, from which we estimate average selection coefficients shown in panel (B) labeled ``Sat" (this method is not applicable in the most conserved third of the genome). The ``Pooled" line refers to harmonic averages of site-specific cost estimates. Error bars indicate 100 bootstraps over patients: note that while error bars are small, there is substantial variation of fitness costs within each diversity quantile. Positions at which putatively adaptive mutations have swept through the population have been excluded.}
\label{fig:sat}
\end{figure}

In first approximation, conservation of a site across global HIV-1 diversity is expected to be a proxy for high fitness cost of mutations at that site, while mutating a site that is observed in many different states probably doesn't affect fitness much.
To quantify the relationship between conservation and fitness cost $s$, we group sites in the HIV-1 genome by global diversity in group M (measured by Shannon entropy of an alignment column, see Methods).
We chose six groups of equal size, i.e., quantiles of global diversity.
Instead of estimating fitness costs for all three possible mutations at a given site, we estimated one fitness cost parameter for each site as the cost of the typical mutation away from the founder virus sequence (a more elaborate model that includes the 12 different mutation rates is described in \FIG{fitness_cost_complex}). 
For each conservation group, we average the frequencies of non-founder nucleotides over all sites and patient samples in 7 time bins.
These average divergences are indicated by dots in \FIG{sat}A along with a nonlinear least square fit of \EQ{sat} to the data of each quantile (each color indicates a conservation group, blue to red by increasing diversity).
The least conserved group accumulates divergence linearly -- this is consistent with our mutation rate estimates above. With increasing conservation, divergence saturates more rapidly and at lower levels.
We set $\mu = 1.2 \cdot 10^{-5}$ per site per day according to our estimate of the neutral mutation rate and fit a single parameter, the fitness cost $s$, for each group. The estimated average costs and their error bars from 100 bootstraps over patients are shown in \FIG{sat}B as a blue line (``Sat").

The fitness cost of mutations in the least conserved 1/6 of the genome is undetectably small, consistent with neutrality. More conserved sites have higher costs, up to about 1\% for sites where the group M alignment entropy of 0.03 bits. For even more conserved sites, saturation is very fast and we estimated the fitness cost using a different averaging procedure (see below).

Notice that for \EQ{sat} to hold, it is essential that the infection is founded by a single founder sequence.
For this reason, patients p3 and p10 were excluded from this part of the analysis since there is evidence indicating that they were infected by more than one viral variant.
Furthermore, it is important to exclude sites subject to immune selection and sites where the initial nucleotide differs from the global consensus. Otherwise, rapid rise of beneficial mutations driven by CTL escape or reversion increase divergence and result in underestimation of the fitness costs.

\subsubsection*{Site-specific fitness costs in the HIV-1 genome}
In addition to averaging mutation trajectories across multiple sites, we also estimated site-specific fitness costs by averaging data from multiple samples during late infection.
Average frequencies at sites where mutations carry large costs saturate rapidly after a time $1/s$. Frequencies of minor variants in different samples are therefore uncorrelated and can be averaged to increase the accuracy of frequency estimates, which then allows direct estimation of site specific costs $s_i$ from the relation $\xsat_i = \mu/s_i$.

\begin{figure}
    \centering
    \includegraphics[width=0.99\columnwidth]{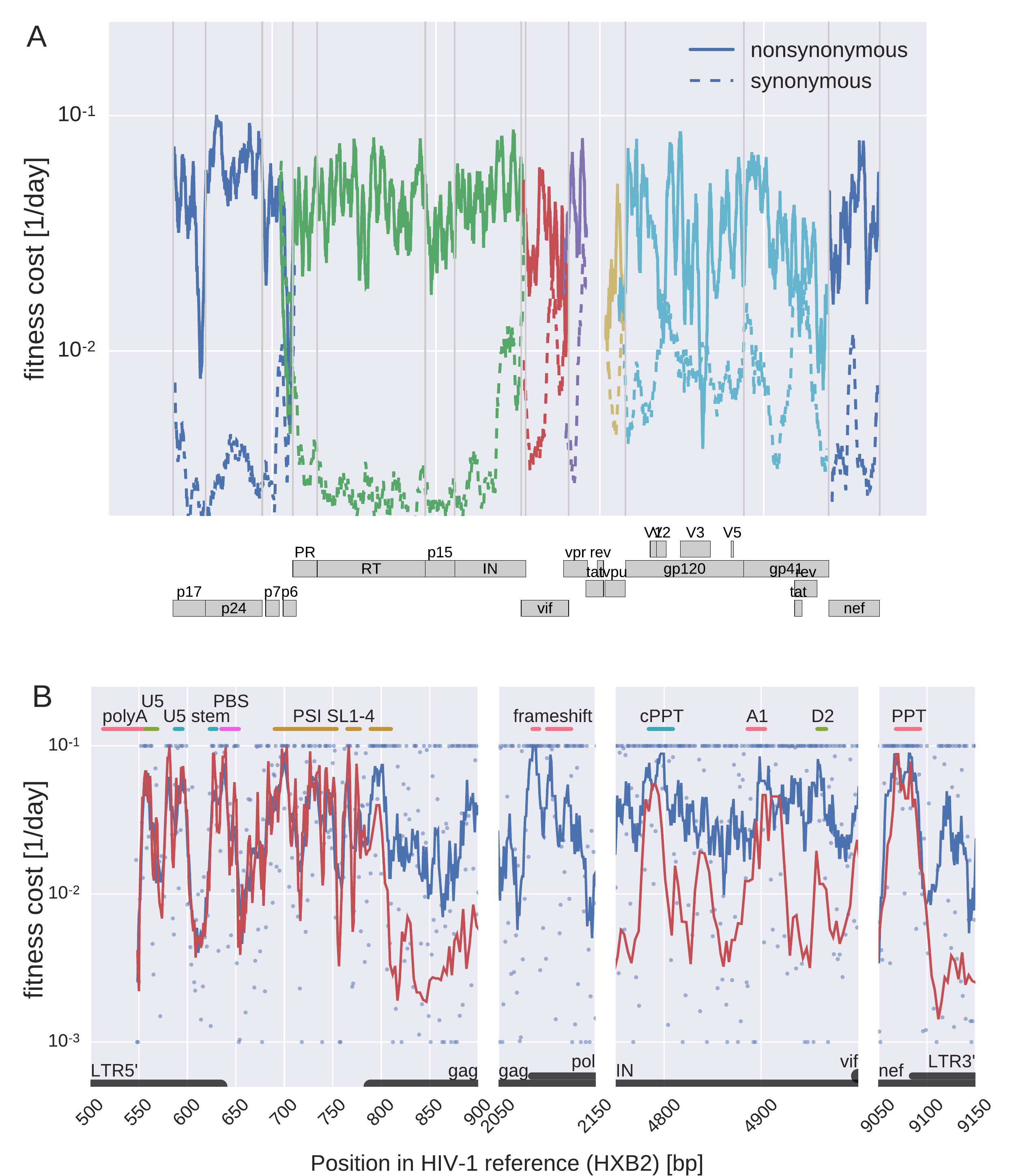}
\caption{{\bf Fitness costs along the HIV-1 genome}. Panel (A) shows fitness costs of synonymous and nonsynonymous mutations in gag, pol, vif, vpu, env, and nef as a geometric sliding average with window size 30. Note that frequency estimates in gp120 are expected to be less accurate due to consistent difficulties amplifying this part of the genome. Panel (B) shows fitness costs in selected regions of the genome that contain important regulatory elements. Blue dots show estimates for individual bases, blue lines show running averages over 8 bases and red lines show running averages excluding bases where mutations cause amino acid changes. PBS: tRNA primer binding site. U5: unique 5' region. SL 1-4 PSI: stem loops of the PSI packaging signal. (c)PPT: (central) poly purine tract. A1, D2: splice sites.}
    \label{fig:along_genome}
\end{figure}

Specifically, we calculate a weighted average frequency from the samples from each patient. The average frequency of nucleotide or amino acid $\alpha$ at position $i$ is then given by
\begin{equation}
\label{eq:xavg}
    \xavg_{i,\alpha} = \frac{1}{\sum_k w_k}\sum_k w_k x_{k,i,\alpha}
\end{equation}
where $x_{k,i,\alpha}$ is the frequency in sample $k$ ($k$ runs over all plasma samples from the patient at least 2 years after infection).
The weight $w_k$ accounts for the variable number of HIV-1 genomes that contributed to the sequencing library as estimated by limiting dilution (see methods below and \citet{zanini_population_2016}).
From individual samples, frequencies above the error rate of 0.002 are assumed to avoid inflation by sequencing and PCR errors (we never observed errors above this level in our control samples).
After averaging samples within patients, we average $\xavg_{i,\alpha}$ over patients and sum all non-consensus nucleotides or amino acids to obtain the average non-consensus frequency $\xavg_i$ for each position $i$ in the HIV-1 genome; the cost at that position is then given by $\mu/\xavg_i$ where $\mu$ is the mutation rate at that position.

As above, we only include data from a particular sample if the majority nucleotide agrees with the global consensus and at which no potential sweep was observed. Without this restriction, the estimated fitness costs would be biased downward by reversions and immune selection.

Notice that although the combined sequencing and PCR error can be up to 0.002 and we don't use counts below this threshold for any \textit{single} sample, pooling many samples allows to estimate much smaller \textit{average} frequencies: if a mutation is present at frequency 0.005 in 10\% of samples, its average frequency is 0.0005. This type of averaging works precisely because frequencies of individual costly mutations are noisy and rare variants are brought to measurable frequencies occasionally by linked selection and sampling.

\FIG{along_genome}A shows fitness costs of mutations at most positions along the HIV-1 genome (including env) separately for synonymous and nonsynonymous mutations: the numerical estimates are available for all sites in the Supplementary Materials. The costs of synonymous and nonsynonymous mutations are clearly different, and distinct peaks are observed at several locations across the genome. Before analyzing these patterns in details (see below), as a consistency check we compared in \FIG{sat}B the average estimates (``Pooled" line) to our previous estimates ``Sat", which take into account the explicit time information of the samples. We found good agreement between the two approaches. To further assess the accuracy of our estimates, in \FIG{fitness_confidence_other_regions} we show the variation in the fitness cost estimate after bootstrapping over patients. The variation is approximately twofold in each direction, so fitness costs above 5\% are clearly separated from costs of 1\% or less. 

Fitness costs estimated from within patient diversity data correlate strongly with global HIV-1 group M diversity (rank correlation $\rho\approx 0.7$ for per site diversity measured by entropy, see \FIG{S_vs_npat}).
Importantly, a particular site contributes to the estimate only if the founder and majority nucleotide in that sample equals the consensus variant. This condition removes any direct signal of cross-sectional diversity.
The correlation increases as intrapatient variation is estimated using more patients (see \FIG{S_vs_npat}), suggesting that fitness costs at individual sites is largely conserved between patients. \FIG{S_vs_npat} also shows scatter plots of global diversity vs fitness costs. 

\begin{figure}
    \centering
    \includegraphics[width=0.95\columnwidth]{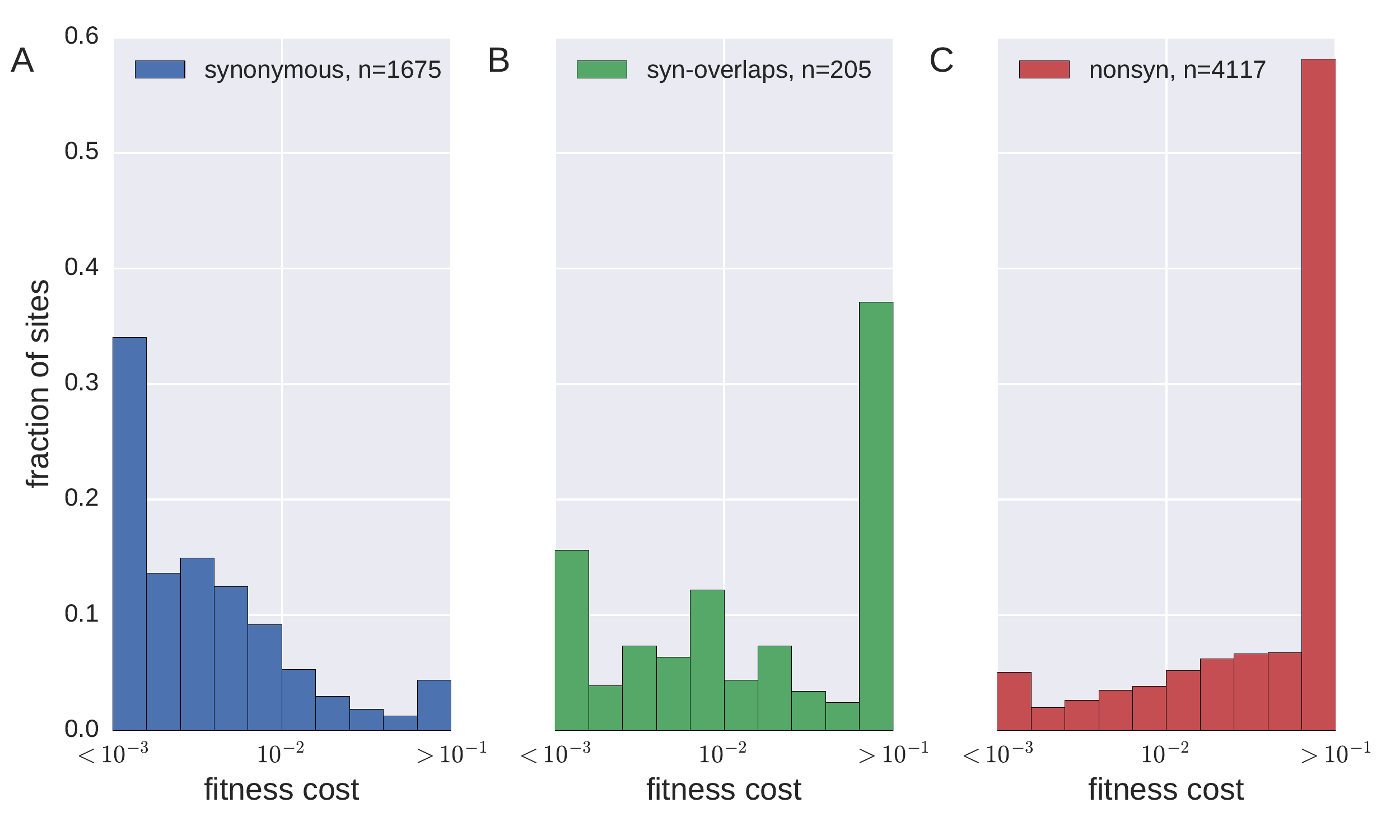}
\caption{{\bf Distributions of fitness costs.} Distributions of (A) synonymous mutations, (B) mutations that are synonymous in one gene but affect another protein in a different reading frame and (C) nonsynonymous mutations (includes codons in gag, pol, vif, vpu, vpr). The extremal bins include all points larger or smaller than the axis boundary. 
}
    \label{fig:codon_selection}
\end{figure}

\subsubsection*{Distributions of fitness costs}
We observe marked differences between the distributions of fitness costs of synonymous and non-synonymous mutations (see \FIG{codon_selection}): about half of all nonsynonymous mutations have estimated fitness costs in excess of 10\%, while the majority of synonymous mutations have fitness costs below 1\%.
The distribution of fitness costs of mutations that are synonymous in one gene, but that affect another gene in a different reading frame, resembles that of nonsynonymous mutations (see \FIG{codon_selection}B).
We estimate about 10\% of synonymous mutations outside env to be highly deleterious; we discuss the specific costs of synonymous mutations in more detail below.

\FIG{gene_distributions} shows the distribution of fitness costs for different genes. In gag and pol, the contrast between synonymous and nonsynonymous mutations is greatest. Synonymous mutations are costly in several isolated regions discussed below but have low fitness effects in much of pol and gag.

The distribution of fitness costs is consistent with those found in other viruses, where typically about 20-40\% of mutations are lethal and another $\sim 40$\% are strongly deleterious with about 30\% being weakly deleterious or neutral  \citep{sanjuan_mutational_2010}. 

\subsubsection*{Fitness costs peak at functional RNA elements}
The HIV-1 genome contains a number of well characterized RNA elements that regulate different stages of the replication cycle. Many of these elements are embedded in protein-coding sequence and because selection reduces genetic diversity \citep{ngandu_extensive_2008,mayrose_synonymous_2013} we expect to estimate higher fitness costs in these regions. Indeed, in \FIG{along_genome}B important regulatory elements are clearly visible as well defined peaks in the running averages of fitness costs along the genome.
In the 5' LTR the largest fitness costs overlap with the hairpin containing the poly-A signal, the U5 sequence \citep{lu_nmr_2011}, the base of the following hairpin, the primer binding site (PBS) and the 1-4 for the PSI element \citep{LANL_annotation}. The frameshift region (slippery sequences plus hairpin), the splice acceptor site A1, and the polypurine tracts (PPT) in integrase and at the 3' LTR show similarly high fitness costs (the TAR element is only partially covered by the sequencing data set and hence not shown here). Mutations within the fourth stem loop of PSI are almost never observed, while synonymous sites are almost free to vary beyond the end of the stem. Synonymous mutations in the RRE are costly, but not as deleterious as those in PPT, the splice acceptor site A1, or the PSI element, indicating a higher evolutionary plasticity. Among the more striking patterns is also the drop in synonymous cost at the beginning of gag. Beyond these known elements, the correlation of fitness costs at synonymous mutations with cross-sectional diversity suggests that there are a number of additional regions with important function on the RNA level, for example a double peak in p24 and three more peaks in pol. While well characterized RNA elements correspond to clear patterns in the estimated fitness costs, RNA secondary structure prediction correlate poorly with fitness costs (see \FIG{correlation_with_RNAstructure} and discussion below).

\subsubsection*{Fitness costs and immune selection}
Among sites that are globally variable (Shannon entropy above 0.1 bits), nonsynonymous mutations are enriched despite having a high fitness cost (cost $>0.03$ per day, odds ratio 15). This enrichment is most pronounced in pol, gag and nef with little enrichment in env. This observation is consistent with host-specific selection pressures (CTL selection) at sites with a large fitness cost; the resulting adaptations revert quickly when transmitted to a new host \citep{li_rapid_2007,friedrich_reversion_2004,leslie_hiv_2004,zanini_population_2016}.

\begin{figure}
    \centering
    \includegraphics[width=0.99\columnwidth]{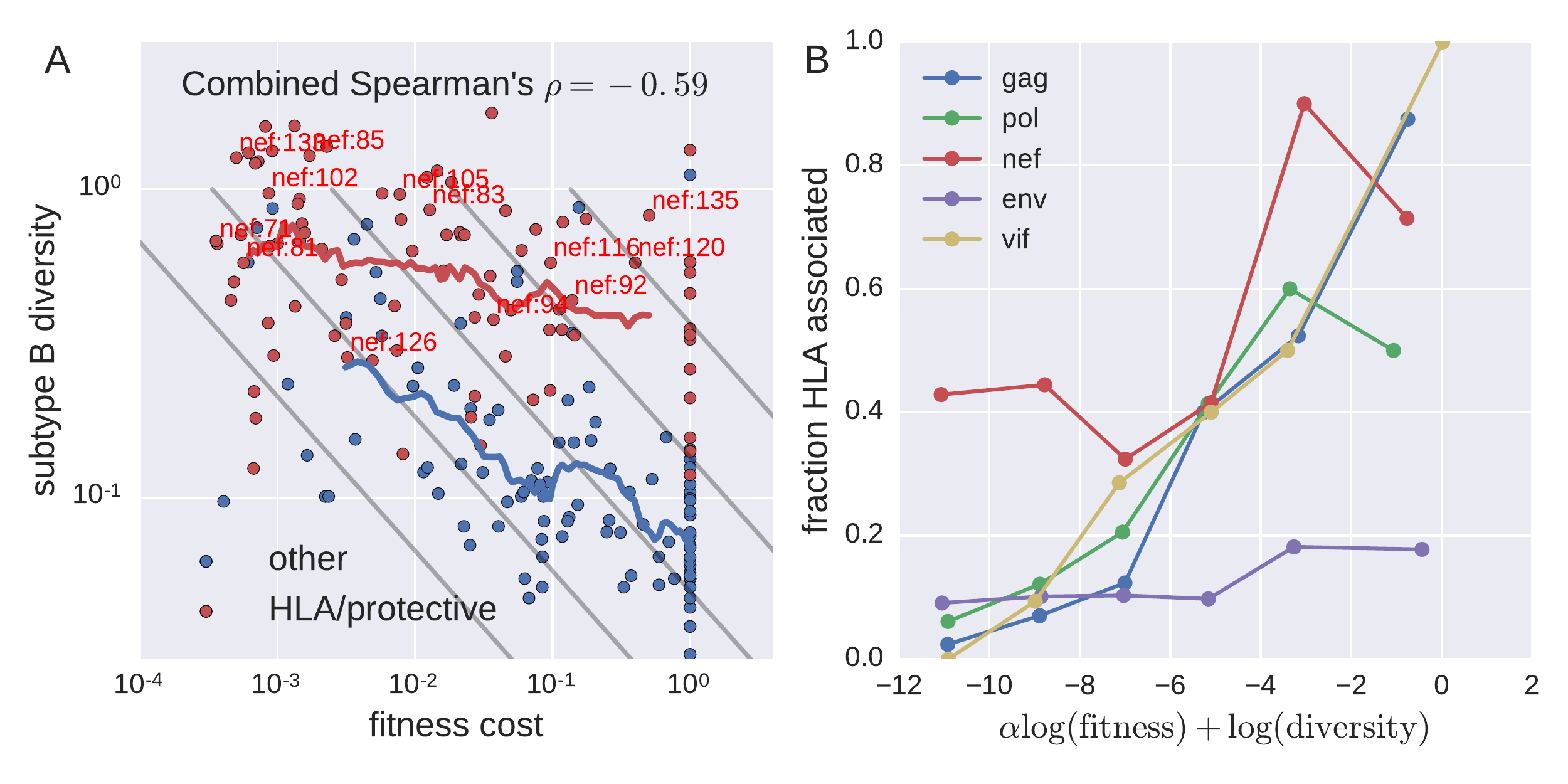}
\caption{{\bf CTL selection blurs the relationship between fitness costs and diversity.} 
(A) The majority of sites in nef with high diversity despite high fitness costs are associated with HLA types (red) \citep{carlson_correlates_2012} or with low viral load (annotated dots) \citep{bartha_genome_genome_2013}. 
(B) Quantification of the fraction of HLA associated sites in bins of increasing diversity and fitness costs (indicated by straight grey lines in (A) with $\alpha=2$). This figure uses data from subtype B patients only.
}
    \label{fig:immune_selection}
\end{figure}

Such patient-specific selection has the potential to blur the relationship between fitness cost and diversity, as shown in \FIG{immune_selection}A for nef (see \FIG{S_vs_npat} for other genes). The majority of sites with high fitness costs and high cross-sectional diversity (upper right corner) have been reported to be associated with HLA type (\citep{carlson_correlates_2012}, shown in red) or with low viral load (\citep{bartha_genome_genome_2013}, annotated dots).
HLA-associated sites that fall into the top right corner of \FIG{immune_selection}A are of particular interest since they are expected to result in virus control \citep{pereyra_HIV_2014}.

To quantify the overrepresentation of HLA associated sites among diverse positions where mutations incur large fitness costs, we plotted the fraction of HLA associated sites in bins indicated by diagonal straight lines in \FIG{immune_selection}A for the genes gag, pol, vif, env, and nef.
Bin boundaries are defined by $\alpha \log(\mathrm{fitness}) + \log(\mathrm{diversity})=\mathrm{const.}$ with $\alpha=2$. 
For all genes test other than env, the fraction of HLA associated sites increases strongly in bins corresponding to high diversity and fitness cost indicating that CTL selection pressure is responsible for global diversity that is deleterious to virus replication.

HLA  associations can only be detected for sites with some global variation. Hence there is a strong ascertainment bias and almost all HLA associated are found in the top half of \FIG{immune_selection}A. Without independent characterization of this bias, a statistical assessment of the relation between CTL selection pressure, fitness cost, and global diversity remains challenging.


\subsubsection*{Fitness costs are weakly correlated with protein disorder and solvent accessibility}
Perturbations to protein structure are expected to reduce virus fitness. Hence mutations that decrease stability, occur in tightly packed regions, or are deeply buried in the protein are expected to incur the greatest fitness costs. Disorder scores and solvent accessibility have been compared to cross-sectional diversity by \citet{li_integrated_2015}.
We correlated these \textit{in-silico} derived scores with intrapatient diversity, finding rank correlation coefficients of about 0.2-0.4 for disorder scores and solvent accessibility. While highly statistically significant, the fraction of variation in diversity explained by these scores is low; this is consistent with previous observations by \citet{meyer_utility_2015}. 
By far the best correlate of fitness cost is cross-sectional conservation, see Table~\ref{tab:correlations}.

The distribution of fitness costs depends strongly on the consensus amino acid. Mutations of cysteins (C), histidines (H), prolines (P), tryptophans (W), and tyrosines (Y) tend to be most costly, while mutations of glutamic acid (E), lysine (K), aspartic acid (D) and arginine (R) are less often very deleterious. These patterns are consistent in gag, pol, and env, see \FIG{amino_acid_costs}. 

\begin{table}[]
    \centering
    \begin{tabular}{|l|c|c|c|c|c|}
\hline  gene &	group M	& subtype B & disorder & accessibility & RNA \\ \hline\hline
gag &	-0.51& -0.59	 &-0.23	 &-0.26  & 0.13 \\
pol &	-0.56& -0.59	 &-0.13	 &-0.31  & 0.09 \\
nef &	-0.54& -0.59	 &-0.30	 &-0.19  & 0.11 \\
env &	-0.47& -0.46 &	0.00 &	0.07     & 0.09 \\
vif &	-0.57& -0.69	 &-0.08	 & -0.16 & 0.06 \\\hline
    \end{tabular}
    \caption{Spearman's correlation coefficients of fitness estimates with cross-sectional diversity (measured as entropy in group M and subtype B alignments), disorder scores and solvent accessibility values obtained from \citep{li_integrated_2015}. The column ``RNA" contains rank correlation coefficients of fitness at synonymous mutations with the pairing probability predicted by \citep{siegfried_rna_2014}. \FIG{S_vs_npat} shows how intrapatient/global diversity correlations improve when basing intrapatient estimates on larger numbers of patients. }
    \label{tab:correlations}
\end{table}

\begin{figure}
    \centering
    \includegraphics[width=\columnwidth]{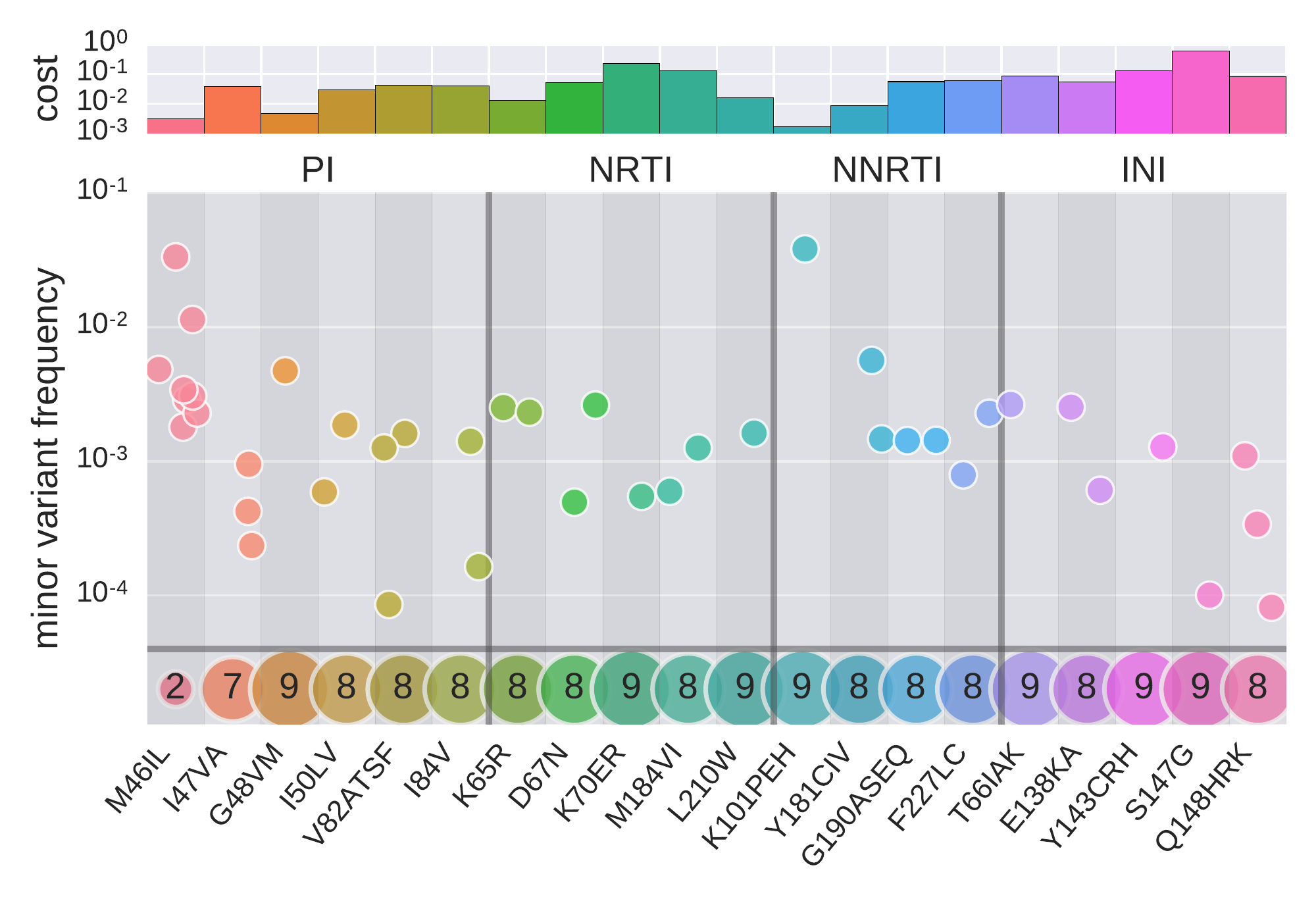}
    \caption{{\bf Pre-existing drug resistance mutations.} Each point shows the time averaged frequency of minor amino acids in individual patients. The bottom row indicates in how many out of 10 patients each mutation is not observed. Most mutations are observed only in a minority of patients suggesting high fitness costs. The top panel shows the estimated fitness costs associated with the mutations. The following mutations were never found at frequencies above 0.1\% in any patient, indicating a large fitness cost: PI: L24I, V32I, I54VTAM, L76V, N88S, L90M; NRTI: M41L, K70ER, L74VI, Y115F, T215YF, K219QE; NNRTI: L100I, K103N, V106AM, E138K, V179DEF, Y188LCH, M230L; INI: E92Q, N155H.}
    \label{fig:drug_resistance}
\end{figure}

\subsubsection*{Most drug resistance mutations have a large fitness cost}
Of particular interest are the fitness costs of mutations that confer resistance against anti-retroviral drugs. The most commonly administered drugs are nucleoside analog reverse transcriptase inhibitors (NRTI), non-nucleoside analog reverse transcriptase inhibitors (NNRTI), protease inhibitors (PI), and integrase inhibitors (INI). Resistance mutations against these drugs are well known \citep{johnson_update_2011}.

Pre-existing low frequency drug resistance mutations have been associated with failing therapy \citep{johnson_minority_2008,li_low-frequency_2011}. Some deep-sequencing studies have characterized such pre-existing variation in treatment-naive patients and found that drug-resistance mutations are usually below the detection limit, suggesting relatively high fitness costs \citep{hedskog_dynamics_2010,gianella_detection_2011,li_low-frequency_2011}. \FIG{drug_resistance} shows estimated frequencies of several drug resistance mutations in the different patients. The majority of mutations are not seen at all, while most of the remainder is observed in only one or two patients (pooled across all time points of each patient). Only the protease mutation M46I is observed consistently across several patients.

The frequency of drug resistance mutations is expected to be inversely proportional to their fitness cost in absence of treatment and of some these costs have been measured in cell cultures (see e.g. \citet{cong_fitness_2007,chow_use_1993,martinez-picado_hiv-1_2008}). Many resistance mutations quickly revert upon treatment interruption suggesting high fitness costs \citep{hedskog_dynamics_2010,joos_hiv_2008,deeks_treatment_2003}. Indeed, for most drug resistance mutations, we estimate fitness costs in excess of 5\% (sites where minor variation is not or only sporadically observed), see top panel in \FIG{drug_resistance}. Note that the costs of very deleterious mutations tend to be underestimated if the mutations are only observed in a small number of patients. For instance, G48VM in the protease and K101PEH in the reverse transcriptase are attributed a low cost but are only observed in one patient, so their actual cost might be larger.

\section*{Discussion}
Sequence evolution of HIV-1 is the determined by the rate and spectrum of mutations as well as their phenotypic effects.
Many studies have focused on beneficial mutations that sweep across the intrapatient HIV-1 population \citep{Asquith_inefficient_2006,neher_recombination_2010,ganusov_fitness_2011,kessinger_inferring_2013}, and we observe similar patterns in our study subjects (see \FIG{sweeps_lines} and \FIG{sweeps_quantification}).
The majority of mutations, however, are deleterious and stay at low frequencies within hosts; selection is constantly pruning deleterious variation from the population to maintain a functional genome. Deleterious mutations contribute substantially to sequence evolution due to their large number: if 5000 sites accumulate deleterious variation at frequencies of 1\%, the typical HIV-1 genome will contain 50 such mutations.
Here, we used longitudinal whole genome deep sequencing data from \citep{zanini_population_2016} to quantify the in vivo mutation rates of HIV-1 and the fitness costs of deleterious mutations.

The accumulation of mutations at approximately neutral sites is consistent with the mutation rates of HIV-1 measured in cell culture using lacZ assays \citep{abram_nature_2010,mansky_lower_1995}. This agreement suggests that the mutation rate of HIV-1, which is the joint rate of the HIV-1 RT, mutagenesis by the innate immune system, and the human DNA-dependent RNA polymerase II, is largely independent of cell type. Because the cell culture studies used an exogenous template while we monitor mutations on the HIV-1 genome itself, it appears also that the mutation rate does not depend, in average, on the nature of the template. The mutation rate at specific genomic sites, however, is likely to depend on the sequence context, similar to other polymerases and as indicated by previous studies \citep{abbotts_mechanism_1993, lewis_uniquely_1999}.
The highest rate is $G\to A$ and transitions are about 5-fold faster than transversions; the lowest rates are between base pairing partners, e.g. $G \leftrightarrow C$, see \FIG{mutation_rate}. If the human RNA pol II has similar error rates as its \textit{C.~elegans} homologue (error rate $4\times 10^{-6}$ per site \citep{gout_large-scale_2013}) roughly a fifth of all mutations observed in HIV are due to the RNA polymerase (assuming an HIV generation time of 1-2 days).

While consistent with cell culture estimates, the rates we estimate are incompatible with those reported by \citet{cuevas_extremely_2015}. Whereas we measure mutations in the population of RNA virions, \citet{cuevas_extremely_2015} counted nonsense mutations in proviral DNA integrated into host cell genomes and estimated a rate of $4\times 10^{-3}$ per site and replication -- more than 100 times higher than our estimate.
Unlike in circulating viral RNA, a large fraction of proviral HIV DNA is heavily hypermutated by enzymes of the APOBEC family \citep{malim_apobec_2009}.
Hypermutation is approximately an all-or-nothing phenomenon in which either a sequence contains dozens of stop codons or none \citep{armitage_apobec3g-induced_2012,cuevas_extremely_2015,delviks-frankenberry_minimal_2016}.

Because of this bimodal nature, hypermutation and reverse transcriptase mutation can not be meaningfully described by one mutation rate matrix. In the former case, a sequence with dozens of stops integrates into the host genome as a defective virus, the in the latter rare independent mutations (about 0.2 per genome) can lead to gradual evolution and adaptation.
Sporadic deamination by APOBEG enzymes might still contribute to the G$\to$A mutation rate and is included in our estimate, but heavily hypermutated sequences are likely ``dead on arrival" and make a minor contribution to genetic diversity, as also argued by others \citep{armitage_apobec3g-induced_2012,delviks-frankenberry_minimal_2016}.

Furthermore, proviral HIV DNA is enriched for hypermutated sequences. While functional proviruses rapidly lead to death of the infected cell, hypermutated proviruses tend to accumulate latently in HIV-1 target cells over the many months a T-cell can live. This accumulation likely results in a multi-fold overrepresentation of hypermutated sequences compared to the probability at which hypermutation happens in a single reverse transcription.
Because we measure mutations from RNA data from plasma, our estimates are not affected by this accumulation bias.

With the time calibrated mutation rate estimates, we estimated absolute fitness costs from mutation selection balance and quantify the relationship between group M diversity and fitness cost. Overall, fitness costs explains about half of the diversity in global alignments of HIV-1 sequences, while the remainder might be linked to patient-specifc processes such as immune escape. The relationship between logarithmic group M diversity (measured as entropy) and logarithmic fitness costs is approximately linear.

Our site-specific fitness landscape highlights a number of known functional elements across the HIV-1 genome, including regulatory elements at the RNA level. Constraints on synonymous mutations appear to be stronger and more prevalent in env than in gag or pol, consistent with earlier results that many synonymous mutations in gp120 tend to be weakly deleterious \citep{zanini_quantifying_2013} and that env recoding results in non-infectious virus \citep{vabret_large-scale_2014}.
However, comparison of our fitness cost estimates with genome wide RNA structure predictions by \citet{siegfried_rna_2014} and \citet{sukosd_full-length_2015} show little correlation. While mutations in validated RNA structure elements are associated with high fitness costs, genome wide predictions of RNA structure explain little variation in fitness costs of synonymous mutations (see \FIG{correlation_with_RNAstructure} and \TAB{correlations}). 
This lack of strong correlation is consistent with the observation that (predicted) pairing patterns evolve rapidly in most of the genome \citep{pollom_comparison_2013} or might reflect inaccuracies in RNA structure prediction: only a minority of pairings agree between the predictions by \citet{siegfried_rna_2014} and \citet{sukosd_full-length_2015}.

Several groups have estimated fitness costs within HIV-1 proteins using experimental approaches~\citep{martinez-picado_hiv-1_2008,thyagarajan_inherent_2014,rihn_uneven_2015}. Our estimates presented here are complementary to those studies in two ways. First, because of the short but dense temporal sampling, cell culture experiments are sensitive to large fitness costs, typically above $>5\%$, while estimates from natural variation are most accurate for effects below a few percent. Second, ex vivo estimates are not affected by the specific conditions of cell culture systems.

Computational methods to estimate fitness landscapes from cross-sectional data have also been proposed \citep{dahirel_coordinate_2011, ferguson_translating_2013}, including a recent effort to include intrapatient diversity via shallow sequencing \cite{hartl_within-patient_2016}.
The relationship between fitness cost and diversity, however, might be blurred since sites that are costly to mutate might still be globally diverse due to frequent escape from CTL pressure.
Indeed, we have shown in \FIG{immune_selection} that globally polymorphic sites that we estimate to have high fitness costs are over-represented among sites known to be HLA associated \citep{carlson_correlates_2012}.
\citet{barton_relative_2016} have shown that the rate of CTL escape depends on fitness costs. 
More generally the cross-sectional inferences and our intra-patient inferences reinforce the notion that HIV-1 evolution is governed by a fitness landscape that consists of a universal component determining the replicative capacity of the virus, and a host specific component responsible of escape mutations \citep{shekhar_spin_2013}. 
Our approach based on longitudinal deep intrapatient data allows to explicitly disentangle these two contributions, since we can condition on the founder sequence and the absence of host-specific selective sweeps. 
Purely cross-sectional inferences of the fitness landscape likely underestimate the fitness cost of mutations at HLA associated positions.


In the future, as whole genome deep sequencing becomes more common, estimates of mutation rates and the fitness landscape could be extended to a higher number of samples. A much larger sample pool might allow site-specific inference of the mutation rates. Furthermore, by providing more accurate minor SNP frequencies, estimates of their associated fitness costs will improve, leading to a deeper understanding of the selective forces that shape viral evolution.

\section*{Materials and Methods}

\subsection*{Code and data availability}
The sequences from the longitudinal samples were taken from \citet{zanini_population_2016} and analyzed using the library \texttt{hivevo\_access} (https://github.com/neherlab/HIVEVO\_access) and custom scripts.

The nucleotide and amino acid cross-sectional alignments of HIV-1 group M were downloaded from the Los Alamos National Laboratory HIV database and filtered for short or otherwise problematic sequences and are available as supplementary material.

Disorder and solvent accessibility scores amino acids for different HIV proteins were provided by the authors of \citep{li_integrated_2015} (available at \url{www.virusface.com}). These scores were mapped to homologous positions in the virus populations via alignments to the reference sequence NL4-3. Positions without scores were discarded. 

Our analysis scripts, as well as the resulting data for the mutation rate and fitness cost estimates, are available online at https://github.com/iosonofabio/HIV\_fitness\_landscape.

\subsection*{Mutation rate estimation}
For each patient, a set of nucleotide sites is identified, for which (i) the entropy in a group M alignment is higher than 0.1 bits and (ii) the consensus nucleotide of the earliest sample corresponds with the HIV-1 group M consensus. Derived alleles at those sites are considered if (i) they are translated in a single reading frame, (ii) they are synonymous changes, (iii) they are outside of known RNA structures or overlapping reading frames. The frequencies of these variable synonymous changes are grouped by mutation (e.g.~$A \to G$) and averaged across the genome and different samples with the following time bins: [0, 500, 1000, 1750, 3000]. Variations of the parameters have been tested and yielded similar results. The time-binned average frequencies are modeled by a linear fit with zero intercept, so the inferred rate $\hat{\mu}$ is:
\[
\hat{\mu} = \frac{\sum_i t_i \cdot x_i}{\sum_i t_i^2},
\]
where $(t_i, x_i)$ are the time and frequency of each point (see \FIG{mutation_rate}A\&B).
Different mutations are estimated (independently) to obtain the entire mutation rate matrix. The whole procedure is repeated for 100 bootstraps over patients to estimate the uncertainty of the rates, shown as $\pm$ errors in \FIG{mutation_rate}C. An error of $\pm 0.0$ means an uncertainty smaller than $\pm 0.1$. See the supplementary script \texttt{mutation\_rate.py} for the estimate implementation.

\subsection*{Estimation of selection coefficients}
The selection coefficients  were estimated using two different approaches, called ``Sat'' and ``Pooled'' in \FIG{sat}B.

\subsubsection*{Nonlinear least squares on saturation curves}
To estimate the fitness costs as in the ``Sat'' curve of \FIG{sat}B, we considered all sites in genomes from viral populations of all patient at which (i) the majority nucleotide at the earliest time point equals the global HIV-1 group M consensus and (ii) the majority nucleotide does not change during the infection. The latter criterion is necessary to ensure we exclude sites under positive selection. At each site, instead of modeling the whole set of 4 possible nucleotides, we used a simplified 2-state model: the subtype M consensus state and the sum of the derived mutations. We collected the frequencies of the derived states from all sites and patients and averaged into two-dimensional bins, by entropy category and time since Estimated Date of Infection (EDI). The averages in each entropy group are shown in \FIG{sat}A as dots: each color indicates a different entropy group (from blue to red, low to high). We fitted those points via nonlinear least squares to equation \eqref{eq:sat} with a single fit parameter, $s$. The resulting fits are shown in \FIG{sat}A and the fitness costs $s$ in \FIG{sat}B.

\subsubsection*{Pooled SNP frequencies from late samples}
To obtain site specific estimates, we averaged SNP frequencies at individual sites according to \EQ{xavg}. The average is weighted to ensure that samples contribute approximately proportionally to the number of template molecules present in the sample.
This weight is calculated from the estimated template input $T_k$ as $w_k = (0.002 + 1/T_k)^{-1}$, where $0.002$ is the combined error rate of RT-PCR and sequencing.
Samples contribute proportionally to the number of RNA templates is small if $T_k$ is small, while for large $T_k$ the sequencing error rate is limiting and the per sample contribution is capped at 500. The weighted average is performed within each patient. To average SNP frequencies further over patients, we use the alignment of each patient to the NL4-3 reference sequence to identify homologous positions to average. As before, we exclude sites that don't agree with the global HIV-1 consensus and sites that sweep (i.e. where the majority state changes during infection). These exclusions are particularly important, since sites from different patients are combined and minor frequencies are only meaningful when measured relative to the same reference nucleotide or amino acid. To determine uncertainties, bootstrap distributions are constructed by resampling the patients contributing the average. Estimates of fitness costs for nucleotide and amino acid mutations where done in very similar ways.

Selection coefficients are estimates via $\mu/\xavg$, where $\mu$ is the sum of mutation rates away from the consensus nucleotide or amino acid estimated above. Amino acid mutation rates are calculated specifically for each patient on the bases of the codon coding for the amino acid in the founder sequence of that patient (amino acid changes requiring two nucleotide changes were ignored).

To determine the uncertainty of fitness cost estimates, we picked sites within small slices of the distribution of selection coefficients and constructed bootstrap distributions for the estimates at each of the positions. \FIG{codon_selection}D shows the combined distributions for each of the positions contained in these initial slices. 

\section*{Acknowledgements}
We thank Lina Thebo and Crista Lanz for excellent technical
assistance and Pleuni Pennings and Nate Cira for helpful comments on the manuscript. This work was supported by the European Research Council through grant Stg.~260686 and partly by grant NSF PHY11-25915 to KITP and the Swedish Research Council through grant K2014-57X-09935-23-5.
\bibliography{manu_arxiv}

\begin{thebibliography}{68}
\expandafter\ifx\csname natexlab\endcsname\relax\def\natexlab#1{#1}\fi
\expandafter\ifx\csname bibnamefont\endcsname\relax
  \def\bibnamefont#1{#1}\fi
\expandafter\ifx\csname bibfnamefont\endcsname\relax
  \def\bibfnamefont#1{#1}\fi
\expandafter\ifx\csname citenamefont\endcsname\relax
  \def\citenamefont#1{#1}\fi
\expandafter\ifx\csname url\endcsname\relax
  \def\url#1{\texttt{#1}}\fi
\expandafter\ifx\csname urlprefix\endcsname\relax\def\urlprefix{URL }\fi
\providecommand{\bibinfo}[2]{#2}
\providecommand{\eprint}[2][]{\url{#2}}

\bibitem[{\citenamefont{Abbotts} \emph{et~al.}(1993)\citenamefont{Abbotts,
  Bebenek, Kunkel, and Wilson}}]{abbotts_mechanism_1993}
\bibinfo{author}{\bibnamefont{Abbotts}, \bibfnamefont{J.}},
  \bibinfo{author}{\bibfnamefont{K.}~\bibnamefont{Bebenek}},
  \bibinfo{author}{\bibfnamefont{T.~A.} \bibnamefont{Kunkel}}, and
  \bibinfo{author}{\bibfnamefont{S.~H.} \bibnamefont{Wilson}},
  \bibinfo{year}{1993}, \bibinfo{journal}{Journal of Biological Chemistry}
  \textbf{\bibinfo{volume}{268}}(\bibinfo{number}{14}), \bibinfo{pages}{10312},
  ISSN \bibinfo{issn}{0021-9258, 1083-351X}.

\bibitem[{\citenamefont{Abram} \emph{et~al.}(2010)\citenamefont{Abram, Ferris,
  Shao, Alvord, and Hughes}}]{abram_nature_2010}
\bibinfo{author}{\bibnamefont{Abram}, \bibfnamefont{M.~E.}},
  \bibinfo{author}{\bibfnamefont{A.~L.} \bibnamefont{Ferris}},
  \bibinfo{author}{\bibfnamefont{W.}~\bibnamefont{Shao}},
  \bibinfo{author}{\bibfnamefont{W.~G.} \bibnamefont{Alvord}}, and
  \bibinfo{author}{\bibfnamefont{S.~H.} \bibnamefont{Hughes}},
  \bibinfo{year}{2010}, \bibinfo{journal}{Journal of virology}
  \textbf{\bibinfo{volume}{84}}(\bibinfo{number}{19}), \bibinfo{pages}{9864}.

\bibitem[{\citenamefont{Acevedo} \emph{et~al.}(2014)\citenamefont{Acevedo,
  Brodsky, and Andino}}]{acevedo_mutational_2014}
\bibinfo{author}{\bibnamefont{Acevedo}, \bibfnamefont{A.}},
  \bibinfo{author}{\bibfnamefont{L.}~\bibnamefont{Brodsky}}, and
  \bibinfo{author}{\bibfnamefont{R.}~\bibnamefont{Andino}},
  \bibinfo{year}{2014}, \bibinfo{journal}{Nature}
  \textbf{\bibinfo{volume}{505}}(\bibinfo{number}{7485}), \bibinfo{pages}{686},
  ISSN \bibinfo{issn}{0028-0836}.

\bibitem[{\citenamefont{Armitage} \emph{et~al.}(2012)\citenamefont{Armitage,
  Deforche, Chang, Wee, Kramer, Welch, Gerstoft, Fugger, McMichael, Rambaut,
  and Iversen}}]{armitage_apobec3g-induced_2012}
\bibinfo{author}{\bibnamefont{Armitage}, \bibfnamefont{A.~E.}},
  \bibinfo{author}{\bibfnamefont{K.}~\bibnamefont{Deforche}},
  \bibinfo{author}{\bibfnamefont{C.-h.} \bibnamefont{Chang}},
  \bibinfo{author}{\bibfnamefont{E.}~\bibnamefont{Wee}},
  \bibinfo{author}{\bibfnamefont{B.}~\bibnamefont{Kramer}},
  \bibinfo{author}{\bibfnamefont{J.~J.} \bibnamefont{Welch}},
  \bibinfo{author}{\bibfnamefont{J.}~\bibnamefont{Gerstoft}},
  \bibinfo{author}{\bibfnamefont{L.}~\bibnamefont{Fugger}},
  \bibinfo{author}{\bibfnamefont{A.}~\bibnamefont{McMichael}},
  \bibinfo{author}{\bibfnamefont{A.}~\bibnamefont{Rambaut}}, and
  \bibinfo{author}{\bibfnamefont{A.~K.~N.} \bibnamefont{Iversen}},
  \bibinfo{year}{2012}, \bibinfo{journal}{PLoS Genet}
  \textbf{\bibinfo{volume}{8}}(\bibinfo{number}{3}), \bibinfo{pages}{e1002550}.

\bibitem[{\citenamefont{Asquith} \emph{et~al.}(2006)\citenamefont{Asquith,
  Edwards, Lipsitch, and McLean}}]{Asquith_inefficient_2006}
\bibinfo{author}{\bibnamefont{Asquith}, \bibfnamefont{B.}},
  \bibinfo{author}{\bibfnamefont{C.~T.~T.} \bibnamefont{Edwards}},
  \bibinfo{author}{\bibfnamefont{M.}~\bibnamefont{Lipsitch}}, and
  \bibinfo{author}{\bibfnamefont{A.~R.} \bibnamefont{McLean}},
  \bibinfo{year}{2006}, \bibinfo{journal}{PLoS Biol}
  \textbf{\bibinfo{volume}{4}}(\bibinfo{number}{4}), \bibinfo{pages}{e90}.

\bibitem[{\citenamefont{Bar} \emph{et~al.}(2012)\citenamefont{Bar, Tsao, Iyer,
  Decker, Yang, Bonsignori, Chen, Hwang, Montefiori, Liao, Hraber, Fischer}
  \emph{et~al.}}]{bar_early_2012}
\bibinfo{author}{\bibnamefont{Bar}, \bibfnamefont{K.~J.}},
  \bibinfo{author}{\bibfnamefont{C.-y.} \bibnamefont{Tsao}},
  \bibinfo{author}{\bibfnamefont{S.~S.} \bibnamefont{Iyer}},
  \bibinfo{author}{\bibfnamefont{J.~M.} \bibnamefont{Decker}},
  \bibinfo{author}{\bibfnamefont{Y.}~\bibnamefont{Yang}},
  \bibinfo{author}{\bibfnamefont{M.}~\bibnamefont{Bonsignori}},
  \bibinfo{author}{\bibfnamefont{X.}~\bibnamefont{Chen}},
  \bibinfo{author}{\bibfnamefont{K.-K.} \bibnamefont{Hwang}},
  \bibinfo{author}{\bibfnamefont{D.~C.} \bibnamefont{Montefiori}},
  \bibinfo{author}{\bibfnamefont{H.-X.} \bibnamefont{Liao}},
  \bibinfo{author}{\bibfnamefont{P.}~\bibnamefont{Hraber}},
  \bibinfo{author}{\bibfnamefont{W.}~\bibnamefont{Fischer}}, \emph{et~al.},
  \bibinfo{year}{2012}, \bibinfo{journal}{PLoS Pathog}
  \textbf{\bibinfo{volume}{8}}(\bibinfo{number}{5}), \bibinfo{pages}{e1002721},
  \urlprefix\url{http://dx.doi.org/10.1371/journal.ppat.1002721}.

\bibitem[{\citenamefont{Bartha} \emph{et~al.}(2013)\citenamefont{Bartha,
  Carlson, Brumme, McLaren, Brumme, John, Haas, Martinez-Picado, Dalmau,
  López-Galíndez, Casado, Rauch} \emph{et~al.}}]{bartha_genome_genome_2013}
\bibinfo{author}{\bibnamefont{Bartha}, \bibfnamefont{I.}},
  \bibinfo{author}{\bibfnamefont{J.~M.} \bibnamefont{Carlson}},
  \bibinfo{author}{\bibfnamefont{C.~J.} \bibnamefont{Brumme}},
  \bibinfo{author}{\bibfnamefont{P.~J.} \bibnamefont{McLaren}},
  \bibinfo{author}{\bibfnamefont{Z.~L.} \bibnamefont{Brumme}},
  \bibinfo{author}{\bibfnamefont{M.}~\bibnamefont{John}},
  \bibinfo{author}{\bibfnamefont{D.~W.} \bibnamefont{Haas}},
  \bibinfo{author}{\bibfnamefont{J.}~\bibnamefont{Martinez-Picado}},
  \bibinfo{author}{\bibfnamefont{J.}~\bibnamefont{Dalmau}},
  \bibinfo{author}{\bibfnamefont{C.}~\bibnamefont{López-Galíndez}},
  \bibinfo{author}{\bibfnamefont{C.}~\bibnamefont{Casado}},
  \bibinfo{author}{\bibfnamefont{A.}~\bibnamefont{Rauch}}, \emph{et~al.},
  \bibinfo{year}{2013}, \bibinfo{journal}{eLife Sciences}
  \textbf{\bibinfo{volume}{2}}, \bibinfo{pages}{e01123}, ISSN
  \bibinfo{issn}{2050-084X}.

\bibitem[{\citenamefont{Barton} \emph{et~al.}(2016)\citenamefont{Barton,
  Goonetilleke, Butler, Walker, McMichael, and
  Chakraborty}}]{barton_relative_2016}
\bibinfo{author}{\bibnamefont{Barton}, \bibfnamefont{J.~P.}},
  \bibinfo{author}{\bibfnamefont{N.}~\bibnamefont{Goonetilleke}},
  \bibinfo{author}{\bibfnamefont{T.~C.} \bibnamefont{Butler}},
  \bibinfo{author}{\bibfnamefont{B.~D.} \bibnamefont{Walker}},
  \bibinfo{author}{\bibfnamefont{A.~J.} \bibnamefont{McMichael}}, and
  \bibinfo{author}{\bibfnamefont{A.~K.} \bibnamefont{Chakraborty}},
  \bibinfo{year}{2016}, \bibinfo{journal}{Nat Commun}
  \textbf{\bibinfo{volume}{7}}, \bibinfo{pages}{11660}.

\bibitem[{\citenamefont{Brodin} \emph{et~al.}(2016)\citenamefont{Brodin,
  Zanini, Thebo, Lanz, Bratt, Neher, and Albert}}]{brodin_establishment_2016}
\bibinfo{author}{\bibnamefont{Brodin}, \bibfnamefont{J.}},
  \bibinfo{author}{\bibfnamefont{F.}~\bibnamefont{Zanini}},
  \bibinfo{author}{\bibfnamefont{L.}~\bibnamefont{Thebo}},
  \bibinfo{author}{\bibfnamefont{C.}~\bibnamefont{Lanz}},
  \bibinfo{author}{\bibfnamefont{G.}~\bibnamefont{Bratt}},
  \bibinfo{author}{\bibfnamefont{R.}~\bibnamefont{Neher}}, and
  \bibinfo{author}{\bibfnamefont{J.}~\bibnamefont{Albert}},
  \bibinfo{year}{2016}, \bibinfo{journal}{bioRxiv} , \bibinfo{pages}{053983}.

\bibitem[{\citenamefont{Carlson} \emph{et~al.}(2012)\citenamefont{Carlson,
  Brumme, Martin, Listgarten, Brockman, Le, Chui, Cotton, Knapp, Riddler,
  Haubrich, Nelson} \emph{et~al.}}]{carlson_correlates_2012}
\bibinfo{author}{\bibnamefont{Carlson}, \bibfnamefont{J.~M.}},
  \bibinfo{author}{\bibfnamefont{C.~J.} \bibnamefont{Brumme}},
  \bibinfo{author}{\bibfnamefont{E.}~\bibnamefont{Martin}},
  \bibinfo{author}{\bibfnamefont{J.}~\bibnamefont{Listgarten}},
  \bibinfo{author}{\bibfnamefont{M.~A.} \bibnamefont{Brockman}},
  \bibinfo{author}{\bibfnamefont{A.~Q.} \bibnamefont{Le}},
  \bibinfo{author}{\bibfnamefont{C.}~\bibnamefont{Chui}},
  \bibinfo{author}{\bibfnamefont{L.~A.} \bibnamefont{Cotton}},
  \bibinfo{author}{\bibfnamefont{D.~J. H.~F.} \bibnamefont{Knapp}},
  \bibinfo{author}{\bibfnamefont{S.~A.} \bibnamefont{Riddler}},
  \bibinfo{author}{\bibfnamefont{R.}~\bibnamefont{Haubrich}},
  \bibinfo{author}{\bibfnamefont{G.}~\bibnamefont{Nelson}}, \emph{et~al.},
  \bibinfo{year}{2012}, \bibinfo{journal}{J. Virol.} ,
  \bibinfo{pages}{JVI.01998}ISSN \bibinfo{issn}{0022-538X, 1098-5514}.

\bibitem[{\citenamefont{Carlson} \emph{et~al.}(2008)\citenamefont{Carlson,
  Brumme, Rousseau, Brumme, Matthews, Kadie, Mullins, Walker, Harrigan,
  Goulder, and Heckerman}}]{carlson_phylogenetic_2008}
\bibinfo{author}{\bibnamefont{Carlson}, \bibfnamefont{J.~M.}},
  \bibinfo{author}{\bibfnamefont{Z.~L.} \bibnamefont{Brumme}},
  \bibinfo{author}{\bibfnamefont{C.~M.} \bibnamefont{Rousseau}},
  \bibinfo{author}{\bibfnamefont{C.~J.} \bibnamefont{Brumme}},
  \bibinfo{author}{\bibfnamefont{P.}~\bibnamefont{Matthews}},
  \bibinfo{author}{\bibfnamefont{C.}~\bibnamefont{Kadie}},
  \bibinfo{author}{\bibfnamefont{J.~I.} \bibnamefont{Mullins}},
  \bibinfo{author}{\bibfnamefont{B.~D.} \bibnamefont{Walker}},
  \bibinfo{author}{\bibfnamefont{P.~R.} \bibnamefont{Harrigan}},
  \bibinfo{author}{\bibfnamefont{P.~J.~R.} \bibnamefont{Goulder}}, and
  \bibinfo{author}{\bibfnamefont{D.}~\bibnamefont{Heckerman}},
  \bibinfo{year}{2008}, \bibinfo{journal}{PLOS Comput Biol}
  \textbf{\bibinfo{volume}{4}}(\bibinfo{number}{11}),
  \bibinfo{pages}{e1000225}, ISSN \bibinfo{issn}{1553-7358}.

\bibitem[{\citenamefont{Carlson} \emph{et~al.}(2014)\citenamefont{Carlson,
  Schaefer, Monaco, Batorsky, Claiborne, Prince, Deymier, Ende, Klatt, DeZiel,
  Lin, Peng} \emph{et~al.}}]{carlson_selection_2014}
\bibinfo{author}{\bibnamefont{Carlson}, \bibfnamefont{J.~M.}},
  \bibinfo{author}{\bibfnamefont{M.}~\bibnamefont{Schaefer}},
  \bibinfo{author}{\bibfnamefont{D.~C.} \bibnamefont{Monaco}},
  \bibinfo{author}{\bibfnamefont{R.}~\bibnamefont{Batorsky}},
  \bibinfo{author}{\bibfnamefont{D.~T.} \bibnamefont{Claiborne}},
  \bibinfo{author}{\bibfnamefont{J.}~\bibnamefont{Prince}},
  \bibinfo{author}{\bibfnamefont{M.~J.} \bibnamefont{Deymier}},
  \bibinfo{author}{\bibfnamefont{Z.~S.} \bibnamefont{Ende}},
  \bibinfo{author}{\bibfnamefont{N.~R.} \bibnamefont{Klatt}},
  \bibinfo{author}{\bibfnamefont{C.~E.} \bibnamefont{DeZiel}},
  \bibinfo{author}{\bibfnamefont{T.-H.} \bibnamefont{Lin}},
  \bibinfo{author}{\bibfnamefont{J.}~\bibnamefont{Peng}}, \emph{et~al.},
  \bibinfo{year}{2014}, \bibinfo{journal}{Science}
  \textbf{\bibinfo{volume}{345}}(\bibinfo{number}{6193}),
  \bibinfo{pages}{1254031}.

\bibitem[{\citenamefont{Chow} \emph{et~al.}(1993)\citenamefont{Chow, Hirsch,
  Merrill, Bechtel, Eron, Kaplan, and D'Aquila}}]{chow_use_1993}
\bibinfo{author}{\bibnamefont{Chow}, \bibfnamefont{Y.-K.}},
  \bibinfo{author}{\bibfnamefont{M.~S.} \bibnamefont{Hirsch}},
  \bibinfo{author}{\bibfnamefont{D.~P.} \bibnamefont{Merrill}},
  \bibinfo{author}{\bibfnamefont{L.~J.} \bibnamefont{Bechtel}},
  \bibinfo{author}{\bibfnamefont{J.~J.} \bibnamefont{Eron}},
  \bibinfo{author}{\bibfnamefont{J.~C.} \bibnamefont{Kaplan}}, and
  \bibinfo{author}{\bibfnamefont{R.~T.} \bibnamefont{D'Aquila}},
  \bibinfo{year}{1993}, \bibinfo{journal}{Nature}
  \textbf{\bibinfo{volume}{361}}(\bibinfo{number}{6413}), \bibinfo{pages}{650}.

\bibitem[{\citenamefont{Cong} \emph{et~al.}(2007)\citenamefont{Cong, Heneine,
  and García-Lerma}}]{cong_fitness_2007}
\bibinfo{author}{\bibnamefont{Cong}, \bibfnamefont{M.-e.}},
  \bibinfo{author}{\bibfnamefont{W.}~\bibnamefont{Heneine}}, and
  \bibinfo{author}{\bibfnamefont{J.~G.} \bibnamefont{García-Lerma}},
  \bibinfo{year}{2007}, \bibinfo{journal}{Journal of Virology}
  \textbf{\bibinfo{volume}{81}}(\bibinfo{number}{6}), \bibinfo{pages}{3037},
  ISSN \bibinfo{issn}{0022-538X, 1098-5514}.

\bibitem[{\citenamefont{Cuevas} \emph{et~al.}(2015)\citenamefont{Cuevas,
  Geller, Garijo, López-Aldeguer, and Sanjuán}}]{cuevas_extremely_2015}
\bibinfo{author}{\bibnamefont{Cuevas}, \bibfnamefont{J.~M.}},
  \bibinfo{author}{\bibfnamefont{R.}~\bibnamefont{Geller}},
  \bibinfo{author}{\bibfnamefont{R.}~\bibnamefont{Garijo}},
  \bibinfo{author}{\bibfnamefont{J.}~\bibnamefont{López-Aldeguer}}, and
  \bibinfo{author}{\bibfnamefont{R.}~\bibnamefont{Sanjuán}},
  \bibinfo{year}{2015}, \bibinfo{journal}{PLoS Biol}
  \textbf{\bibinfo{volume}{13}}(\bibinfo{number}{9}),
  \bibinfo{pages}{e1002251}.

\bibitem[{\citenamefont{Dahirel} \emph{et~al.}(2011)\citenamefont{Dahirel,
  Shekhar, Pereyra, Miura, Artyomov, Talsania, Allen, Altfeld, Carrington,
  Irvine, Walker, and Chakraborty}}]{dahirel_coordinate_2011}
\bibinfo{author}{\bibnamefont{Dahirel}, \bibfnamefont{V.}},
  \bibinfo{author}{\bibfnamefont{K.}~\bibnamefont{Shekhar}},
  \bibinfo{author}{\bibfnamefont{F.}~\bibnamefont{Pereyra}},
  \bibinfo{author}{\bibfnamefont{T.}~\bibnamefont{Miura}},
  \bibinfo{author}{\bibfnamefont{M.}~\bibnamefont{Artyomov}},
  \bibinfo{author}{\bibfnamefont{S.}~\bibnamefont{Talsania}},
  \bibinfo{author}{\bibfnamefont{T.~M.} \bibnamefont{Allen}},
  \bibinfo{author}{\bibfnamefont{M.}~\bibnamefont{Altfeld}},
  \bibinfo{author}{\bibfnamefont{M.}~\bibnamefont{Carrington}},
  \bibinfo{author}{\bibfnamefont{D.~J.} \bibnamefont{Irvine}},
  \bibinfo{author}{\bibfnamefont{B.~D.} \bibnamefont{Walker}}, and
  \bibinfo{author}{\bibfnamefont{A.~K.} \bibnamefont{Chakraborty}},
  \bibinfo{year}{2011}, \bibinfo{journal}{PNAS}
  \textbf{\bibinfo{volume}{108}}(\bibinfo{number}{28}), \bibinfo{pages}{11530},
  ISSN \bibinfo{issn}{0027-8424, 1091-6490},
  \urlprefix\url{http://www.pnas.org/content/108/28/11530}.

\bibitem[{\citenamefont{Deeks}(2003)}]{deeks_treatment_2003}
\bibinfo{author}{\bibnamefont{Deeks}, \bibfnamefont{S.~G.}},
  \bibinfo{year}{2003}, \bibinfo{journal}{Lancet}
  \textbf{\bibinfo{volume}{362}}(\bibinfo{number}{9400}),
  \bibinfo{pages}{2002}, ISSN \bibinfo{issn}{1474-547X}.

\bibitem[{\citenamefont{Delviks-Frankenberry}
  \emph{et~al.}(2016)\citenamefont{Delviks-Frankenberry, Nikolaitchik, Burdick,
  Gorelick, Keele, Hu, and Pathak}}]{delviks-frankenberry_minimal_2016}
\bibinfo{author}{\bibnamefont{Delviks-Frankenberry}, \bibfnamefont{K.~A.}},
  \bibinfo{author}{\bibfnamefont{O.~A.} \bibnamefont{Nikolaitchik}},
  \bibinfo{author}{\bibfnamefont{R.~C.} \bibnamefont{Burdick}},
  \bibinfo{author}{\bibfnamefont{R.~J.} \bibnamefont{Gorelick}},
  \bibinfo{author}{\bibfnamefont{B.~F.} \bibnamefont{Keele}},
  \bibinfo{author}{\bibfnamefont{W.-S.} \bibnamefont{Hu}}, and
  \bibinfo{author}{\bibfnamefont{V.~K.} \bibnamefont{Pathak}},
  \bibinfo{year}{2016}, \bibinfo{journal}{PLOS Pathog}
  \textbf{\bibinfo{volume}{12}}(\bibinfo{number}{5}), ISSN
  \bibinfo{issn}{1553-7374}.

\bibitem[{\citenamefont{Doud} \emph{et~al.}(2015)\citenamefont{Doud, Ashenberg,
  and Bloom}}]{doud_site-specific_2015}
\bibinfo{author}{\bibnamefont{Doud}, \bibfnamefont{M.~B.}},
  \bibinfo{author}{\bibfnamefont{O.}~\bibnamefont{Ashenberg}}, and
  \bibinfo{author}{\bibfnamefont{J.~D.} \bibnamefont{Bloom}},
  \bibinfo{year}{2015}, \bibinfo{journal}{Mol Biol Evol}
  \textbf{\bibinfo{volume}{32}}(\bibinfo{number}{11}), \bibinfo{pages}{2944},
  ISSN \bibinfo{issn}{0737-4038, 1537-1719}.

\bibitem[{\citenamefont{Ferguson} \emph{et~al.}(2013)\citenamefont{Ferguson,
  Mann, Omarjee, Ndung’u, Walker, and
  Chakraborty}}]{ferguson_translating_2013}
\bibinfo{author}{\bibnamefont{Ferguson}, \bibfnamefont{A.}},
  \bibinfo{author}{\bibfnamefont{J.}~\bibnamefont{Mann}},
  \bibinfo{author}{\bibfnamefont{S.}~\bibnamefont{Omarjee}},
  \bibinfo{author}{\bibfnamefont{T.}~\bibnamefont{Ndung’u}},
  \bibinfo{author}{\bibfnamefont{B.}~\bibnamefont{Walker}}, and
  \bibinfo{author}{\bibfnamefont{A.}~\bibnamefont{Chakraborty}},
  \bibinfo{year}{2013}, \bibinfo{journal}{Immunity}
  \textbf{\bibinfo{volume}{38}}(\bibinfo{number}{3}), \bibinfo{pages}{606},
  ISSN \bibinfo{issn}{1074-7613},
  \urlprefix\url{http://www.sciencedirect.com/science/article/pii/S1074761313001076}.

\bibitem[{\citenamefont{Friedrich} \emph{et~al.}(2004)\citenamefont{Friedrich,
  Dodds, Yant, Vojnov, Rudersdorf, Cullen, Evans, Desrosiers, Mothé, Sidney,
  Sette, Kunstman} \emph{et~al.}}]{friedrich_reversion_2004}
\bibinfo{author}{\bibnamefont{Friedrich}, \bibfnamefont{T.~C.}},
  \bibinfo{author}{\bibfnamefont{E.~J.} \bibnamefont{Dodds}},
  \bibinfo{author}{\bibfnamefont{L.~J.} \bibnamefont{Yant}},
  \bibinfo{author}{\bibfnamefont{L.}~\bibnamefont{Vojnov}},
  \bibinfo{author}{\bibfnamefont{R.}~\bibnamefont{Rudersdorf}},
  \bibinfo{author}{\bibfnamefont{C.}~\bibnamefont{Cullen}},
  \bibinfo{author}{\bibfnamefont{D.~T.} \bibnamefont{Evans}},
  \bibinfo{author}{\bibfnamefont{R.~C.} \bibnamefont{Desrosiers}},
  \bibinfo{author}{\bibfnamefont{B.~R.} \bibnamefont{Mothé}},
  \bibinfo{author}{\bibfnamefont{J.}~\bibnamefont{Sidney}},
  \bibinfo{author}{\bibfnamefont{A.}~\bibnamefont{Sette}},
  \bibinfo{author}{\bibfnamefont{K.}~\bibnamefont{Kunstman}}, \emph{et~al.},
  \bibinfo{year}{2004}, \bibinfo{journal}{Nat Med}
  \textbf{\bibinfo{volume}{10}}(\bibinfo{number}{3}), \bibinfo{pages}{275}.

\bibitem[{\citenamefont{Ganusov} \emph{et~al.}(2011)\citenamefont{Ganusov,
  Goonetilleke, Liu, Ferrari, Shaw, {McMichael}, Borrow, Korber, and
  Perelson}}]{ganusov_fitness_2011}
\bibinfo{author}{\bibnamefont{Ganusov}, \bibfnamefont{V.~V.}},
  \bibinfo{author}{\bibfnamefont{N.}~\bibnamefont{Goonetilleke}},
  \bibinfo{author}{\bibfnamefont{M.~K.~P.} \bibnamefont{Liu}},
  \bibinfo{author}{\bibfnamefont{G.}~\bibnamefont{Ferrari}},
  \bibinfo{author}{\bibfnamefont{G.~M.} \bibnamefont{Shaw}},
  \bibinfo{author}{\bibfnamefont{A.~J.} \bibnamefont{{McMichael}}},
  \bibinfo{author}{\bibfnamefont{P.}~\bibnamefont{Borrow}},
  \bibinfo{author}{\bibfnamefont{B.~T.} \bibnamefont{Korber}}, and
  \bibinfo{author}{\bibfnamefont{A.~S.} \bibnamefont{Perelson}},
  \bibinfo{year}{2011}, \bibinfo{journal}{J.~Virol}
  \textbf{\bibinfo{volume}{85}}(\bibinfo{number}{20}), \bibinfo{pages}{10518}.

\bibitem[{\citenamefont{Gianella} \emph{et~al.}(2011)\citenamefont{Gianella,
  Delport, Pacold, Young, Choi, Little, Richman, Pond, and
  Smith}}]{gianella_detection_2011}
\bibinfo{author}{\bibnamefont{Gianella}, \bibfnamefont{S.}},
  \bibinfo{author}{\bibfnamefont{W.}~\bibnamefont{Delport}},
  \bibinfo{author}{\bibfnamefont{M.~E.} \bibnamefont{Pacold}},
  \bibinfo{author}{\bibfnamefont{J.~A.} \bibnamefont{Young}},
  \bibinfo{author}{\bibfnamefont{J.~Y.} \bibnamefont{Choi}},
  \bibinfo{author}{\bibfnamefont{S.~J.} \bibnamefont{Little}},
  \bibinfo{author}{\bibfnamefont{D.~D.} \bibnamefont{Richman}},
  \bibinfo{author}{\bibfnamefont{S.~L.~K.} \bibnamefont{Pond}}, and
  \bibinfo{author}{\bibfnamefont{D.~M.} \bibnamefont{Smith}},
  \bibinfo{year}{2011}, \bibinfo{journal}{J. Virol.}
  \textbf{\bibinfo{volume}{85}}(\bibinfo{number}{16}), \bibinfo{pages}{8359},
  ISSN \bibinfo{issn}{0022-538X, 1098-5514}.

\bibitem[{\citenamefont{Goonetilleke}
  \emph{et~al.}(2009)\citenamefont{Goonetilleke, Liu, Salazar-Gonzalez,
  Ferrari, Giorgi, Ganusov, Keele, Learn, Turnbull, Salazar, Weinhold, Moore}
  \emph{et~al.}}]{goonetilleke_first_2009}
\bibinfo{author}{\bibnamefont{Goonetilleke}, \bibfnamefont{N.}},
  \bibinfo{author}{\bibfnamefont{M.~K.~P.} \bibnamefont{Liu}},
  \bibinfo{author}{\bibfnamefont{J.~F.} \bibnamefont{Salazar-Gonzalez}},
  \bibinfo{author}{\bibfnamefont{G.}~\bibnamefont{Ferrari}},
  \bibinfo{author}{\bibfnamefont{E.}~\bibnamefont{Giorgi}},
  \bibinfo{author}{\bibfnamefont{V.~V.} \bibnamefont{Ganusov}},
  \bibinfo{author}{\bibfnamefont{B.~F.} \bibnamefont{Keele}},
  \bibinfo{author}{\bibfnamefont{G.~H.} \bibnamefont{Learn}},
  \bibinfo{author}{\bibfnamefont{E.~L.} \bibnamefont{Turnbull}},
  \bibinfo{author}{\bibfnamefont{M.~G.} \bibnamefont{Salazar}},
  \bibinfo{author}{\bibfnamefont{K.~J.} \bibnamefont{Weinhold}},
  \bibinfo{author}{\bibfnamefont{S.}~\bibnamefont{Moore}}, \emph{et~al.},
  \bibinfo{year}{2009}, \bibinfo{journal}{J Exp Med}
  \textbf{\bibinfo{volume}{206}}(\bibinfo{number}{6}), \bibinfo{pages}{1253},
  ISSN \bibinfo{issn}{0022-1007, 1540-9538}.

\bibitem[{\citenamefont{Gout} \emph{et~al.}(2013)\citenamefont{Gout, Thomas,
  Smith, Okamoto, and Lynch}}]{gout_large-scale_2013}
\bibinfo{author}{\bibnamefont{Gout}, \bibfnamefont{J.-F.}},
  \bibinfo{author}{\bibfnamefont{W.~K.} \bibnamefont{Thomas}},
  \bibinfo{author}{\bibfnamefont{Z.}~\bibnamefont{Smith}},
  \bibinfo{author}{\bibfnamefont{K.}~\bibnamefont{Okamoto}}, and
  \bibinfo{author}{\bibfnamefont{M.}~\bibnamefont{Lynch}},
  \bibinfo{year}{2013}, \bibinfo{journal}{PNAS}
  \textbf{\bibinfo{volume}{110}}(\bibinfo{number}{46}), \bibinfo{pages}{18584},
  ISSN \bibinfo{issn}{0027-8424, 1091-6490}.

\bibitem[{\citenamefont{Haigh}(1978)}]{haigh_accumulation_1978}
\bibinfo{author}{\bibnamefont{Haigh}, \bibfnamefont{J.}}, \bibinfo{year}{1978},
  \bibinfo{journal}{Theoretical Population Biology}
  \textbf{\bibinfo{volume}{14}}(\bibinfo{number}{2}), \bibinfo{pages}{251},
  ISSN \bibinfo{issn}{0040-5809}.

\bibitem[{\citenamefont{Haldane}(1937)}]{haldane_effect_1937}
\bibinfo{author}{\bibnamefont{Haldane}, \bibfnamefont{J.~B.~S.}},
  \bibinfo{year}{1937}, \bibinfo{journal}{The American Naturalist}
  \textbf{\bibinfo{volume}{71}}(\bibinfo{number}{735}), \bibinfo{pages}{337},
  ISSN \bibinfo{issn}{0003-0147, 1537-5323}.

\bibitem[{\citenamefont{Hartl} \emph{et~al.}(2016)\citenamefont{Hartl, Theys,
  Feder, Gelbart, Stern, and Pennings}}]{hartl_within-patient_2016}
\bibinfo{author}{\bibnamefont{Hartl}, \bibfnamefont{M.}},
  \bibinfo{author}{\bibfnamefont{K.}~\bibnamefont{Theys}},
  \bibinfo{author}{\bibfnamefont{A.}~\bibnamefont{Feder}},
  \bibinfo{author}{\bibfnamefont{M.}~\bibnamefont{Gelbart}},
  \bibinfo{author}{\bibfnamefont{A.}~\bibnamefont{Stern}}, and
  \bibinfo{author}{\bibfnamefont{P.~S.} \bibnamefont{Pennings}},
  \bibinfo{year}{2016}, \bibinfo{journal}{bioRxiv} , \bibinfo{pages}{057026}.

\bibitem[{\citenamefont{Hedskog} \emph{et~al.}(2010)\citenamefont{Hedskog,
  Mild, Jernberg, Sherwood, Bratt, Leitner, Lundeberg, Andersson, and
  Albert}}]{hedskog_dynamics_2010}
\bibinfo{author}{\bibnamefont{Hedskog}, \bibfnamefont{C.}},
  \bibinfo{author}{\bibfnamefont{M.}~\bibnamefont{Mild}},
  \bibinfo{author}{\bibfnamefont{J.}~\bibnamefont{Jernberg}},
  \bibinfo{author}{\bibfnamefont{E.}~\bibnamefont{Sherwood}},
  \bibinfo{author}{\bibfnamefont{G.}~\bibnamefont{Bratt}},
  \bibinfo{author}{\bibfnamefont{T.}~\bibnamefont{Leitner}},
  \bibinfo{author}{\bibfnamefont{J.}~\bibnamefont{Lundeberg}},
  \bibinfo{author}{\bibfnamefont{B.}~\bibnamefont{Andersson}}, and
  \bibinfo{author}{\bibfnamefont{J.}~\bibnamefont{Albert}},
  \bibinfo{year}{2010}, \bibinfo{journal}{PLoS ONE}
  \textbf{\bibinfo{volume}{5}}(\bibinfo{number}{7}), \bibinfo{pages}{e11345},
  \urlprefix\url{http://dx.doi.org/10.1371/journal.pone.0011345}.

\bibitem[{\citenamefont{Hinkley} \emph{et~al.}(2011)\citenamefont{Hinkley,
  Martins, Chappey, Haddad, Stawiski, Whitcomb, Petropoulos, and
  Bonhoeffer}}]{hinkley_systems_2011}
\bibinfo{author}{\bibnamefont{Hinkley}, \bibfnamefont{T.}},
  \bibinfo{author}{\bibfnamefont{J.}~\bibnamefont{Martins}},
  \bibinfo{author}{\bibfnamefont{C.}~\bibnamefont{Chappey}},
  \bibinfo{author}{\bibfnamefont{M.}~\bibnamefont{Haddad}},
  \bibinfo{author}{\bibfnamefont{E.}~\bibnamefont{Stawiski}},
  \bibinfo{author}{\bibfnamefont{J.~M.} \bibnamefont{Whitcomb}},
  \bibinfo{author}{\bibfnamefont{C.~J.} \bibnamefont{Petropoulos}}, and
  \bibinfo{author}{\bibfnamefont{S.}~\bibnamefont{Bonhoeffer}},
  \bibinfo{year}{2011}, \bibinfo{journal}{Nat Genet}
  \textbf{\bibinfo{volume}{43}}(\bibinfo{number}{5}), \bibinfo{pages}{487},
  ISSN \bibinfo{issn}{1061-4036}.

\bibitem[{\citenamefont{Johnson} \emph{et~al.}(2008)\citenamefont{Johnson, Li,
  Wei, Lipscomb, Irlbeck, Craig, Smith, Bennett, Monsour, Sandstrom, Lanier,
  and Heneine}}]{johnson_minority_2008}
\bibinfo{author}{\bibnamefont{Johnson}, \bibfnamefont{J.~A.}},
  \bibinfo{author}{\bibfnamefont{J.-F.} \bibnamefont{Li}},
  \bibinfo{author}{\bibfnamefont{X.}~\bibnamefont{Wei}},
  \bibinfo{author}{\bibfnamefont{J.}~\bibnamefont{Lipscomb}},
  \bibinfo{author}{\bibfnamefont{D.}~\bibnamefont{Irlbeck}},
  \bibinfo{author}{\bibfnamefont{C.}~\bibnamefont{Craig}},
  \bibinfo{author}{\bibfnamefont{A.}~\bibnamefont{Smith}},
  \bibinfo{author}{\bibfnamefont{D.~E.} \bibnamefont{Bennett}},
  \bibinfo{author}{\bibfnamefont{M.}~\bibnamefont{Monsour}},
  \bibinfo{author}{\bibfnamefont{P.}~\bibnamefont{Sandstrom}},
  \bibinfo{author}{\bibfnamefont{E.~R.} \bibnamefont{Lanier}}, and
  \bibinfo{author}{\bibfnamefont{W.}~\bibnamefont{Heneine}},
  \bibinfo{year}{2008}, \bibinfo{journal}{PLoS Med}
  \textbf{\bibinfo{volume}{5}}(\bibinfo{number}{7}), \bibinfo{pages}{e158}.

\bibitem[{\citenamefont{Johnson} \emph{et~al.}(2011)\citenamefont{Johnson,
  Calvez, Günthard, Paredes, Pillay, Shafer, Wensing, and
  Richman}}]{johnson_update_2011}
\bibinfo{author}{\bibnamefont{Johnson}, \bibfnamefont{V.}},
  \bibinfo{author}{\bibfnamefont{V.}~\bibnamefont{Calvez}},
  \bibinfo{author}{\bibfnamefont{H.}~\bibnamefont{Günthard}},
  \bibinfo{author}{\bibfnamefont{R.}~\bibnamefont{Paredes}},
  \bibinfo{author}{\bibfnamefont{D.}~\bibnamefont{Pillay}},
  \bibinfo{author}{\bibfnamefont{R.}~\bibnamefont{Shafer}},
  \bibinfo{author}{\bibfnamefont{A.}~\bibnamefont{Wensing}}, and
  \bibinfo{author}{\bibfnamefont{D.}~\bibnamefont{Richman}},
  \bibinfo{year}{2011}, \bibinfo{journal}{Top Antivir Med}
  \textbf{\bibinfo{volume}{19}}(\bibinfo{number}{4}), \bibinfo{pages}{156},
  ISSN \bibinfo{issn}{2161-5861},
  \urlprefix\url{http://europepmc.org/abstract/med/22156218}.

\bibitem[{\citenamefont{Joos} \emph{et~al.}(2008)\citenamefont{Joos, Fischer,
  Kuster, Pillai, Wong, Böni, Hirschel, Weber, Trkola, Günthard, and
  Study2}}]{joos_hiv_2008}
\bibinfo{author}{\bibnamefont{Joos}, \bibfnamefont{B.}},
  \bibinfo{author}{\bibfnamefont{M.}~\bibnamefont{Fischer}},
  \bibinfo{author}{\bibfnamefont{H.}~\bibnamefont{Kuster}},
  \bibinfo{author}{\bibfnamefont{S.~K.} \bibnamefont{Pillai}},
  \bibinfo{author}{\bibfnamefont{J.~K.} \bibnamefont{Wong}},
  \bibinfo{author}{\bibfnamefont{J.}~\bibnamefont{Böni}},
  \bibinfo{author}{\bibfnamefont{B.}~\bibnamefont{Hirschel}},
  \bibinfo{author}{\bibfnamefont{R.}~\bibnamefont{Weber}},
  \bibinfo{author}{\bibfnamefont{A.}~\bibnamefont{Trkola}},
  \bibinfo{author}{\bibfnamefont{H.~F.} \bibnamefont{Günthard}}, and
  \bibinfo{author}{\bibfnamefont{T.~S. H.~C.} \bibnamefont{Study2}},
  \bibinfo{year}{2008}, \bibinfo{journal}{PNAS}
  \textbf{\bibinfo{volume}{105}}(\bibinfo{number}{43}), \bibinfo{pages}{16725},
  ISSN \bibinfo{issn}{0027-8424, 1091-6490}.

\bibitem[{\citenamefont{Kessinger} \emph{et~al.}(2013)\citenamefont{Kessinger,
  Perelson, and Neher}}]{kessinger_inferring_2013}
\bibinfo{author}{\bibnamefont{Kessinger}, \bibfnamefont{T.~A.}},
  \bibinfo{author}{\bibfnamefont{A.~S.} \bibnamefont{Perelson}}, and
  \bibinfo{author}{\bibfnamefont{R.~A.} \bibnamefont{Neher}},
  \bibinfo{year}{2013}, \bibinfo{journal}{Front. Immunol.}
  \textbf{\bibinfo{volume}{4}}, \bibinfo{pages}{252}.

\bibitem[{\citenamefont{Kimura}(1955)}]{Kimura_stochastic_1955}
\bibinfo{author}{\bibnamefont{Kimura}, \bibfnamefont{M.}},
  \bibinfo{year}{1955}, \bibinfo{journal}{Cold Spring Harb Symp Quant Biol}
  \textbf{\bibinfo{volume}{20}}, \bibinfo{pages}{33}.

\bibitem[{\citenamefont{Kimura}(1968)}]{kimura_evolutionary_1968}
\bibinfo{author}{\bibnamefont{Kimura}, \bibfnamefont{M.}},
  \bibinfo{year}{1968}, \bibinfo{journal}{Nature}
  \textbf{\bibinfo{volume}{217}}(\bibinfo{number}{5129}), \bibinfo{pages}{624}.

\bibitem[{\citenamefont{{LANL HIV sequence data base}}(2016)}]{LANL_annotation}
\bibinfo{author}{\bibnamefont{{LANL HIV sequence data base}}},
  \bibinfo{year}{2016}, \bibinfo{title}{{HXB2} genome annotation},
  \urlprefix\url{http://www.hiv.lanl.gov/content/sequence/HIV/MAP/annotation.html}.

\bibitem[{\citenamefont{Leslie} \emph{et~al.}(2004)\citenamefont{Leslie,
  Pfafferott, Chetty, Draenert, Addo, Feeney, Tang, Holmes, Allen, Prado,
  Altfeld, Brander} \emph{et~al.}}]{leslie_hiv_2004}
\bibinfo{author}{\bibnamefont{Leslie}, \bibfnamefont{A.~J.}},
  \bibinfo{author}{\bibfnamefont{K.~J.} \bibnamefont{Pfafferott}},
  \bibinfo{author}{\bibfnamefont{P.}~\bibnamefont{Chetty}},
  \bibinfo{author}{\bibfnamefont{R.}~\bibnamefont{Draenert}},
  \bibinfo{author}{\bibfnamefont{M.~M.} \bibnamefont{Addo}},
  \bibinfo{author}{\bibfnamefont{M.}~\bibnamefont{Feeney}},
  \bibinfo{author}{\bibfnamefont{Y.}~\bibnamefont{Tang}},
  \bibinfo{author}{\bibfnamefont{E.~C.} \bibnamefont{Holmes}},
  \bibinfo{author}{\bibfnamefont{T.}~\bibnamefont{Allen}},
  \bibinfo{author}{\bibfnamefont{J.~G.} \bibnamefont{Prado}},
  \bibinfo{author}{\bibfnamefont{M.}~\bibnamefont{Altfeld}},
  \bibinfo{author}{\bibfnamefont{C.}~\bibnamefont{Brander}}, \emph{et~al.},
  \bibinfo{year}{2004}, \bibinfo{journal}{Nat. Med.}
  \textbf{\bibinfo{volume}{10}}(\bibinfo{number}{3}), \bibinfo{pages}{282}.

\bibitem[{\citenamefont{Lewis} \emph{et~al.}(1999)\citenamefont{Lewis, Bebenek,
  Beard, Wilson, and Kunkel}}]{lewis_uniquely_1999}
\bibinfo{author}{\bibnamefont{Lewis}, \bibfnamefont{D.~A.}},
  \bibinfo{author}{\bibfnamefont{K.}~\bibnamefont{Bebenek}},
  \bibinfo{author}{\bibfnamefont{W.~A.} \bibnamefont{Beard}},
  \bibinfo{author}{\bibfnamefont{S.~H.} \bibnamefont{Wilson}}, and
  \bibinfo{author}{\bibfnamefont{T.~A.} \bibnamefont{Kunkel}},
  \bibinfo{year}{1999}, \bibinfo{journal}{Journal of Biological Chemistry}
  \textbf{\bibinfo{volume}{274}}(\bibinfo{number}{46}), \bibinfo{pages}{32924},
  ISSN \bibinfo{issn}{0021-9258, 1083-351X}.

\bibitem[{\citenamefont{Li} \emph{et~al.}(2007)\citenamefont{Li, Gladden,
  Altfeld, Kaldor, Cooper, Kelleher, and Allen}}]{li_rapid_2007}
\bibinfo{author}{\bibnamefont{Li}, \bibfnamefont{B.}},
  \bibinfo{author}{\bibfnamefont{A.~D.} \bibnamefont{Gladden}},
  \bibinfo{author}{\bibfnamefont{M.}~\bibnamefont{Altfeld}},
  \bibinfo{author}{\bibfnamefont{J.~M.} \bibnamefont{Kaldor}},
  \bibinfo{author}{\bibfnamefont{D.~A.} \bibnamefont{Cooper}},
  \bibinfo{author}{\bibfnamefont{A.~D.} \bibnamefont{Kelleher}}, and
  \bibinfo{author}{\bibfnamefont{T.~M.} \bibnamefont{Allen}},
  \bibinfo{year}{2007}, \bibinfo{journal}{J. Virol.}
  \textbf{\bibinfo{volume}{81}}(\bibinfo{number}{1}), \bibinfo{pages}{193}.

\bibitem[{\citenamefont{Li} \emph{et~al.}(2015)\citenamefont{Li, Piampongsant,
  Faria, Voet, Pineda-Peña, Khouri, Lemey, Vandamme, and
  Theys}}]{li_integrated_2015}
\bibinfo{author}{\bibnamefont{Li}, \bibfnamefont{G.}},
  \bibinfo{author}{\bibfnamefont{S.}~\bibnamefont{Piampongsant}},
  \bibinfo{author}{\bibfnamefont{N.~R.} \bibnamefont{Faria}},
  \bibinfo{author}{\bibfnamefont{A.}~\bibnamefont{Voet}},
  \bibinfo{author}{\bibfnamefont{A.~a.-C.} \bibnamefont{Pineda-Peña}},
  \bibinfo{author}{\bibfnamefont{R.}~\bibnamefont{Khouri}},
  \bibinfo{author}{\bibfnamefont{P.}~\bibnamefont{Lemey}},
  \bibinfo{author}{\bibfnamefont{A.-M.} \bibnamefont{Vandamme}}, and
  \bibinfo{author}{\bibfnamefont{K.}~\bibnamefont{Theys}},
  \bibinfo{year}{2015}, \bibinfo{journal}{Retrovirology}
  \textbf{\bibinfo{volume}{12}}(\bibinfo{number}{1}), \bibinfo{pages}{18}, ISSN
  \bibinfo{issn}{1742-4690}.

\bibitem[{\citenamefont{Li} \emph{et~al.}(2011)\citenamefont{Li, Paredes,
  Ribaudo, Svarovskaia, Metzner, Kozal, Hullsiek, Balduin, Jakobsen, Geretti,
  Thiebaut, Ostergaard} \emph{et~al.}}]{li_low-frequency_2011}
\bibinfo{author}{\bibnamefont{Li}, \bibfnamefont{J.~Z.}},
  \bibinfo{author}{\bibfnamefont{R.}~\bibnamefont{Paredes}},
  \bibinfo{author}{\bibfnamefont{H.~J.} \bibnamefont{Ribaudo}},
  \bibinfo{author}{\bibfnamefont{E.~S.} \bibnamefont{Svarovskaia}},
  \bibinfo{author}{\bibfnamefont{K.~J.} \bibnamefont{Metzner}},
  \bibinfo{author}{\bibfnamefont{M.~J.} \bibnamefont{Kozal}},
  \bibinfo{author}{\bibfnamefont{K.~H.} \bibnamefont{Hullsiek}},
  \bibinfo{author}{\bibfnamefont{M.}~\bibnamefont{Balduin}},
  \bibinfo{author}{\bibfnamefont{M.~R.} \bibnamefont{Jakobsen}},
  \bibinfo{author}{\bibfnamefont{A.~M.} \bibnamefont{Geretti}},
  \bibinfo{author}{\bibfnamefont{R.}~\bibnamefont{Thiebaut}},
  \bibinfo{author}{\bibfnamefont{L.}~\bibnamefont{Ostergaard}}, \emph{et~al.},
  \bibinfo{year}{2011}, \bibinfo{journal}{JAMA}
  \textbf{\bibinfo{volume}{305}}(\bibinfo{number}{13}), \bibinfo{pages}{1327}.

\bibitem[{\citenamefont{Lu} \emph{et~al.}(2011)\citenamefont{Lu, Heng, Garyu,
  Monti, Garcia, Kharytonchyk, Dorjsuren, Kulandaivel, Jones, Hiremath,
  Divakaruni, LaCotti} \emph{et~al.}}]{lu_nmr_2011}
\bibinfo{author}{\bibnamefont{Lu}, \bibfnamefont{K.}},
  \bibinfo{author}{\bibfnamefont{X.}~\bibnamefont{Heng}},
  \bibinfo{author}{\bibfnamefont{L.}~\bibnamefont{Garyu}},
  \bibinfo{author}{\bibfnamefont{S.}~\bibnamefont{Monti}},
  \bibinfo{author}{\bibfnamefont{E.~L.} \bibnamefont{Garcia}},
  \bibinfo{author}{\bibfnamefont{S.}~\bibnamefont{Kharytonchyk}},
  \bibinfo{author}{\bibfnamefont{B.}~\bibnamefont{Dorjsuren}},
  \bibinfo{author}{\bibfnamefont{G.}~\bibnamefont{Kulandaivel}},
  \bibinfo{author}{\bibfnamefont{S.}~\bibnamefont{Jones}},
  \bibinfo{author}{\bibfnamefont{A.}~\bibnamefont{Hiremath}},
  \bibinfo{author}{\bibfnamefont{S.~S.} \bibnamefont{Divakaruni}},
  \bibinfo{author}{\bibfnamefont{C.}~\bibnamefont{LaCotti}}, \emph{et~al.},
  \bibinfo{year}{2011}, \bibinfo{journal}{Science}
  \textbf{\bibinfo{volume}{334}}(\bibinfo{number}{6053}), \bibinfo{pages}{242},
  ISSN \bibinfo{issn}{1095-9203}.

\bibitem[{\citenamefont{Malim}(2009)}]{malim_apobec_2009}
\bibinfo{author}{\bibnamefont{Malim}, \bibfnamefont{M.~H.}},
  \bibinfo{year}{2009}, \bibinfo{journal}{Philosophical Transactions of the
  Royal Society of London B: Biological Sciences}
  \textbf{\bibinfo{volume}{364}}(\bibinfo{number}{1517}), \bibinfo{pages}{675},
  ISSN \bibinfo{issn}{0962-8436, 1471-2970}.

\bibitem[{\citenamefont{Mansky and Temin}(1995)}]{mansky_lower_1995}
\bibinfo{author}{\bibnamefont{Mansky}, \bibfnamefont{L.~M.}}, and
  \bibinfo{author}{\bibfnamefont{H.~M.} \bibnamefont{Temin}},
  \bibinfo{year}{1995}, \bibinfo{journal}{J. Virol.}
  \textbf{\bibinfo{volume}{69}}(\bibinfo{number}{8}), \bibinfo{pages}{5087},
  ISSN \bibinfo{issn}{0022-538X, 1098-5514}.

\bibitem[{\citenamefont{Martinez-Picado and
  Martinez}(2008)}]{martinez-picado_hiv-1_2008}
\bibinfo{author}{\bibnamefont{Martinez-Picado}, \bibfnamefont{J.}}, and
  \bibinfo{author}{\bibfnamefont{M.~A.} \bibnamefont{Martinez}},
  \bibinfo{year}{2008}, \bibinfo{journal}{Virus Research}
  \textbf{\bibinfo{volume}{134}}(\bibinfo{number}{1–2}),
  \bibinfo{pages}{104}, ISSN \bibinfo{issn}{0168-1702}.

\bibitem[{\citenamefont{Mayrose} \emph{et~al.}(2013)\citenamefont{Mayrose,
  Stern, Burdelova, Sabo, Laham-Karam, Zamostiano, Bacharach, and
  Pupko}}]{mayrose_synonymous_2013}
\bibinfo{author}{\bibnamefont{Mayrose}, \bibfnamefont{I.}},
  \bibinfo{author}{\bibfnamefont{A.}~\bibnamefont{Stern}},
  \bibinfo{author}{\bibfnamefont{E.~O.} \bibnamefont{Burdelova}},
  \bibinfo{author}{\bibfnamefont{Y.}~\bibnamefont{Sabo}},
  \bibinfo{author}{\bibfnamefont{N.}~\bibnamefont{Laham-Karam}},
  \bibinfo{author}{\bibfnamefont{R.}~\bibnamefont{Zamostiano}},
  \bibinfo{author}{\bibfnamefont{E.}~\bibnamefont{Bacharach}}, and
  \bibinfo{author}{\bibfnamefont{T.}~\bibnamefont{Pupko}},
  \bibinfo{year}{2013}, \bibinfo{journal}{BMC Evolutionary Biology}
  \textbf{\bibinfo{volume}{13}}, \bibinfo{pages}{164}, ISSN
  \bibinfo{issn}{1471-2148}.

\bibitem[{\citenamefont{Meyer and Wilke}(2015)}]{meyer_utility_2015}
\bibinfo{author}{\bibnamefont{Meyer}, \bibfnamefont{A.~G.}}, and
  \bibinfo{author}{\bibfnamefont{C.~O.} \bibnamefont{Wilke}},
  \bibinfo{year}{2015}, \bibinfo{journal}{Journal of The Royal Society
  Interface} \textbf{\bibinfo{volume}{12}}(\bibinfo{number}{111}),
  \bibinfo{pages}{20150579}, ISSN \bibinfo{issn}{1742-5689, 1742-5662}.

\bibitem[{\citenamefont{Neher}(2013)}]{neher_genetic_2013}
\bibinfo{author}{\bibnamefont{Neher}, \bibfnamefont{R.~A.}},
  \bibinfo{year}{2013}, \bibinfo{journal}{Annual Review of Ecology, Evolution,
  and Systematics} \textbf{\bibinfo{volume}{44}}(\bibinfo{number}{1}),
  \bibinfo{pages}{null}.

\bibitem[{\citenamefont{Neher and Leitner}(2010)}]{neher_recombination_2010}
\bibinfo{author}{\bibnamefont{Neher}, \bibfnamefont{R.~A.}}, and
  \bibinfo{author}{\bibfnamefont{T.}~\bibnamefont{Leitner}},
  \bibinfo{year}{2010}, \bibinfo{journal}{PLoS Comput Biol}
  \textbf{\bibinfo{volume}{6}}(\bibinfo{number}{1}), \bibinfo{pages}{e1000660}.

\bibitem[{\citenamefont{Ngandu} \emph{et~al.}(2008)\citenamefont{Ngandu,
  Scheffler, Moore, Woodman, Martin, and Seoighe}}]{ngandu_extensive_2008}
\bibinfo{author}{\bibnamefont{Ngandu}, \bibfnamefont{N.~K.}},
  \bibinfo{author}{\bibfnamefont{K.}~\bibnamefont{Scheffler}},
  \bibinfo{author}{\bibfnamefont{P.}~\bibnamefont{Moore}},
  \bibinfo{author}{\bibfnamefont{Z.}~\bibnamefont{Woodman}},
  \bibinfo{author}{\bibfnamefont{D.}~\bibnamefont{Martin}}, and
  \bibinfo{author}{\bibfnamefont{C.}~\bibnamefont{Seoighe}},
  \bibinfo{year}{2008}, \bibinfo{journal}{Virology Journal}
  \textbf{\bibinfo{volume}{5}}, \bibinfo{pages}{160}, ISSN
  \bibinfo{issn}{1743-422X}.

\bibitem[{\citenamefont{Parera} \emph{et~al.}(2007)\citenamefont{Parera,
  Fernàndez, Clotet, and Martínez}}]{parera_hiv-1_2007}
\bibinfo{author}{\bibnamefont{Parera}, \bibfnamefont{M.}},
  \bibinfo{author}{\bibfnamefont{G.}~\bibnamefont{Fernàndez}},
  \bibinfo{author}{\bibfnamefont{B.}~\bibnamefont{Clotet}}, and
  \bibinfo{author}{\bibfnamefont{M.~A.} \bibnamefont{Martínez}},
  \bibinfo{year}{2007}, \bibinfo{journal}{Mol Biol Evol}
  \textbf{\bibinfo{volume}{24}}(\bibinfo{number}{2}), \bibinfo{pages}{382},
  ISSN \bibinfo{issn}{0737-4038, 1537-1719}.

\bibitem[{\citenamefont{Pereyra} \emph{et~al.}(2014)\citenamefont{Pereyra,
  Heckerman, Carlson, Kadie, Soghoian, Karel, Goldenthal, Davis, DeZiel, Lin,
  Peng, Piechocka} \emph{et~al.}}]{pereyra_HIV_2014}
\bibinfo{author}{\bibnamefont{Pereyra}, \bibfnamefont{F.}},
  \bibinfo{author}{\bibfnamefont{D.}~\bibnamefont{Heckerman}},
  \bibinfo{author}{\bibfnamefont{J.~M.} \bibnamefont{Carlson}},
  \bibinfo{author}{\bibfnamefont{C.}~\bibnamefont{Kadie}},
  \bibinfo{author}{\bibfnamefont{D.~Z.} \bibnamefont{Soghoian}},
  \bibinfo{author}{\bibfnamefont{D.}~\bibnamefont{Karel}},
  \bibinfo{author}{\bibfnamefont{A.}~\bibnamefont{Goldenthal}},
  \bibinfo{author}{\bibfnamefont{O.~B.} \bibnamefont{Davis}},
  \bibinfo{author}{\bibfnamefont{C.~E.} \bibnamefont{DeZiel}},
  \bibinfo{author}{\bibfnamefont{T.}~\bibnamefont{Lin}},
  \bibinfo{author}{\bibfnamefont{J.}~\bibnamefont{Peng}},
  \bibinfo{author}{\bibfnamefont{A.}~\bibnamefont{Piechocka}}, \emph{et~al.},
  \bibinfo{year}{2014}, \bibinfo{journal}{J. Virol.}
  \textbf{\bibinfo{volume}{88}}(\bibinfo{number}{22}), \bibinfo{pages}{12937},
  ISSN \bibinfo{issn}{0022-538X, 1098-5514}.

\bibitem[{\citenamefont{Petropoulos}
  \emph{et~al.}(2000)\citenamefont{Petropoulos, Parkin, Limoli, Lie, Wrin,
  Huang, Tian, Smith, Winslow, Capon, and Whitcomb}}]{petropoulos_novel_2000}
\bibinfo{author}{\bibnamefont{Petropoulos}, \bibfnamefont{C.~J.}},
  \bibinfo{author}{\bibfnamefont{N.~T.} \bibnamefont{Parkin}},
  \bibinfo{author}{\bibfnamefont{K.~L.} \bibnamefont{Limoli}},
  \bibinfo{author}{\bibfnamefont{Y.~S.} \bibnamefont{Lie}},
  \bibinfo{author}{\bibfnamefont{T.}~\bibnamefont{Wrin}},
  \bibinfo{author}{\bibfnamefont{W.}~\bibnamefont{Huang}},
  \bibinfo{author}{\bibfnamefont{H.}~\bibnamefont{Tian}},
  \bibinfo{author}{\bibfnamefont{D.}~\bibnamefont{Smith}},
  \bibinfo{author}{\bibfnamefont{G.~A.} \bibnamefont{Winslow}},
  \bibinfo{author}{\bibfnamefont{D.~J.} \bibnamefont{Capon}}, and
  \bibinfo{author}{\bibfnamefont{J.~M.} \bibnamefont{Whitcomb}},
  \bibinfo{year}{2000}, \bibinfo{journal}{Antimicrob. Agents Chemother.}
  \textbf{\bibinfo{volume}{44}}(\bibinfo{number}{4}), \bibinfo{pages}{920},
  ISSN \bibinfo{issn}{0066-4804, 1098-6596}.

\bibitem[{\citenamefont{Pollom} \emph{et~al.}(2013)\citenamefont{Pollom, Dang,
  Potter, Gorelick, Burch, Weeks, and Swanstrom}}]{pollom_comparison_2013}
\bibinfo{author}{\bibnamefont{Pollom}, \bibfnamefont{E.}},
  \bibinfo{author}{\bibfnamefont{K.~K.} \bibnamefont{Dang}},
  \bibinfo{author}{\bibfnamefont{E.~L.} \bibnamefont{Potter}},
  \bibinfo{author}{\bibfnamefont{R.~J.} \bibnamefont{Gorelick}},
  \bibinfo{author}{\bibfnamefont{C.~L.} \bibnamefont{Burch}},
  \bibinfo{author}{\bibfnamefont{K.~M.} \bibnamefont{Weeks}}, and
  \bibinfo{author}{\bibfnamefont{R.}~\bibnamefont{Swanstrom}},
  \bibinfo{year}{2013}, \bibinfo{journal}{PLOS Pathog}
  \textbf{\bibinfo{volume}{9}}(\bibinfo{number}{4}), \bibinfo{pages}{e1003294},
  ISSN \bibinfo{issn}{1553-7374}.

\bibitem[{\citenamefont{Rihn} \emph{et~al.}(2015)\citenamefont{Rihn, Hughes,
  Wilson, and Bieniasz}}]{rihn_uneven_2015}
\bibinfo{author}{\bibnamefont{Rihn}, \bibfnamefont{S.~J.}},
  \bibinfo{author}{\bibfnamefont{J.}~\bibnamefont{Hughes}},
  \bibinfo{author}{\bibfnamefont{S.~J.} \bibnamefont{Wilson}}, and
  \bibinfo{author}{\bibfnamefont{P.~D.} \bibnamefont{Bieniasz}},
  \bibinfo{year}{2015}, \bibinfo{journal}{Journal of Virology}
  \textbf{\bibinfo{volume}{89}}(\bibinfo{number}{1}), \bibinfo{pages}{552},
  ISSN \bibinfo{issn}{0022-538X, 1098-5514}.

\bibitem[{\citenamefont{Sanju\'an}(2010)}]{sanjuan_mutational_2010}
\bibinfo{author}{\bibnamefont{Sanju\'an}, \bibfnamefont{R.}},
  \bibinfo{year}{2010}, \bibinfo{journal}{Philosophical Transactions of the
  Royal Society of London B: Biological Sciences}
  \textbf{\bibinfo{volume}{365}}(\bibinfo{number}{1548}),
  \bibinfo{pages}{1975}.

\bibitem[{\citenamefont{Schneidewind}
  \emph{et~al.}(2009)\citenamefont{Schneidewind, Brumme, Brumme, Power, Reyor,
  O'Sullivan, Gladden, Hempel, Kuntzen, Wang, Oniangue-Ndza, Jessen}
  \emph{et~al.}}]{schneidewind_transmission_2009}
\bibinfo{author}{\bibnamefont{Schneidewind}, \bibfnamefont{A.}},
  \bibinfo{author}{\bibfnamefont{Z.~L.} \bibnamefont{Brumme}},
  \bibinfo{author}{\bibfnamefont{C.~J.} \bibnamefont{Brumme}},
  \bibinfo{author}{\bibfnamefont{K.~A.} \bibnamefont{Power}},
  \bibinfo{author}{\bibfnamefont{L.~L.} \bibnamefont{Reyor}},
  \bibinfo{author}{\bibfnamefont{K.}~\bibnamefont{O'Sullivan}},
  \bibinfo{author}{\bibfnamefont{A.}~\bibnamefont{Gladden}},
  \bibinfo{author}{\bibfnamefont{U.}~\bibnamefont{Hempel}},
  \bibinfo{author}{\bibfnamefont{T.}~\bibnamefont{Kuntzen}},
  \bibinfo{author}{\bibfnamefont{Y.~E.} \bibnamefont{Wang}},
  \bibinfo{author}{\bibfnamefont{C.}~\bibnamefont{Oniangue-Ndza}},
  \bibinfo{author}{\bibfnamefont{H.}~\bibnamefont{Jessen}}, \emph{et~al.},
  \bibinfo{year}{2009}, \bibinfo{journal}{J. Virol.}
  \textbf{\bibinfo{volume}{83}}(\bibinfo{number}{8}), \bibinfo{pages}{3993},
  ISSN \bibinfo{issn}{0022-538X, 1098-5514},
  \urlprefix\url{http://jvi.asm.org/content/83/8/3993}.

\bibitem[{\citenamefont{Seifert} \emph{et~al.}(2015)\citenamefont{Seifert,
  Di~Giallonardo, Metzner, Günthard, and
  Beerenwinkel}}]{seifert_framework_2015}
\bibinfo{author}{\bibnamefont{Seifert}, \bibfnamefont{D.}},
  \bibinfo{author}{\bibfnamefont{F.}~\bibnamefont{Di~Giallonardo}},
  \bibinfo{author}{\bibfnamefont{K.~J.} \bibnamefont{Metzner}},
  \bibinfo{author}{\bibfnamefont{H.~F.} \bibnamefont{Günthard}}, and
  \bibinfo{author}{\bibfnamefont{N.}~\bibnamefont{Beerenwinkel}},
  \bibinfo{year}{2015}, \bibinfo{journal}{Genetics}
  \textbf{\bibinfo{volume}{199}}(\bibinfo{number}{1}), \bibinfo{pages}{191},
  ISSN \bibinfo{issn}{1943-2631}.

\bibitem[{\citenamefont{Shekhar} \emph{et~al.}(2013)\citenamefont{Shekhar,
  Ruberman, Ferguson, Barton, Kardar, and Chakraborty}}]{shekhar_spin_2013}
\bibinfo{author}{\bibnamefont{Shekhar}, \bibfnamefont{K.}},
  \bibinfo{author}{\bibfnamefont{C.~F.} \bibnamefont{Ruberman}},
  \bibinfo{author}{\bibfnamefont{A.~L.} \bibnamefont{Ferguson}},
  \bibinfo{author}{\bibfnamefont{J.~P.} \bibnamefont{Barton}},
  \bibinfo{author}{\bibfnamefont{M.}~\bibnamefont{Kardar}}, and
  \bibinfo{author}{\bibfnamefont{A.~K.} \bibnamefont{Chakraborty}},
  \bibinfo{year}{2013}, \bibinfo{journal}{Phys. Rev. E}
  \textbf{\bibinfo{volume}{88}}(\bibinfo{number}{6}), \bibinfo{pages}{062705}.

\bibitem[{\citenamefont{Siegfried} \emph{et~al.}(2014)\citenamefont{Siegfried,
  Busan, Rice, Nelson, and Weeks}}]{siegfried_rna_2014}
\bibinfo{author}{\bibnamefont{Siegfried}, \bibfnamefont{N.~A.}},
  \bibinfo{author}{\bibfnamefont{S.}~\bibnamefont{Busan}},
  \bibinfo{author}{\bibfnamefont{G.~M.} \bibnamefont{Rice}},
  \bibinfo{author}{\bibfnamefont{J.~A.~E.} \bibnamefont{Nelson}}, and
  \bibinfo{author}{\bibfnamefont{K.~M.} \bibnamefont{Weeks}},
  \bibinfo{year}{2014}, \bibinfo{journal}{Nat Meth}
  \textbf{\bibinfo{volume}{11}}, \bibinfo{pages}{959}, ISSN
  \bibinfo{issn}{1548-7091}.

\bibitem[{\citenamefont{S\"uk\"osd}
  \emph{et~al.}(2015)\citenamefont{S\"uk\"osd, Andersen, Seemann, Jensen,
  Hansen, Gorodkin, and Kjems}}]{sukosd_full-length_2015}
\bibinfo{author}{\bibnamefont{S\"uk\"osd}, \bibfnamefont{Z.}},
  \bibinfo{author}{\bibfnamefont{E.~S.} \bibnamefont{Andersen}},
  \bibinfo{author}{\bibfnamefont{S.~E.} \bibnamefont{Seemann}},
  \bibinfo{author}{\bibfnamefont{M.~K.} \bibnamefont{Jensen}},
  \bibinfo{author}{\bibfnamefont{M.}~\bibnamefont{Hansen}},
  \bibinfo{author}{\bibfnamefont{J.}~\bibnamefont{Gorodkin}}, and
  \bibinfo{author}{\bibfnamefont{J.}~\bibnamefont{Kjems}},
  \bibinfo{year}{2015}, \bibinfo{journal}{Nucl. Acids Res.} ,
  \bibinfo{pages}{gkv1039}.

\bibitem[{\citenamefont{Thyagarajan and
  Bloom}(2014)}]{thyagarajan_inherent_2014}
\bibinfo{author}{\bibnamefont{Thyagarajan}, \bibfnamefont{B.}}, and
  \bibinfo{author}{\bibfnamefont{J.~D.} \bibnamefont{Bloom}},
  \bibinfo{year}{2014}, \bibinfo{journal}{eLife Sciences}
  \textbf{\bibinfo{volume}{3}}, \bibinfo{pages}{e03300}, ISSN
  \bibinfo{issn}{2050-084X}.

\bibitem[{\citenamefont{Vabret} \emph{et~al.}(2014)\citenamefont{Vabret,
  Bailly-Bechet, Lepelley, Najburg, Schwartz, Verrier, and
  Tangy}}]{vabret_large-scale_2014}
\bibinfo{author}{\bibnamefont{Vabret}, \bibfnamefont{N.}},
  \bibinfo{author}{\bibfnamefont{M.}~\bibnamefont{Bailly-Bechet}},
  \bibinfo{author}{\bibfnamefont{A.}~\bibnamefont{Lepelley}},
  \bibinfo{author}{\bibfnamefont{V.}~\bibnamefont{Najburg}},
  \bibinfo{author}{\bibfnamefont{O.}~\bibnamefont{Schwartz}},
  \bibinfo{author}{\bibfnamefont{B.}~\bibnamefont{Verrier}}, and
  \bibinfo{author}{\bibfnamefont{F.}~\bibnamefont{Tangy}},
  \bibinfo{year}{2014}, \bibinfo{journal}{J. Virol.}
  \textbf{\bibinfo{volume}{88}}(\bibinfo{number}{8}), \bibinfo{pages}{4161},
  ISSN \bibinfo{issn}{1098-5514}.

\bibitem[{\citenamefont{de~Visser and Krug}(2014)}]{de_visser_empirical_2014}
\bibinfo{author}{\bibnamefont{de~Visser}, \bibfnamefont{J.~A. G.~M.}}, and
  \bibinfo{author}{\bibfnamefont{J.}~\bibnamefont{Krug}}, \bibinfo{year}{2014},
  \bibinfo{journal}{Nat Rev Genet}
  \textbf{\bibinfo{volume}{15}}(\bibinfo{number}{7}), \bibinfo{pages}{480},
  ISSN \bibinfo{issn}{1471-0056}.

\bibitem[{\citenamefont{Walker and McMichael}(2012)}]{walker_t-cell_2012}
\bibinfo{author}{\bibnamefont{Walker}, \bibfnamefont{B.}}, and
  \bibinfo{author}{\bibfnamefont{A.}~\bibnamefont{McMichael}},
  \bibinfo{year}{2012}, \bibinfo{journal}{Cold Spring Harb Perspect Med}
  \textbf{\bibinfo{volume}{2}}(\bibinfo{number}{11}).

\bibitem[{\citenamefont{Zanini} \emph{et~al.}(2016)\citenamefont{Zanini,
  Brodin, Thebo, Lanz, Bratt, Albert, and Neher}}]{zanini_population_2016}
\bibinfo{author}{\bibnamefont{Zanini}, \bibfnamefont{F.}},
  \bibinfo{author}{\bibfnamefont{J.}~\bibnamefont{Brodin}},
  \bibinfo{author}{\bibfnamefont{L.}~\bibnamefont{Thebo}},
  \bibinfo{author}{\bibfnamefont{C.}~\bibnamefont{Lanz}},
  \bibinfo{author}{\bibfnamefont{G.}~\bibnamefont{Bratt}},
  \bibinfo{author}{\bibfnamefont{J.}~\bibnamefont{Albert}}, and
  \bibinfo{author}{\bibfnamefont{R.~A.} \bibnamefont{Neher}},
  \bibinfo{year}{2016}, \bibinfo{journal}{eLife Sciences}
  \textbf{\bibinfo{volume}{4}}, \bibinfo{pages}{e11282}, ISSN
  \bibinfo{issn}{2050-084X},
  \urlprefix\url{http://elifesciences.org/content/4/e11282}.

\bibitem[{\citenamefont{Zanini and Neher}(2013)}]{zanini_quantifying_2013}
\bibinfo{author}{\bibnamefont{Zanini}, \bibfnamefont{F.}}, and
  \bibinfo{author}{\bibfnamefont{R.~A.} \bibnamefont{Neher}},
  \bibinfo{year}{2013}, \bibinfo{journal}{J. Virol.}
  \textbf{\bibinfo{volume}{87}}(\bibinfo{number}{21}), \bibinfo{pages}{11843},
  ISSN \bibinfo{issn}{0022-538X, 1098-5514},
  \urlprefix\url{http://jvi.asm.org/content/87/21/11843}.

\end{thebibliography}

\clearpage
\onecolumngrid
\appendix
\renewcommand{\thefigure}{S\arabic{figure}}
\setcounter{figure}{0}
\section*{Supplementary material}

\begin{figure}[h!]
    \centering
    \includegraphics[width=0.48\columnwidth]{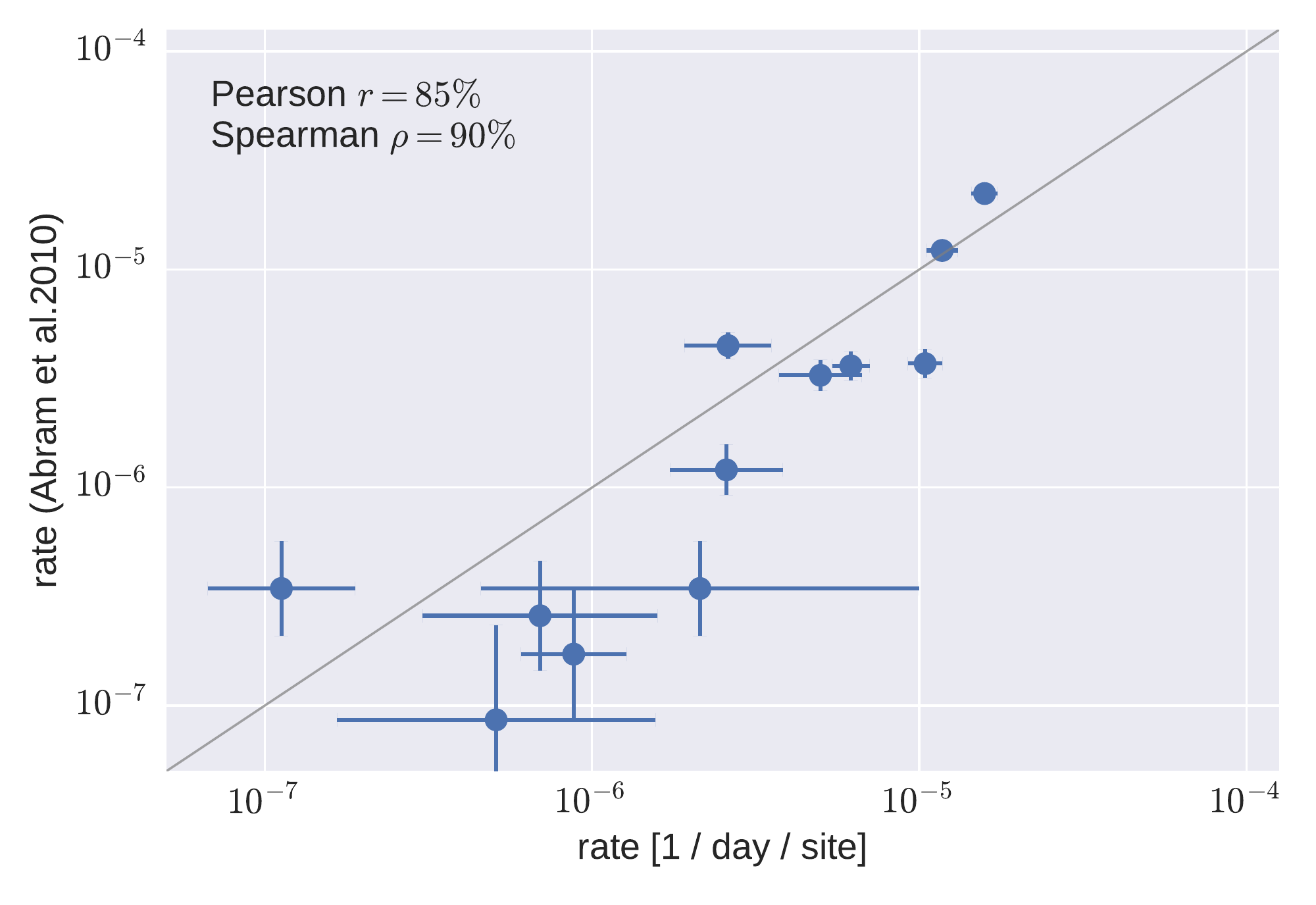}
    \caption{
    Comparison of our estimates for the neutral mutation rates to \textit{in vitro} estimates by \citet{abram_nature_2010}. Error bars for the estimates are standard deviations over 100 patient bootstraps. Error bars for the values from Abram et al.\ (2010) are standard deviations of binomial sampling noise (low-frequency mutations were observed 1-2 times only in that study).
    }
    \label{fig:mutation_rate_comparison}
\end{figure}

\begin{figure}[h!]
    \centering
    \includegraphics[width=0.48\columnwidth]{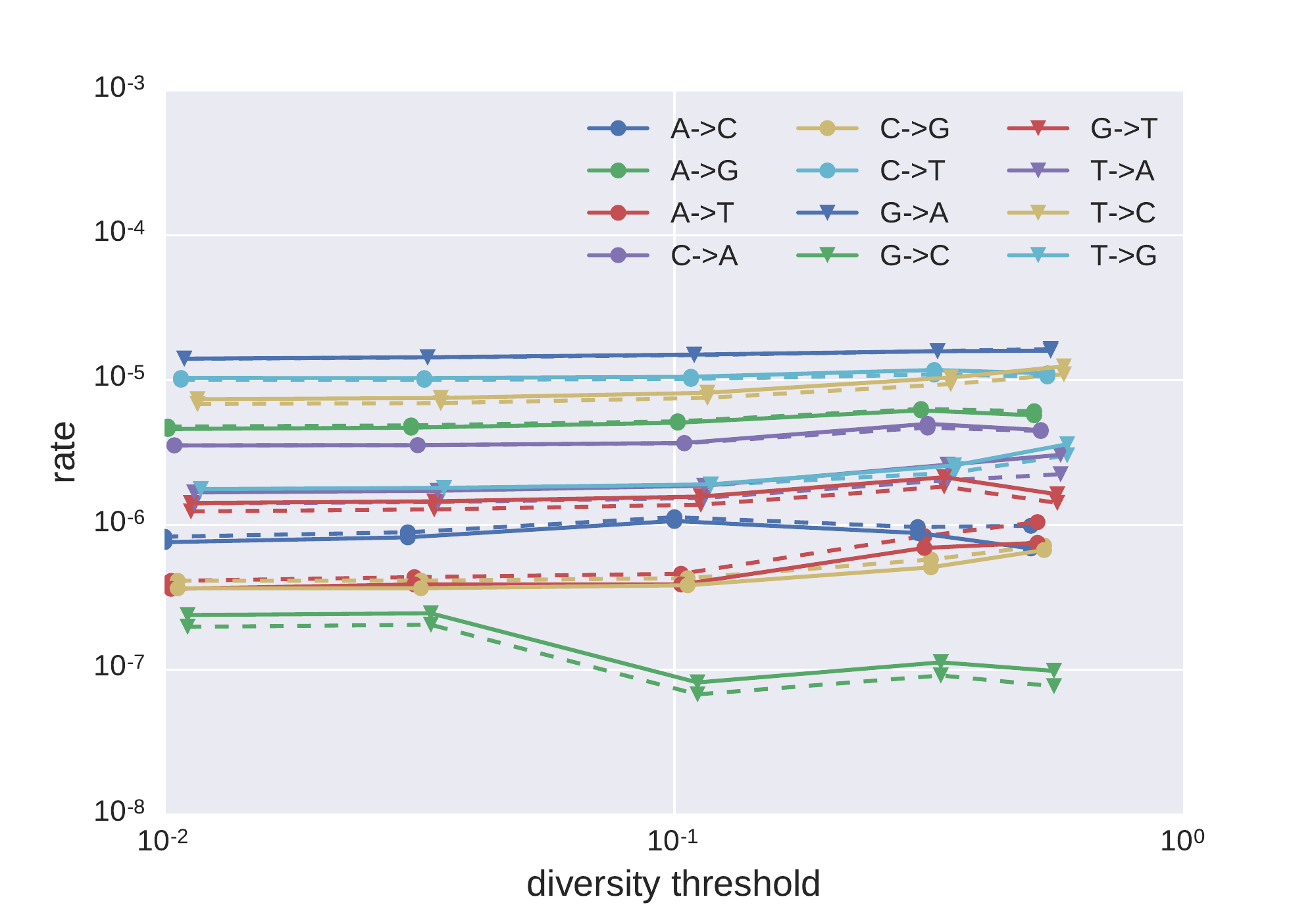}
    \includegraphics[width=0.48\columnwidth]{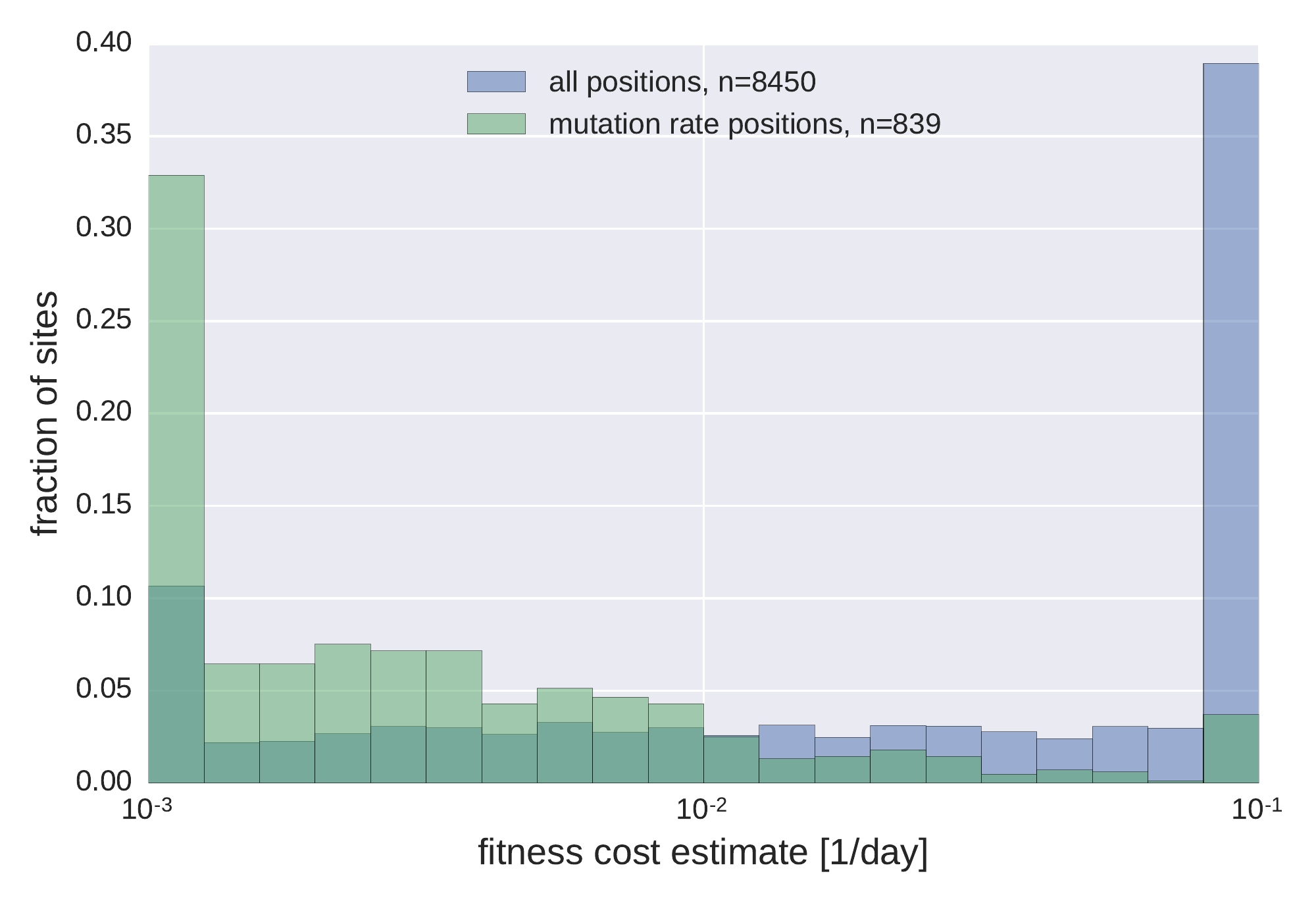}
    \caption{Sensitivity of mutation rate estimates on the criteria used to define the set of approximately neutral positions.
    (A) Mutation rate estimates depend only weakly on the threshold used to define the approximately neutral set of positions or whether gp120 is included or not. (B) The positions chosen to estimate the neutral mutation rate are among the most neutral positions as estimated by the saturation of intrapatient frequencies. Note that frequencies of neutral mutations don't saturate and can be less diverse than expected due to linked selection and drift; this is not a problem for our estimates as we do not infer site-specific mutation rates.
    }
    \label{fig:mutation_rate_sensitivity}
\end{figure}

\begin{figure}
    \centering
    \includegraphics[width=0.7\columnwidth]{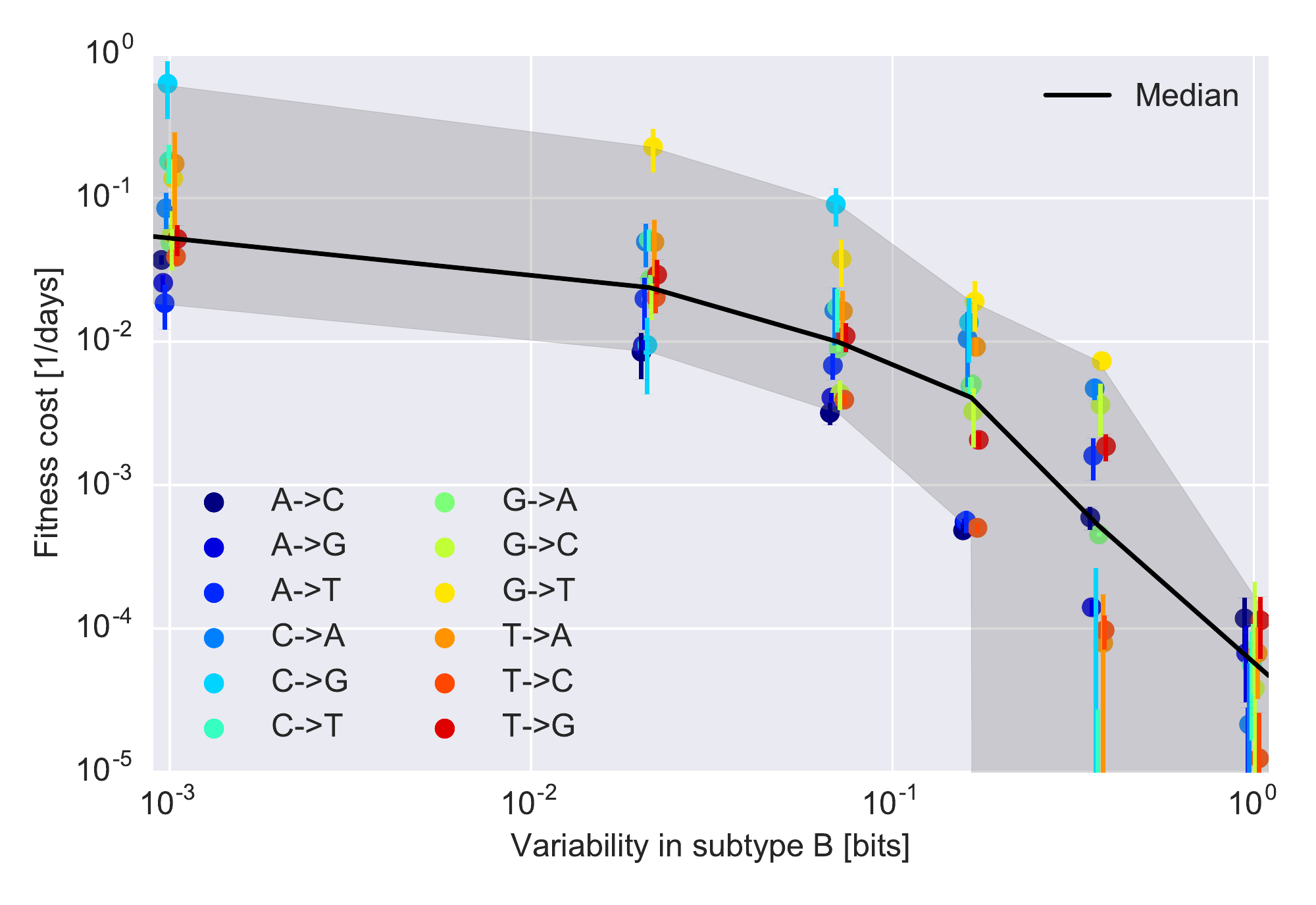}
    \caption{Fitness cost estimates as a function of subtype conservation, from saturation curves similar to \FIG{sat} "Sat" but separate for each of the 12 mutations. The general picture is the same like shown in \FIG{sat}, but some mutations appear to be slightly more or less suppressed than the average.}
    \label{fig:fitness_cost_complex}
\end{figure}

\begin{figure}
    \centering
    \includegraphics[width=0.64\columnwidth]{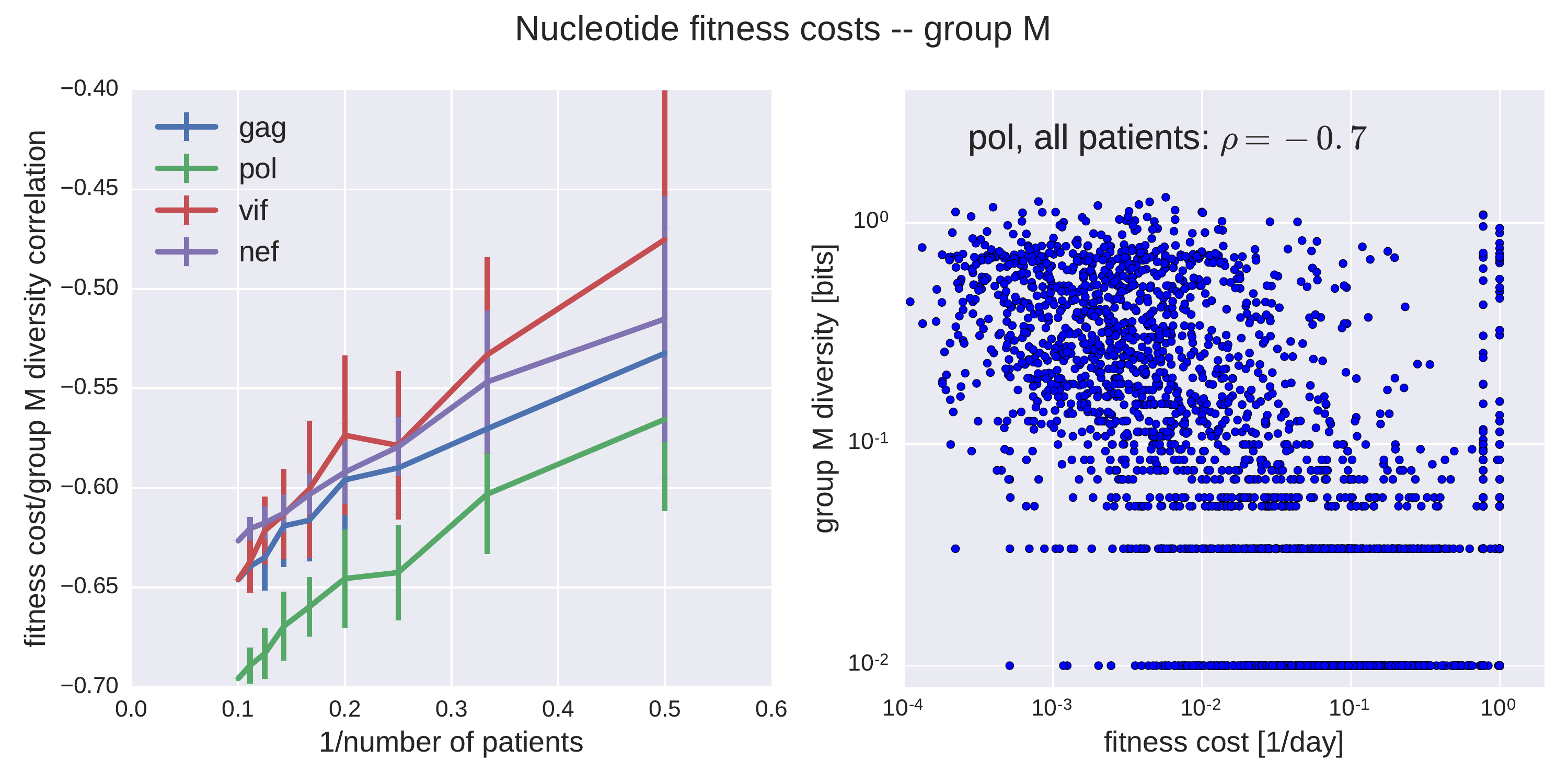}
    \includegraphics[width=0.64\columnwidth]{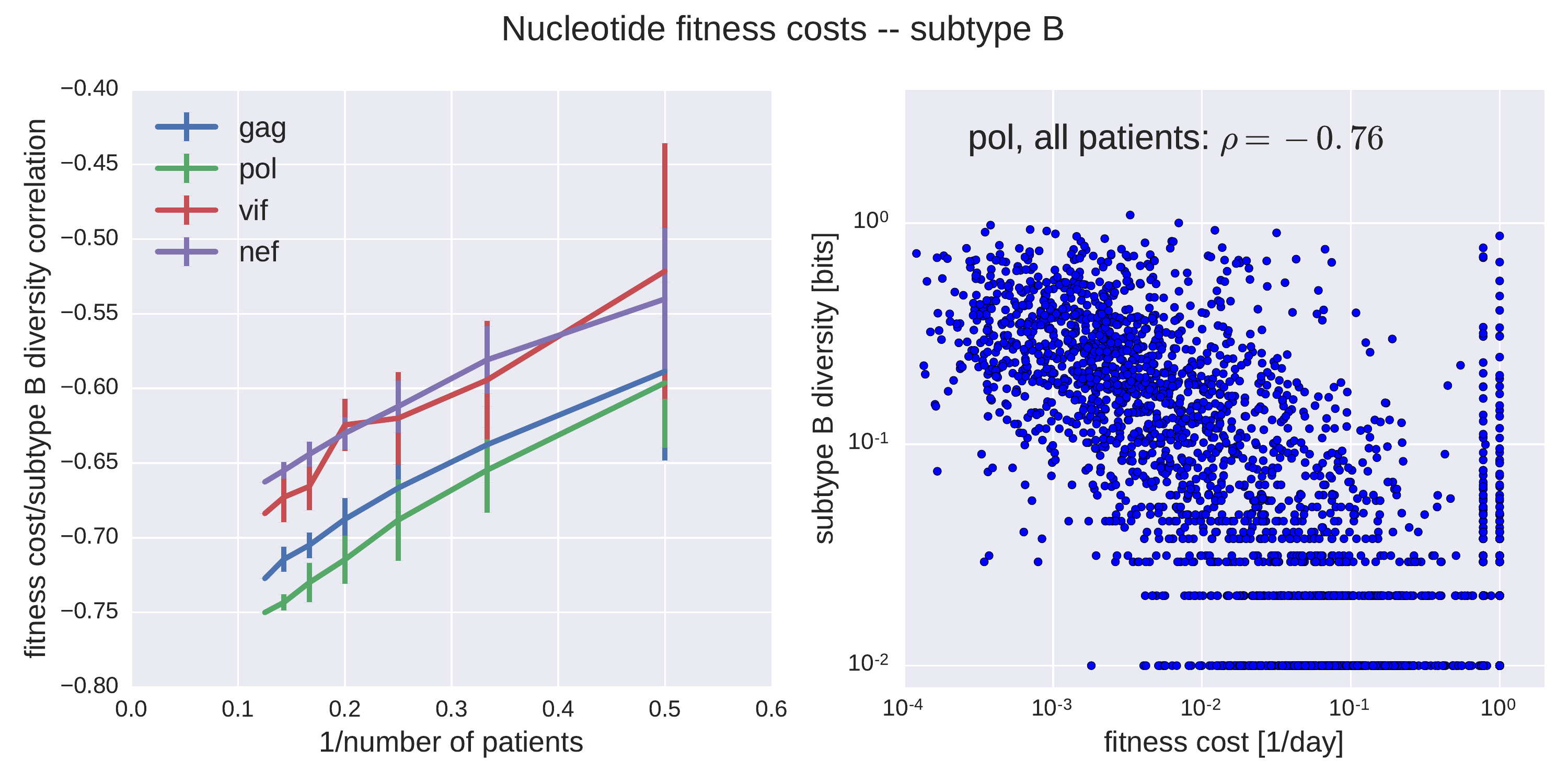}
    \includegraphics[width=0.64\columnwidth]{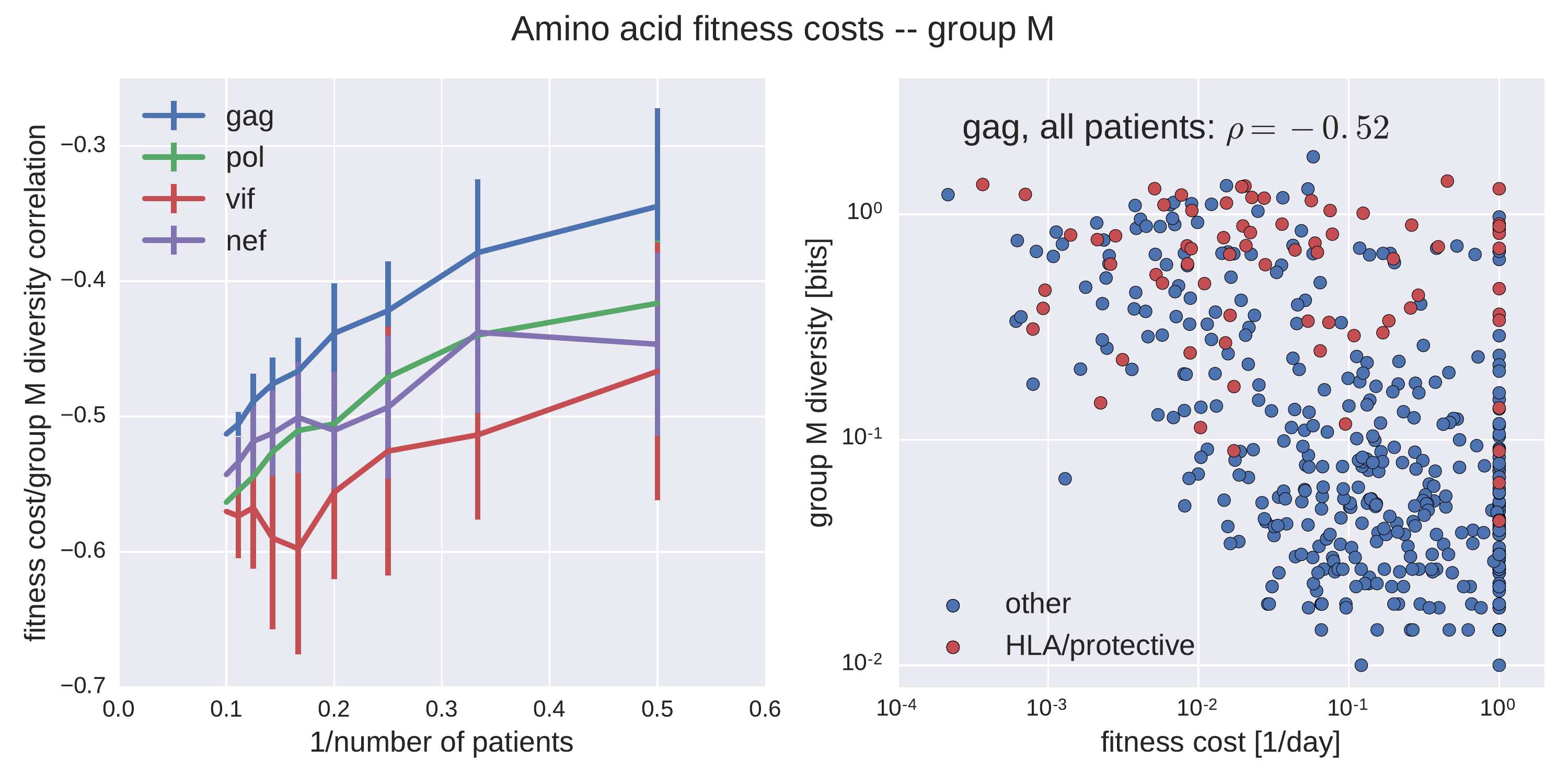}
    \includegraphics[width=0.64\columnwidth]{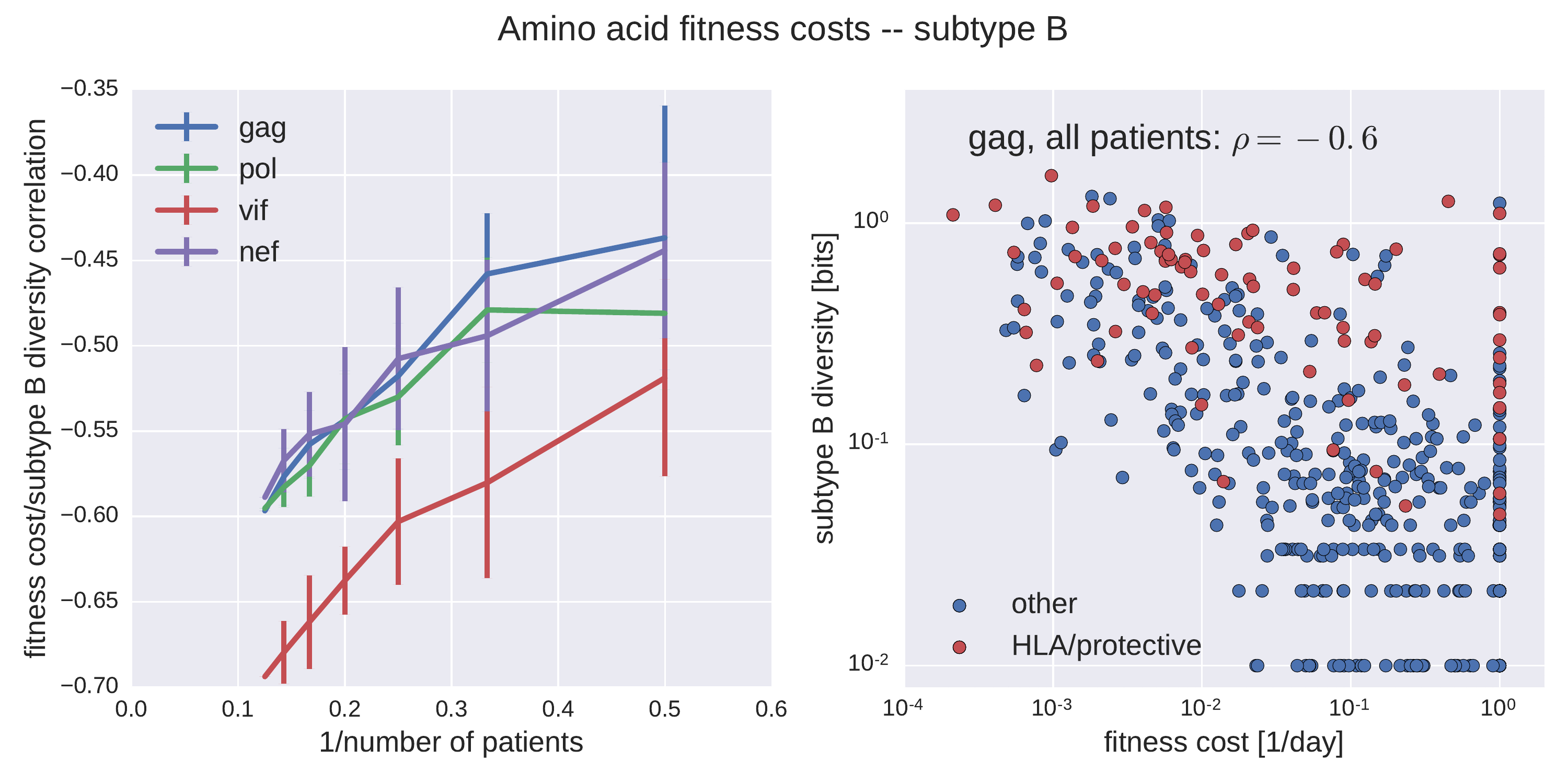}
    \caption{{\bf Correlation of fitness cost with global cross-sectional diversity.} The left panels show how correlation improves as fitness costs are estimates using data from more and more patients. The right panels show a scatter plot of fitness cost vs cross-sectional diversity using data from all patients for one of the proteins. The top panels show costs for nucleotide mutations, the bottom panels for amino acid mutations (and highlight HLA associated of protective sites, \citep{carlson_correlates_2012,bartha_genome_genome_2013}). 
    }
    \label{fig:S_vs_npat}
\end{figure}

\begin{figure}
    \centering
    \includegraphics[width=\textwidth]{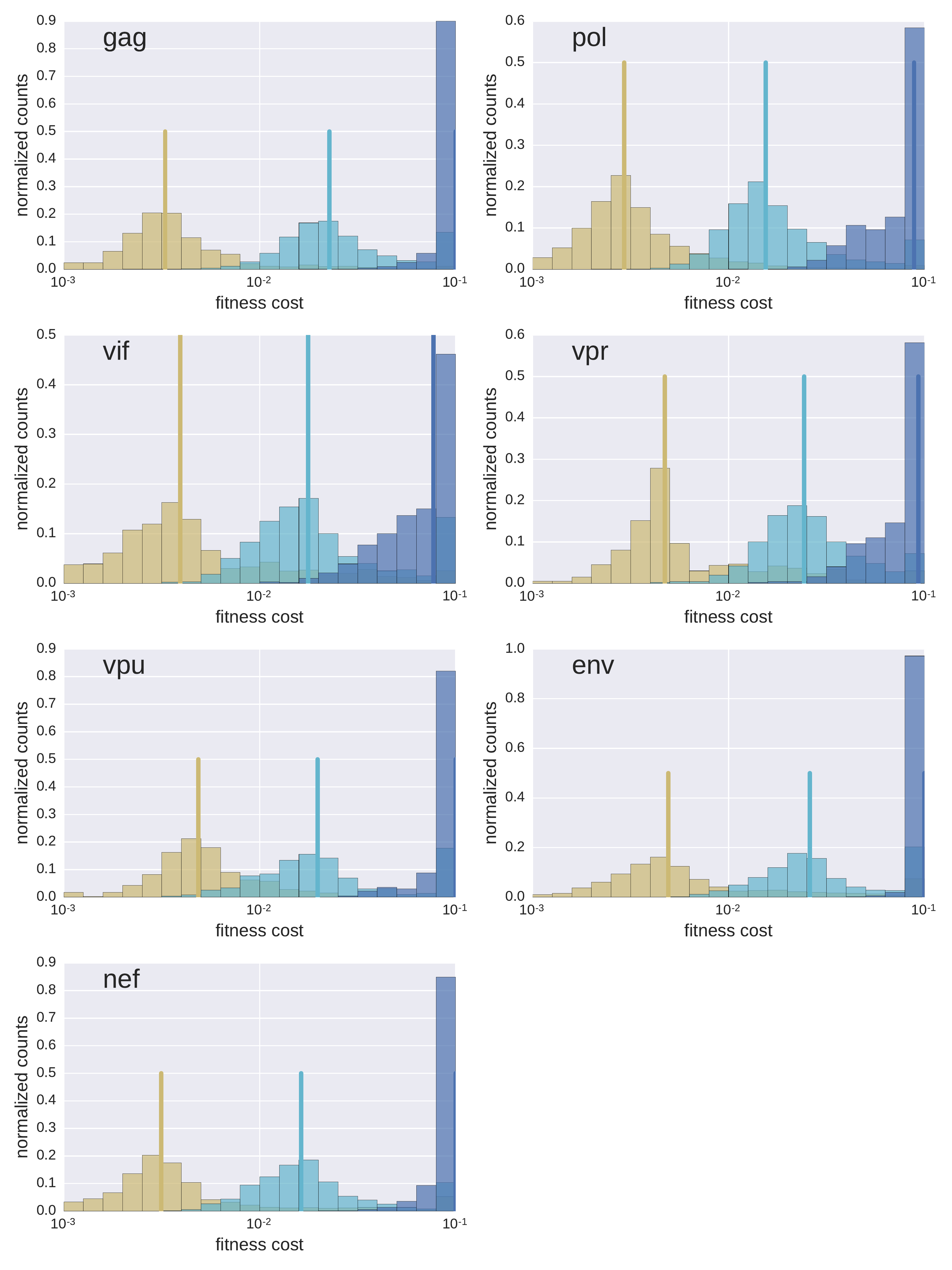}
    \caption{Bootstrap confidence on fitness costs for various regions of the genome.}
    \label{fig:fitness_confidence_other_regions}
\end{figure}

\begin{figure}
    \centering
    \includegraphics[width=0.7\columnwidth]{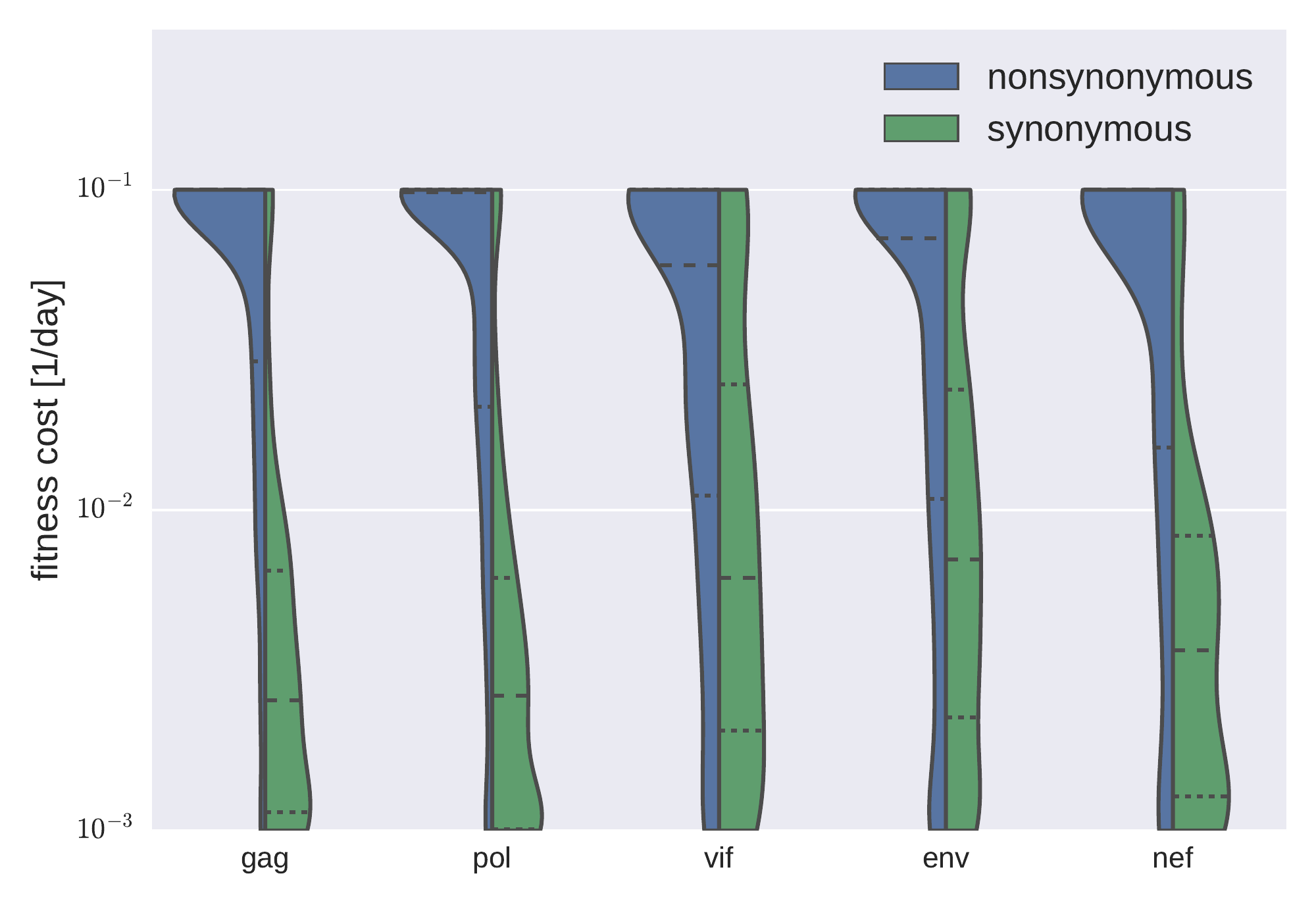}
\caption{{\bf Fitness costs in different genes.} Distribution of fitness costs of synonymous and nonsynonymous mutations in different genes. Note that frequency estimates in gp120 are expected to be less accurate due to consistent difficulties amplifying this part of the genome.}
    \label{fig:gene_distributions}
\end{figure}

\begin{figure}
    \centering
    \includegraphics[width=0.99\columnwidth]{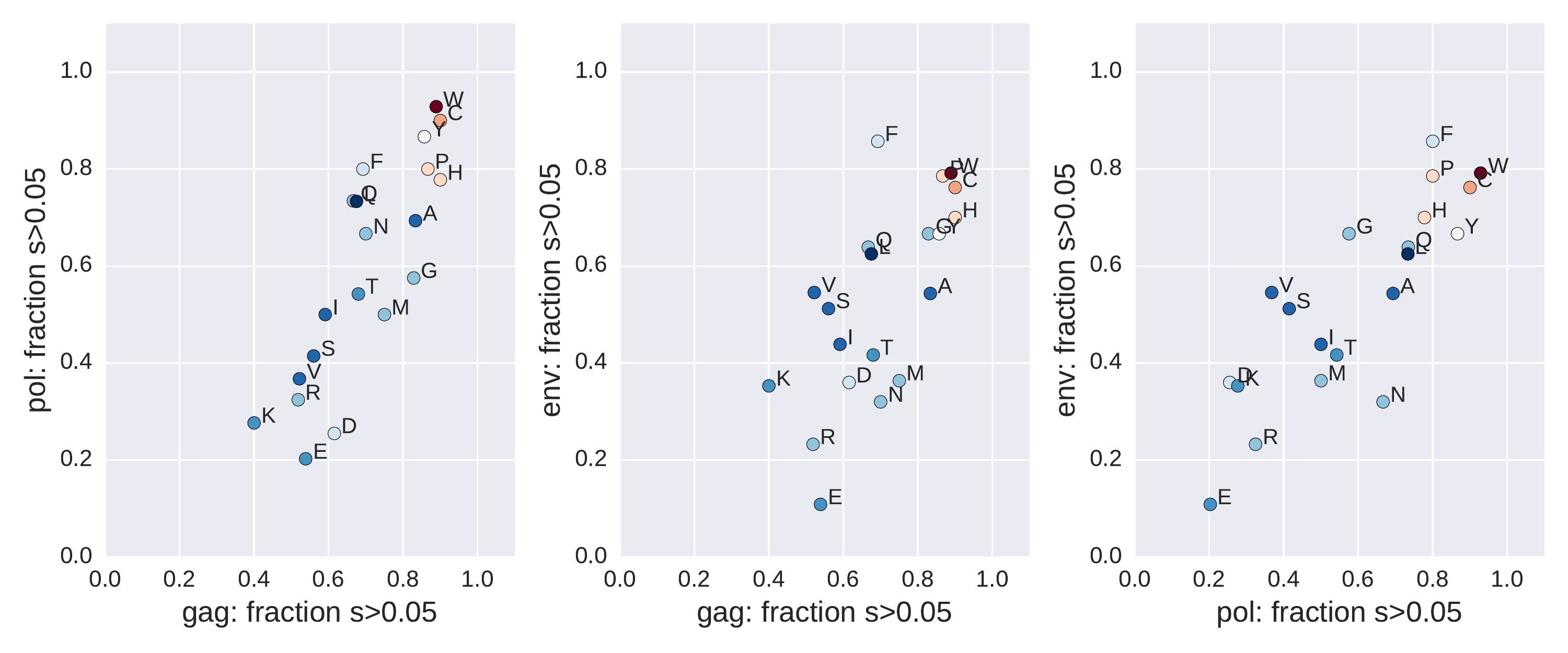}
\caption{{\bf Fitness costs and consensus amino acid.} The fraction of sites with fitness costs $>0.05$ per day depends consistently on the consensus amino acid. Mutations of cysteins (C), histidines (H), prolines (P), tryptophans (W), and tyrosines (Y) tend to be most costly. Points are colored according to the diagonal of the BLOSSUM80 matrix, from blue to white to red, indicating a fair degree of agreement, especially for the most costly residues.}
    \label{fig:amino_acid_costs}
\end{figure}

\begin{figure}
    \centering
    \includegraphics[width=\columnwidth]{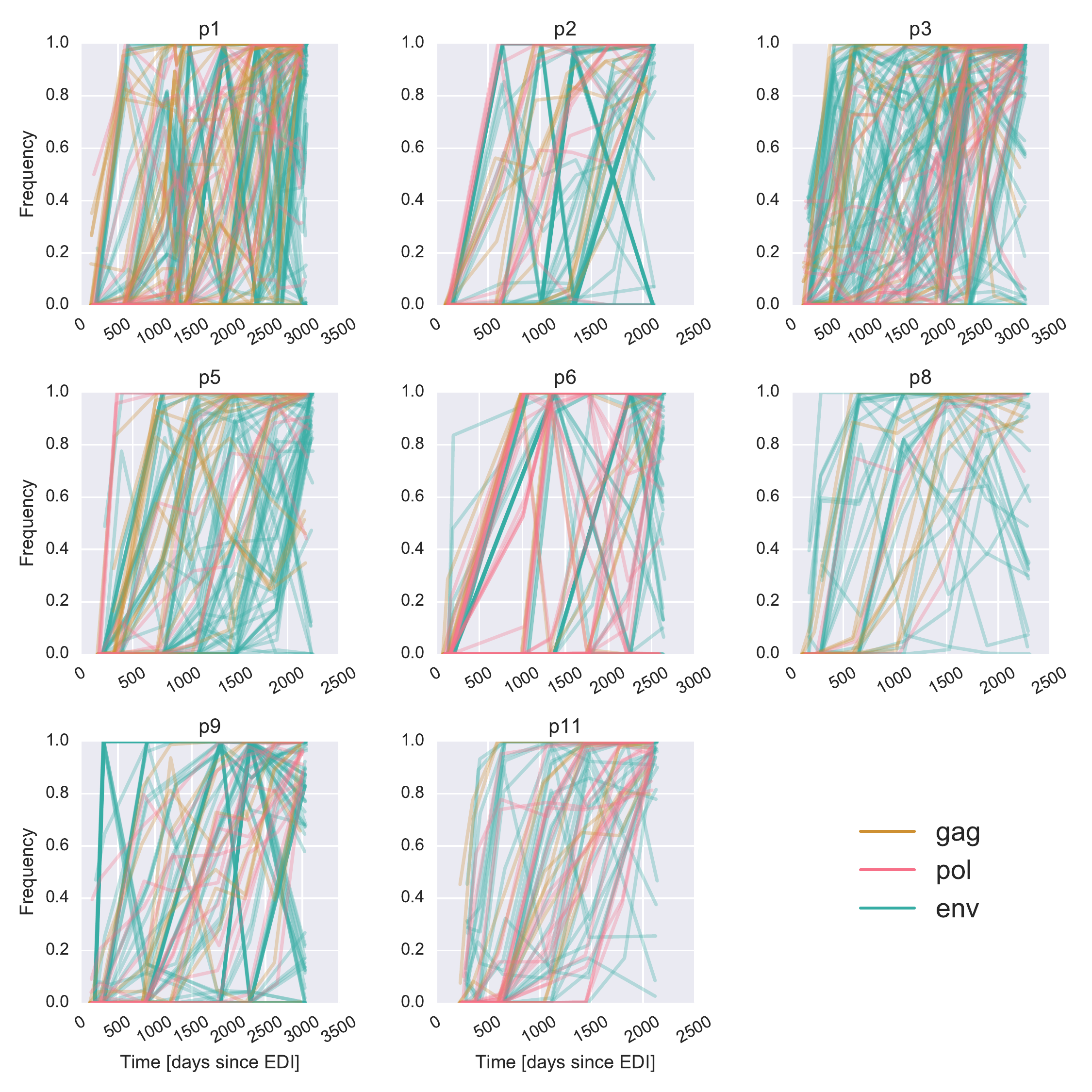}
    \caption{Many mutations sweep across the viral population at the same time. Each panel shows the trajectories of putative selective sweeps in a study patient, i.e. mutations that reach 90\% frequency at least at one time point. These trajectories include not only driver mutations, i.e. beneficially selected for, but also linked passenger mutations, e.g. synonymous mutations or mutations that carry little additional cost. The number of sweeps observed across a whole infection in our patients is as follows: p1, 145; p2, 95; p3, 147; p5, 95; p6, 94; p8, 41; p9, 111; p11, 63. p7 did not yield early samples and is not shown.}
    \label{fig:sweeps_lines}
\end{figure}

\begin{figure}
    \centering
    \includegraphics[width=\columnwidth]{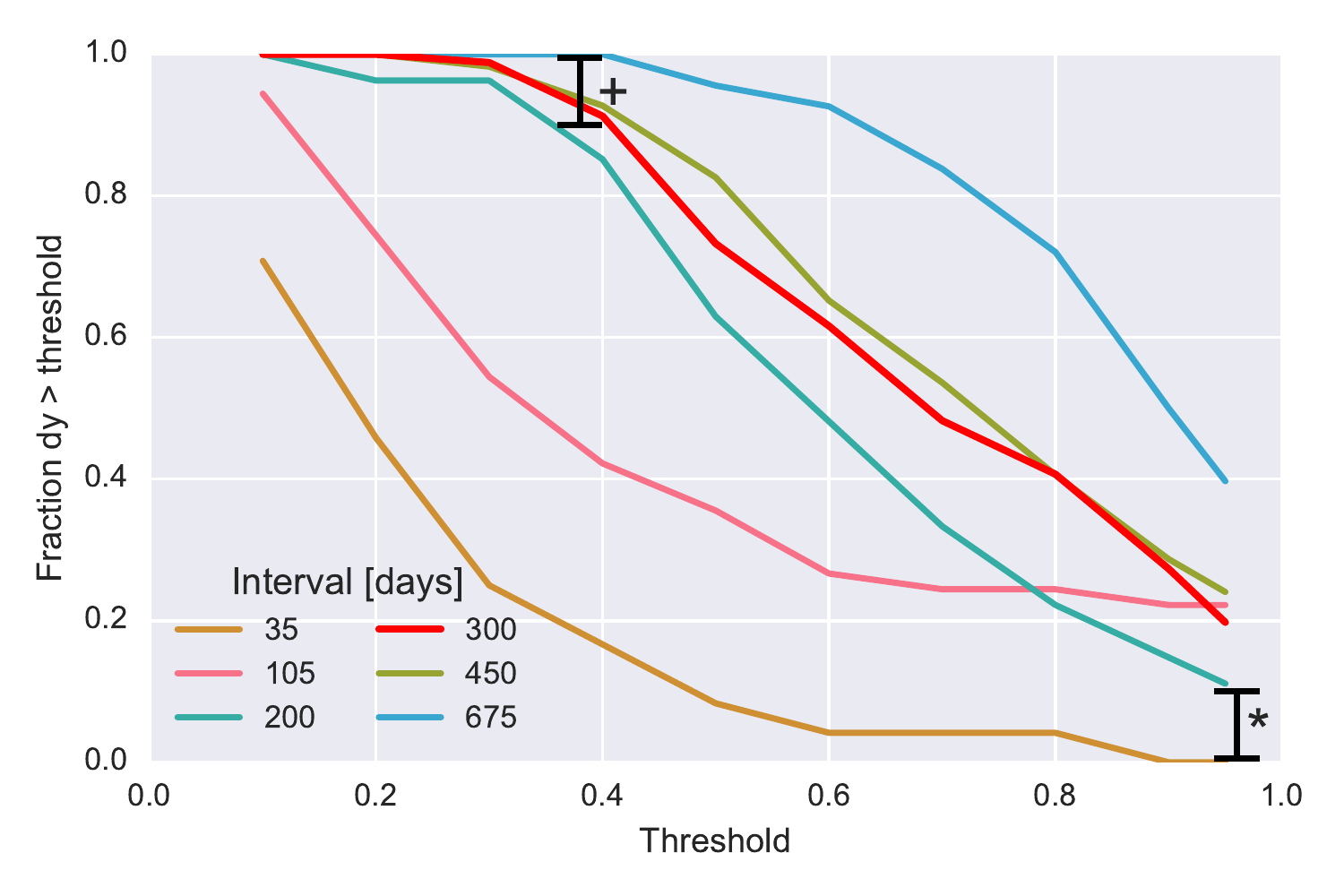}
    \caption{Most sweeps suggest a fitness benefit around 1\% per day. The increase in allele frequency between two consecutive samples is termed \textit{dy}, and the fraction of sweeps with an increase larger than a threshold is shown for different thresholds (x axis) and for pairs of samples at different temporal distance (each line refers to one temporal distance category, as shown in the legend). Around 10\% of sweeps happen in much faster than 200 days (fraction *), around 10\% are much slower than 400 days (fraction +), and 80\% of sweeps take around 300 days, which suggests a fitness benefit of around 1 to 2\% per day, in agreement with previous estimates about chronic infection \citep{neher_recombination_2010}. A similar result is obtained if only nonsynonymous changes are considered.}
    \label{fig:sweeps_quantification}
\end{figure}

\begin{figure}
    \centering
    \includegraphics[width=0.8\columnwidth]{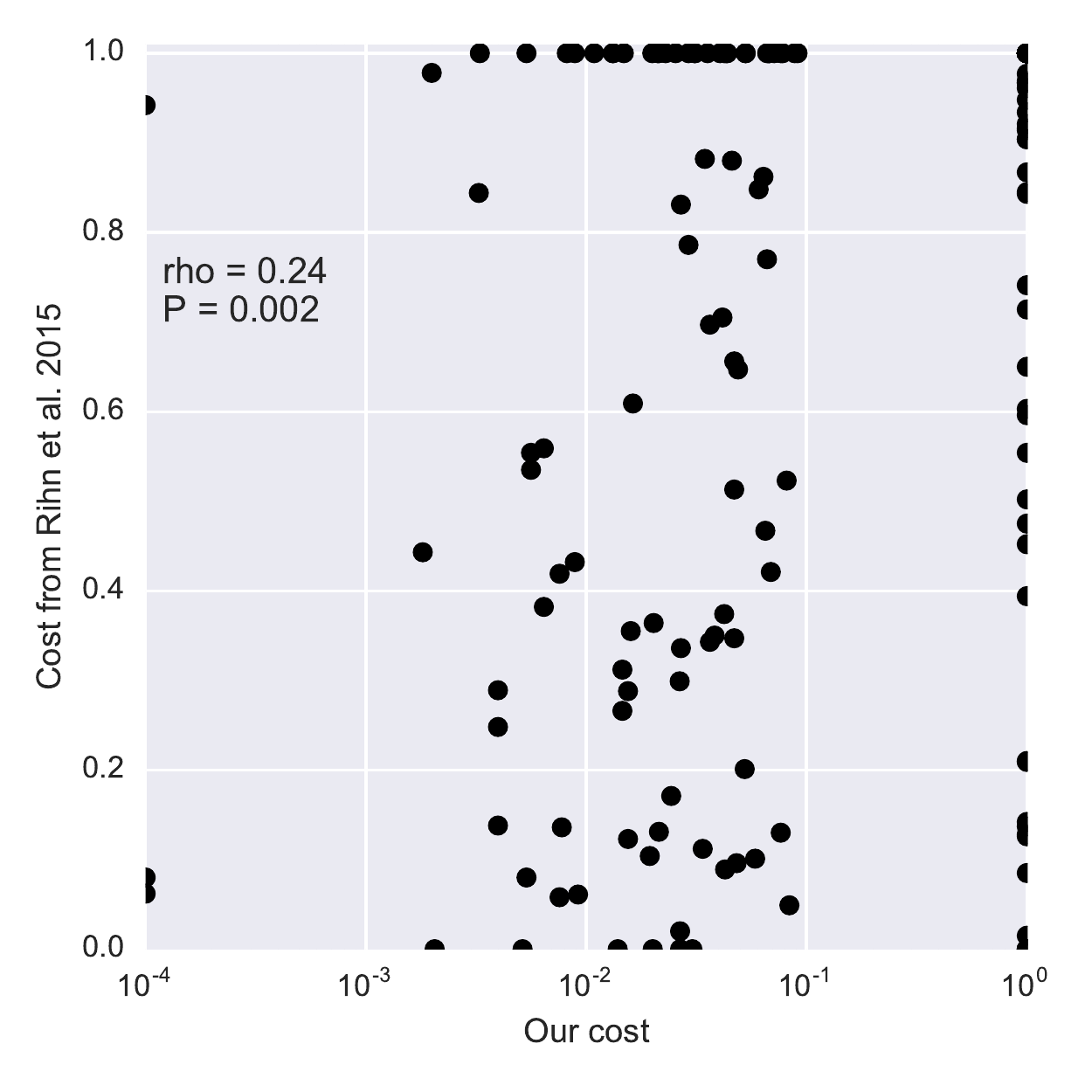}
    \caption{Fitness costs in integrase are correlated with published \textit{in vitro} experiments \citep{rihn_uneven_2015}. The rank correlation coefficient is 0.24 (P-value = 0.002), which indicates a partial agreement between our results and Rihn et al. (2015). There are three reasons why no perfect correlation is expected. First, cell culture fitness determinations are sensitive to costs above 3-5\% whereas our \textit{in vivo} method is accurate between 0.1\% and 10\% approximately. This makes the two approaches nicely complementary in scope. Second, cell cultures are not perfect models of the viral dynamics in a patient, hence some selective pressures might differ. Third, one limitation of our study is that for each site we do not test for a specific mutation, so a few discrepancies might be due to this methodological difference. In cases when \citet{rihn_uneven_2015} tested more than one mutation at a site, the same cost from our table was reused. To further test the significance of the correlation, we repeated the correlation analysis several times after reshuffling sites and costs and found no significant correlation in those cases.}
    \label{fig:comparison_Rihn}
\end{figure}

\begin{figure}
    \centering
    \includegraphics[width=\linewidth]{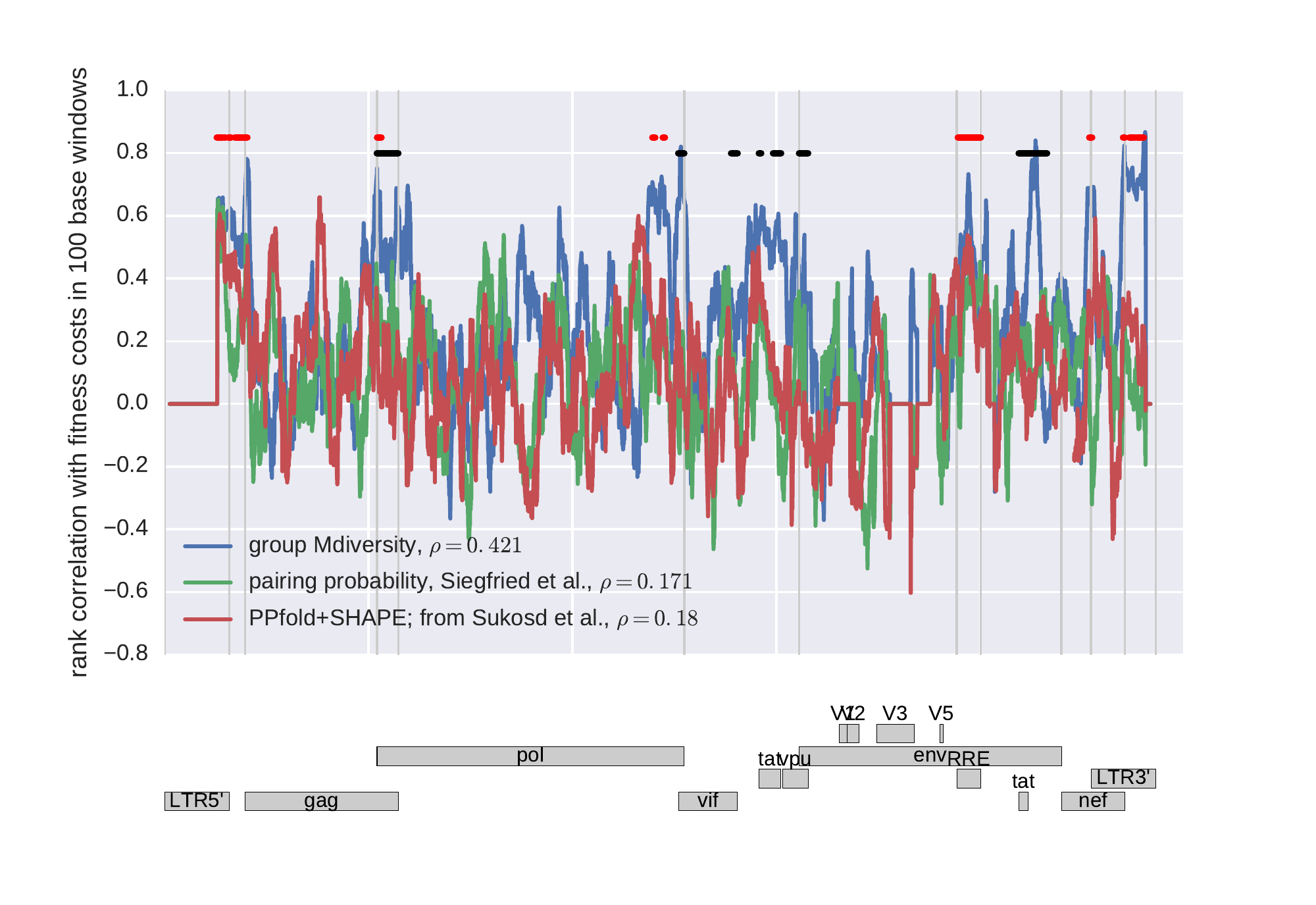}
    \caption{Fitness estimates are at synonymous$^{*}$ sites are well correlated (in sliding 100 bp windows) with group M diversity, but correlation with RNA structure prediction by \citet{siegfried_rna_2014} and \citet{sukosd_full-length_2015} is weaker and limited to a few regions.
    Pronounced peaks of the correlation between diversity and fitness costs at synonymous positions coincide with overlapping reading frames (marked in black in the top part of the figure) and known regulatory elements (marked in red). The strongest correlation is observed in the central and 3' poly purine tracts, around the overlap of gag and pol, and in the 3' LTR.
    The genome wide correlation (given in the legend) is highly significant in all cases but low for RNA structure predictions. $^{*}$ synonymous sites are defined here as those at least the transition does not result in in an amino acid change in \textit{gag}, \textit{pol}, \textit{vif}, \textit{vpu}, \textit{env}, and \textit{nef}.}
    \label{fig:correlation_with_RNAstructure}
\end{figure}

\begin{figure}
    \centering
    \includegraphics[width=0.6\linewidth]{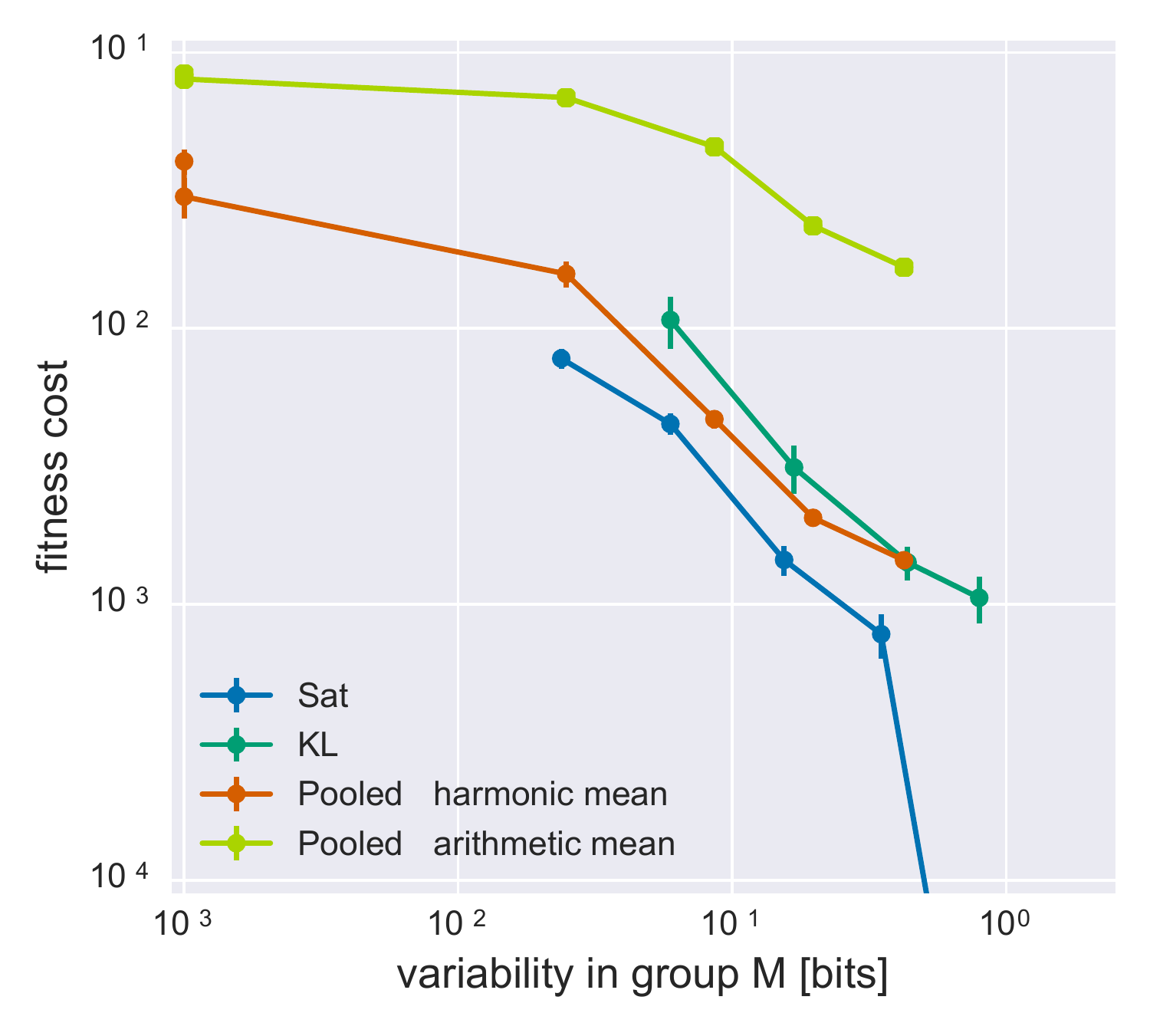}
    \caption{Fitness cost estimates based on temporal correlations of allele frequencies are consistent with the saturation and the pooled estimates. The "Sat" and "Pooled" curves are like in \FIG{sat}, the "KL" curve uses the estimate method based on minimization of Kullback-Leibler divergence (see below). The arithmetic mean of the "Pooled" estimate is higher than the harmonic mean, as expected; this quantity basically describes the fraction of polymorphisms within each group of sites.
    }
    \label{fig:sat_KL_pooled}
\end{figure}

\clearpage
\subsection*{Estimation of selection coefficients by Kullback-Leibler divergence minimization}
In addition to the two modelling methods presented in \FIG{sat}, "Sat" and "Pooled", we tested a third approach that exploits the time information of samples (like the "Sat" method) but also models the temporal correlations of SNP frequencies (see \FIG{sat_KL_pooled}). These correlations are not accounted for in the ``Sat'' fitting procedure which simply fits average values for each bin.

We capture the correlation structure of the SNP frequency trajectories by modelling the full probability distribution $P(\mathbf{x})$ of observing all SNPs from all times at a certain combination of frequencies:
\[
\mathbf{x} = (x_{t,i} \dots),
\]
where $t$ indicates each time point and $i$ each conservation group. We combine all SNP trajectories (summed minor derived states) of all sites within one conservation quantile into $\mathbf{x}$, separately for each patient. We approximate the joint probability distribution $P(\mathbf{x})$ by a theoretical distribution $W(\mathbf{x})$ that is the solution of the stochastic equation \eqref{eq:Langevin} with a constant diffusive noise term $\eta(t)$ to make it mathematically tractable
\[
\left< \eta^2(t) \right> \propto Dt.
\]
where $D$ defines the noise intensity. The solution of eq. \eqref{eq:Langevin} under these simplifying assumptions is a multivariate Gaussian distribution: 
\bq
W(\mathbf{x}) =
\frac{\exp\left[-\frac{1}{2}(\mathbf{x}-\langle \mathbf{x}\rangle)^T
K^{-1}(\mathbf{x}-\langle \mathbf{x}\rangle)\right]}{\sqrt{(2\pi)^{N}\det{K}}},
\label{eq:W}
\eq
where $K$ is the covariance matrix of SNP frequencies. Mean and covariance of $W(\mathbf{x})$ are given respectively by
\bn 
\langle x(t)\rangle &=& \frac{\mu}{s}\left(1-e^{-st}\right),\nonumber\\
K(t,t') &=& \frac{D}{s}\left[e^{-s|t-t'|} -e^{-s(t+t')}\right],
\en
We now want to estimate the parameters $s$ and $D$ from the data while keeping $\mu$, the mutation rate, fixed at the measured value $1.2 \cdot 10^{-5}$ per day per site. To this end, we construct an empirical distribution of SNP frequency trajectories as a multivariate Gaussian with mean and covariances obtained by averaging the data across sites:
\bn 
\xavg(t) &=& \frac{1}{L}\sum_k x_k(t),\\
\kappa (t_i,t_j) &=& 
\frac{1}{L-1}\sum_k \left[x_k(t_i) - \xavg(t_i)\right] 
\left[x_k(t_j) - \xavg(t_j)\right].\nonumber
\en
Here $k$ is the site/position index, the $\xavg$ designates average minor SNP frequency in the conservation quantile analysed, $t_i$ and $t_j$ are time points along the trajectory, and $L$ is the number of sites used in the average.

Mean and covariance fully determine the empirical Gaussian distribution, so we can extract the best model parameters by minimizing the distance of this distribution and the theoretical one. A convenient measure of the divergence between the two distributions 
is so-called Kullback-Leibler divergence, defined as
\bq
KL = \int  
P(\mathbf{x})\log\left[\frac{P(\mathbf{x})}{W(\mathbf{x})}\right] d\mathbf{x}.
\eq
Averaging over the empirical distribution $P(\mathbf{x})$ is now equivalent
to averaging over sites, which allows us to write the Kullback-Leibler divergence (KL) as
\bn\label{eq:KL}
KL&=& C - \frac{1}{L}\log W(\mathbf{x})= 
C+ \log{\sqrt{(2\pi)^N\det K}}  \nonumber\\
&& + \frac{1}{2} \sum_{i,j}\left\{ \left[\xavg(t_i) - \langle x(t_i)\rangle\right] 
(K^{-1})_{ij}\left[\xavg(t_j) - \langle x(t_j)\rangle\right] +
(K^{-1})_{ij}\kappa_{ji}\right\}.
\en
Finally, we notice that for different conservation groups, the KL is additive. We can thus sum over all conservation groups to esitmate all $s$ and $D$ parameters simultaneously (one $s$ and one $D$ per group). The resulting values for $s$ are shown in \FIG{sat_KL_pooled} as the "KL" curve and is in good agreement with the two previous methods used to estimate average fitness costs.

\end{document}